\documentclass[aip,amsmath,amssymb,reprint]{revtex4-2}

\usepackage{graphicx}
\usepackage{dcolumn}
\usepackage{bm}

\usepackage[version=3]{mhchem} 
\usepackage{xcolor}
\usepackage{svg}
\usepackage{amsmath}
\usepackage{tabularray}

\usepackage[utf8]{inputenc}
\usepackage[T1]{fontenc}
\usepackage{mathptmx}
\usepackage{etoolbox}

\makeatletter
\def\@email#1#2{%
 \endgroup
 \patchcmd{\titleblock@produce}
  {\frontmatter@RRAPformat}
  {\frontmatter@RRAPformat{\produce@RRAP{*#1\href{mailto:#2}{#2}}}\frontmatter@RRAPformat}
  {}{}
}%
\makeatother

\begin{document}

\preprint{xxx}

\title[NMR modes]{Theory and modeling of molecular modes in the NMR relaxation of fluids$^\dag$}

\author{Thiago J. Pinheiro dos Santos}
\affiliation{Department of Chemical and Biomolecular Engineering, Rice University, Houston, Texas 77005, USA.}
\author{Betul Orcan-Ekmekci}
\affiliation{Department of Mathematics, Rice University, Houston, Texas 77005, USA.}
\author{Walter G. Chapman}
\affiliation{Department of Chemical and Biomolecular Engineering, Rice University, Houston, Texas 77005, USA.}
\author{Philip M. Singer}
\affiliation{Department of Chemical and Biomolecular Engineering, Rice University, Houston, Texas 77005, USA.}
\email{ps41@rice.edu}
\author{Dilipkumar N. Asthagiri}
\affiliation{Oak Ridge National Laboratory, Oak Ridge, Tennessee 37830, USA.}
\email{asthagiridn@ornl.gov}

\date{\today}

\begin{abstract} 
Traditional theories of the NMR autocorrelation function for intramolecular dipole pairs assume single-exponential decay, yet the calculated autocorrelation of realistic systems display a rich, multi-exponential behavior resulting in anomalous NMR relaxation dispersion (i.e., frequency dependence). We develop an approach to model and interpret the multi-exponential autocorrelation using simple, physical models within a rigorous statistical mechanical development that encompasses both rotational and translational diffusion in the same framework. We recast the problem of evaluating the autocorrelation in terms of averaging over a diffusion propagator whose evolution is described by a Fokker-Planck equation. The time-independent part admits an eigenfunction expansion, allowing us to write the propagator as a sum over modes. Each mode has a spatial part that depends on the specified eigenfunction, and a temporal part that depends on the corresponding eigenvalue (i.e., correlation time) with a simple, exponential decay. The spatial part is a probability distribution of the dipole-pair, analogous to the stationary states of a quantum harmonic oscillator. Drawing inspiration from the idea of inherent structures in liquids, we interpret each of the spatial contributions as a specific molecular mode. These modes can be used to model and predict NMR dipole-dipole relaxation dispersion of fluids by incorporating phenomena on the molecular level. We validate our statistical mechanical description of the distribution in molecular modes with molecular dynamics simulations interpreted without any relaxation models or adjustable parameters: the most important poles in the Pad{\'e}-Laplace transform of the simulated autocorrelation agree with the eigenvalues predicted by the theory.
\newline

\noindent \footnotesize{$^\dag$Notice: This manuscript has been authored by UT-Battelle, LLC, under contract DE-AC05-00OR22725 with the US Department of Energy (DOE). The US government retains and the publisher, by accepting the article for publication, acknowledges that the US government retains a nonexclusive, paid-up, irrevocable, worldwide license to publish or reproduce the published form of this manuscript, or allow others to do so, for US government purposes. DOE will provide public access to these results of federally sponsored research in accordance with the DOE Public Access Plan (http://energy.gov/downloads/doe-public-access-plan).}
\end{abstract}

\maketitle
 
\section{Introduction}

Atomic nuclei with non-zero spin magnetic moment ($I \geq 1/2$) can be excited and driven out of equilibrium by a resonating radio-frequency electromagnetic radiation in the presence of an external magnetic field $B_0$. When the radio-frequency perturbation is removed, the relaxation of the spins back to the initial equilibrium with $B_0$ provides important insights into the structure and behavior of matter. This process of nuclear magnetic resonance (NMR) relaxation has thus emerged as a powerful tool in probing matter non-destructively, with applications spanning hydrocarbon recovery to material science \cite{Bruce2000,Edwin2000,Price2009}. It is known that the NMR relaxation at different frequencies $\omega_0 = \gamma_I B_0$ (a.k.a., NMR relaxation dispersion) is governed by  phenomena at different length and time scales \cite{Moniz1963,Price2009}. The NMR dipole-dipole relaxation that is of interest here -- either ``like spins'' (e.g., $^1$H-$^1$H pairs) or ``unlike spins'' (e.g., $^{13}$C-$^1$H pairs, or alternatively e$^{-}$-$^1$H pairs) -- is sensitively dependent on the translational and rotational diffusion of the species in the system. Further, since the perturbation by the external applied field is small relative to thermal energies, it is possible to use the unperturbed Hamiltonian of the system to model the evolution of the system and then use that information to study the NMR relaxation. Molecular dynamics (MD) simulations have proven successful in predicting NMR dipole-dipole relaxation of bulk, viscous, and confined fluids in good to excellent agreement with experiments for both like spins \cite{Singer2017,Singer2018,singer:jcp2018b,asthagiri:jpcb2020,Singer2020,Parambathu2020,valiyaparambathu:jpcl2023} and unlike spins (at high frequencies) \cite{Singer2021,Pinheiro2022}, all without using adjustable parameters to interpret the simulations or by coarse-graining the dynamics. Besides us, other groups have also presented compelling studies exploring NMR relaxation using molecular simulations \cite{bruschweiler:jacs1992,lindgren:pccp2009,faux:pre2015,becher:jcp2021,madhavi:jml2021, beckmann:JPCB2022,rybin:pre2022, wang:CS2023, philips:jpcb2023, estrada:ef2023, gravelle:ef2023,wang:jpcc2023,wang:jpcc2023b}. 

The original theory of intramolecular NMR relaxation, due to Bloembergen, Purcell, and Pound (BPP) \cite{bloembergen:pr1948}, assumes that relaxation arises from the rotational diffusion of the molecule. BPP assumed that the molecules behaved as hard-spheres and their rotational motion in the liquid was random. By estimating the rotational correlation time $\tau_d$ based on an extension of Debye's theory of dielectric dispersion, BPP predicted that the NMR autocorrelation function $G(t)$ for dipole pairs decays mono-exponentially, resulting in a specific functional form (namely, a Lorenztian) for NMR relaxation dispersion. From early on, it was well-appreciated that the BPP model or its derivatives were oversimplified, and efforts were made to address the shortcomings  \cite{Woessner1965,Steele1963a,Brownstein1979,Lipari1982,Lopez1987}.  In practice, however, to circumvent the shortcomings it is more common to construct an empirical (i.e., phenomenological) model for the NMR relaxation dispersion, such as the Kohlrausch-Williams-Watts function \cite{williams:tfs1970}, Cole-Davidson function \cite{davidson:jcp1951} or Lipari-Szabo \cite{lipari:jacs1982} model for like spins, or, the extended Solomon-Bloembergen-Morgan model \cite{solomon:pr1955,Bloembergen1961,wahsner:cr2019} for unlike spins.

Meanwhile, Torrey \cite{Torrey1953} addressed the role of translational diffusion on intermolecular NMR relaxation by using a random walk model. This led to a single characteristic translational relaxation time $\tau_T$ to be associated with the diffusion correlation time $\tau_D$. Later, Hwang and Freed improved Torrey's model by taking the structure of the fluid into account\cite{Hwang1975}; specifically, they considered a hard-sphere fluid or a dilute electrolyte for which models of pair-correlations are available. 

Woessner took the pioneering first steps to understand the role of molecular conformations to induce multi-exponential behavior. He studied the diffusional motion of spherical and ellipsoidal objects within which the magnetic dipoles (of either fixed or varying separation and orientation) were embedded \cite{Woessner1962a,Woessner1962b}. Woessner attributed the multiple correlation times to different internal conformations, although by construction his formalism would be restricted to a few discrete configurations \cite{Woessner1962a,Woessner1962b,Woessner1965,Woessner1972}.
Steele used a classical mechanical approach to model molecular reorientations, specifically by modeling the torque on molecules by a constant friction plus a random contribution
\cite{Steele1963a,Steele1963b,Moniz1963}. A different approach based on using an analytical diffusion propagator was successfully applied to derive pulsed gradient spin-echo attenuation equations \cite{Ghadirian2013}, NMR relaxation under anomalous diffusion \cite{Zavada1999}, and relaxation of dipole pairs in planar \cite{Price2009,Neuman1974}, cylindrical \cite{Ghadirian2013,Price2009}, and spherical pores \cite{Neuman1974,Ghadirian2013,Grebenkov2008,Price2009} with different boundary conditions.

In our earlier studies, instead of fitting the NMR autocorrelation function $G(t)$ to an empirical model with adjustable parameters, we sought to expand $G(t)$ for the case of regular diffusion into multiple ``modes" 
\begin{equation}
    G(t) = \int_0^{\infty} \exp \left( -\frac{t}{\tau} \right) P(\tau) d\tau ,
    \label{eq:expansion}
\end{equation}
where each ``mode" $\tau$ has a mono-exponential decay and $P(\tau)$ is the probability density distribution of the time constants of the ``modes". The intuition guiding this approach was that there are inherent dynamical structures in the fluid that define each mode. Note also that with such an expansion, the NMR relaxation dispersion is given by BPP-like terms weighted by $P(\tau)$, and can potentially give physical insights into how specific molecular motions affect relaxation. 

In this work, we establish a theoretical framework for the expansion (Eq.~\ref{eq:expansion}) and derive a clearer physical meaning into the ``modes." To this end, we describe the NMR relaxation between two dipoles using classical statistical mechanics. Instead of assuming discrete characteristic conformations, we solve the Fokker-Planck equation to obtain the phase distribution function, from which we calculate the autocorrelation function $G(t)$ and NMR relaxation dispersion. For specific cases, we can obtain analytical solutions and these turn out to obey the multi-exponential behavior defined above. We use MD simulations to validate the theory and guide the physical interpretation of the simulation results. For dipoles that are constrained to remain at a fixed separation, the multi-exponential behavior collapses into a  mono-exponential behavior, in agreement with the traditional BPP theory.

\section{Theory}

Let us assume two dipoles undergoing relaxation in the presence of an external magnetic field $B_0$ in a viscous fluid. For convenience, we fix the first dipole at the center of the coordinates, and the other dipole is free to move around the first. 
The NMR autocorrelation function $G^m(t)$ resulting from fluctuating magnetic fields of dipole pairs is given by \cite{McConnell2009,Singer2017,Singer2021}
\begin{equation}
    G^m(t) = \left< \frac{Y_2^m \left( \theta(t), \phi (t) \right)}{r^3(t)} \frac{Y_2^m \left( \theta(0), \phi (0) \right)}{r^3(0)} \right>, \label{eq:Gm}
\end{equation}    
where $\theta(t)$ and $\phi(t)$ are the angles formed by the two magnetic dipoles and the external magnetic field $B_0$. A constant factor depending on the spins has been suppressed for clarity. We will return to this in the Section III, Simulation Methodology. $Y_2^m$ is the spherical harmonic of degree 2, $r(t)$ is the distance between particles $I$ and $S$.
For isotropic systems, as is the case in this work, $G^m(t)$ is independent of the order $m$, which is equivalent to saying that the direction of the applied static magnetic field $B_0$ is arbitrary. Without loss of generality, we therefore choose to compute only the $m=0$ case, and from here on refer only to $G(t)$ where $m=0$ is implied. Also, in the following, for concision, we will use $Y^0_2(t) \equiv Y^0_2(\theta(t),\phi(t))$. 

The Fourier transform of $G(t)$, for different Larmor frequencies $\omega_0 = \gamma_I B_0$, gives the spectral density function $J(\omega)$ through the relationship \cite{Singer2017}
\begin{equation}
    J(\omega) = 2 \int_0^{\infty} G(t) \cos (\omega t) dt, \label{eq:Jomega}
\end{equation}
which is used to calculate the relaxation times $T_1$ and $T_2$ at different NMR frequencies $\omega_0 = \gamma_I B_0$, a.k.a. relaxation dispersion, for like spins \cite{Singer2020} or unlike spins \cite{Pinheiro2022}.

The auto-correlation in Eq. \eqref{eq:Gm} can be obtained once we know the diffusion propagator 
\begin{equation}
    \begin{split}
    \rho(\textbf{r},\textbf{r}_0,t) = \rho &\left( \textbf{r},t|\textbf{r}_0,0 \right) p(\textbf{r}_0),
    \end{split}
\end{equation}
for the probability of traversing $\textbf{r}_0$ to $\textbf{r}$ in time $t$; here, $\textbf{r}=\{r,\theta,\phi\}$ (in spherical coordinates). Notice that $p(\textbf{r}_0)$ is the prior probability of the initial state, where $\rho \left( \textbf{r},t|\textbf{r}_0,0 \right)$ is the transition probability \cite{Bruce2000,Ghadirian2013}. In the NMR literature, $\rho(\textbf{r},t|\textbf{r}_0,0)$ is also referred as the diffusion propagator or the Green function \cite{Price2009,Ziener2015}. Observe that the quantity $\rho(\textbf{r},\textbf{r}_0,t)$ is hence a phase probability density function, and $\rho(\textbf{r},\textbf{r}_0,t) d\textbf{r} d\textbf{r}_0$ is the joint probability of having the dipole with configuration $\textbf{r}_0$ in $d\textbf{r}_0$ initially and with configuration $\textbf{r}$ in $d\textbf{r}$ at any time $t$ \cite{Bruce2000}.

Once $\rho(\textbf{r},\textbf{r}_0,t)$ is available, we can rewrite the ensemble average in Eq.\ \eqref{eq:Gm} as an integration over the diffusion propagator as \cite{Bruce2000,Ziener2015}
\begin{equation}
    \begin{split}
        G(t) = \int d\textbf{r} \int d\textbf{r}_0 \frac{Y_2^0(t)}{r^3(t)} \frac{Y_2^0(0)}{r^3(0)} \rho(\textbf{r},\textbf{r}_0,t)\, , \label{eq:Gm_ACF} 
    \end{split}
\end{equation}
where, $d\textbf{r} = r^2 \sin\theta dr\, d\theta\, d\phi$ (and likewise for $d\textbf{r}_0$), and the limits of integration over $r$ or $r_0$ ranges from $r_i$ (the initial radial location) to $r_f$ (the final radial location). 

The evolution of $\rho(\textbf{r}, \textbf{r}_0,t)$ can be described by the Fokker-Planck equation \cite{Caldas2014,Ghadirian2013,Hwang1975}, in the form that already assumes the fluctuation–dissipation theorem. For notational simplicity, denoting $\rho(\textbf{r},t)\equiv \rho(\textbf{r},\textbf{r}_0,t)$, we have 
\begin{equation}
    \frac{\partial}{\partial t} \rho(\textbf{r},t) = \nabla^2 \left [ D \rho(\textbf{r},t) \right] - \nabla \cdot \left[ \rho(\textbf{r},t) \frac{D}{k_B T} F(\textbf{r},t) \right], \label{eq:Gen_FP}
\end{equation}
where $D$ is the diffusion coefficient of the bulk fluid (i.e., without confinement), $k_B$ is the Boltzmann constant, $T$ is temperature, and $F(\textbf{r},t)$ is the force acting on the second dipole. Here we assume the Einstein-Smoluchowski relation for the friction forces acting on the dipole \cite{Piasecki2007}. 
For the particular case of a stationary potential of interaction $U(\textbf{r})$  between the dipoles, we know that
\begin{equation}
    F(\textbf{r},t) = - \nabla U(\textbf{r}),
\end{equation}
which allows us to rewrite Eq. \eqref{eq:Gen_FP} as
\begin{equation}
    \frac{\partial}{\partial t} \rho(\textbf{r},t) = \nabla^2 \left [ D \rho(\textbf{r},t) \right] + \nabla \cdot \left[ \rho(\textbf{r},t) \frac{D}{k_B T} \nabla U(\textbf{r}) \right]. \label{eq:Gen_FP_2}
\end{equation}
This equation should be solved subject to periodic boundary conditions $\rho(r,\theta,\phi,t)=\rho(r,\theta,\phi+2\pi,t)$, proper radial boundary conditions  for the inner radius $r_i$ and outer radius $r_f$, impenetrability condition $\rho(0,\theta,\phi,t)=0$, and initial condition $\rho(r,\theta,\phi,0)=p(r_0,\theta_0,\phi_0)$.
At this point, we observe that we can express the solution to the problem in terms of the matrix exponential operator
\begin{equation}
        \rho(\textbf{r},t) = \exp \left[ - Dt \nabla^2 + \frac{Dt}{k_B T} \nabla \cdot \left( \nabla U(r,\theta,\phi) \right) \right] \rho(\textbf{r},0). \label{eq:Gen_sol}
\end{equation}
where $\rho(\textbf{r},0)$ is the initial distribution.
Substituting Eq.\ \eqref{eq:Gen_sol} into  Eq.\ \eqref{eq:Gm_ACF}, we have 
\begin{widetext}
\begin{equation}
        G(t) = \int d\textbf{r} \int d\textbf{r}_0 \frac{Y_2^0(t)}{r^3(t)}  \frac{Y_2^0(0)}{r^3(0)} \exp \left[ -Dt\nabla^2 + \frac{Dt}{k_B T} \nabla \cdot \left( \nabla U(\textbf{r}) \right) \right] \rho(\textbf{r},0). \label{eq:Gm_ACF_2}
\end{equation}    
\end{widetext}

Notice that explicit time dependence only appears in the exponential factor, which then emphasizes that 
$G(t)$ can be represented by the generic form
\begin{equation}
    \begin{split}
        G(t) &= \sum_{k=1}^{\infty} P(\tau_k) \exp \left( - \frac{t}{\tau_k} \right), \label{eq:Ptau}
    \end{split}
\end{equation}
in which each characteristic correlation time $\tau_k = \tau_k(r,\theta,\phi,r_0,\theta_0,\phi_0)$. However, except for very simple cases, the analytical transformation from Eq. \eqref{eq:Gm_ACF_2} to Eq. \eqref{eq:Ptau} is non-trivial. 
Further, $P(\tau)$ can be multi-modal. However, in principle, each mode with characteristic time $\tau$ contains key information about the structure and the dynamics of the system. 

It is worth emphasizing here that for the case of $^1$H NMR relaxation of water in biological cells, Brownstein and Tarr had already identified that a multi-exponential time decay of the NMR magnetization $M(t)$ arises as a consequence of an eigenvalue problem, which is dependent on the size, shape, and surface relaxivity of the cells \cite{Brownstein1979}. Further advances were also made in the literature \cite{Bernatowicz2006,Ghadirian2013,mitchell:jcp2019,yan:pra2023}. Although our formalism is independent from these works, and furthermore our formalism addresses the NMR autocorrelation function $G(t)$ rather than the NMR magnetization decay $M(t)$,  the fact that we obtained similar conclusions regarding the arising of multi-exponential decays supports the multi-eigenvalue approach for NMR relaxation in general.

In the following sections, we will employ the idea of solving for the phase distribution function $\rho(\textbf{r},t)$ and finding the corresponding NMR autocorrelation function $G(t)$ for dipole pairs. This will also enable analytical expressions for the underlying distribution $P(\tau)$ of molecular modes of relaxation.

\subsection{Fixed distance and non-interacting dipoles}

\begin{figure}[!t]
    \centering
        \includegraphics[width=0.50\textwidth]{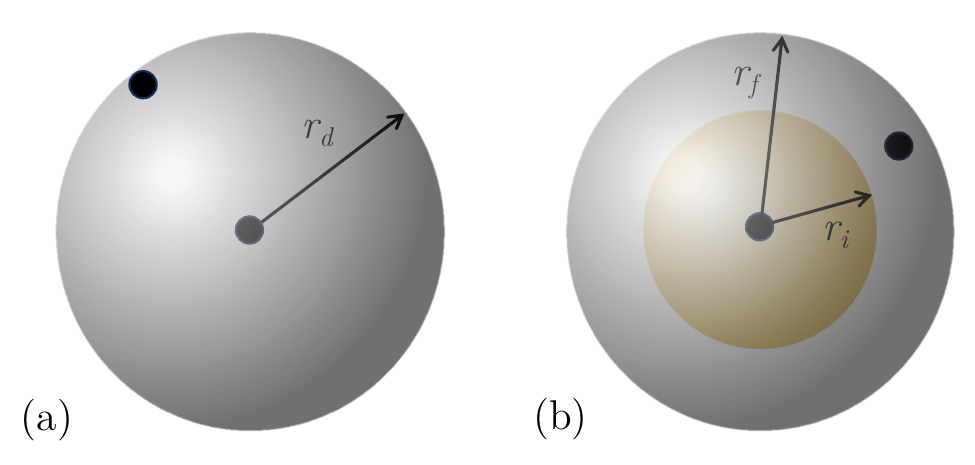}
    \caption{Configuration of the two magnetic dipoles in which $(a)$ they are separated by a constant distance $r_d$ (discrete spherical shell) and $(b)$ they are separated by a non-constant distance bounded by an inner radius $r_i$ and an outer radius $r_f$ (thick spherical shell)}
    \label{fig:Fig1}
\end{figure}

Consider first two non-interacting dipoles at fixed separation $r_d$ (Figure \ref{fig:Fig1}(a)).
For convenience, we fixed one of the dipoles at the center of coordinates, and we aim to solve for the phase distribution function of the second dipole with respect to the central one.

In spherical coordinates, $\textbf{r}_d = \{ r_d, \theta, \phi \}$, $U(\textbf{r}_d)=0$, and thus Eq. \eqref{eq:Gen_FP_2} can be written as 
\begin{equation}
\begin{split}
    & \frac{\partial}{\partial t} \rho(\textbf{r}_d,t) = \\ & D \left[ \frac{1}{r_d^2 \sin \theta} \frac{\partial}{\partial \theta} \left( \sin \theta \frac{\partial}{\partial \theta} \right)  + \frac{1}{r_d^2 \sin^2 \theta} \frac{\partial^2}{\partial \phi^2} \right] \rho(\textbf{r}_d,t). \label{eq:FP_BPP}
\end{split}
\end{equation}
where $D$ is a diffusion coefficient of the dipole. We assume normal diffusion, i.e.\ $D$ is a constant. The 
initial condition of the mobile dipole is 
\begin{equation}
    \begin{split}
        \rho(\textbf{r}_d,0) &= \frac{\delta(\theta - \theta_0) \delta(\phi-\phi_0)}{\sin \theta} p(\textbf{r}_{d,0}),
    \end{split}
\end{equation}
where $p(\textbf{r}_{d,0})$ is the probability of having that initial discrete position, $\textbf{r}_{d,0}$. Further, 
for an isotropic system, the system obeys periodicity in $\theta$ and $\phi$, and then one can show that 
\begin{equation}
        \rho(\textbf{r}_d,t) = \frac{1}{4\pi}  Y_{2}^{0}(t) Y_{2}^{0}(0) \exp \left( -\frac{6Dt}{r_d^2} \right) \, .
\end{equation}
Using this result in Eq.\ \eqref{eq:Gm_ACF} without integrating over the constant $r_d$, and by the orthogonality of the spherical harmonics, we obtain that
\begin{equation}
    \begin{split}
         G(t) = \frac{1}{4\pi r_d^6} \exp \left( -\frac{6Dt}{r_d^2} \right). \label{eq:BPP}
    \end{split}
\end{equation}

This result is simply the mono-exponential decay from the traditional BPP theory and has been derived in different ways \cite{Bruce2000,Bloembergen1961}. Additional details of the analytical solution are given in the Supplementary Information. Finally, observe from Eq. \eqref{eq:BPP} that the characteristic time for this mono-exponential decay is given by
\begin{equation}
    \tau_d = \frac{r_d^2}{6D}, \label{eq:tau_BPP}
\end{equation}
and, after restoring the constant factors from Eq.~\eqref{eq:Gm}, the corresponding amplitude can be obtained by identifying the second moment contribution to $G(t)$ at $t=0$, such that 
\begin{equation}
    \begin{split}
        & P(\tau_d) = \frac{1}{4\pi r_d^6} \, . \label{eq:Ptau_BPP}
    \end{split}
\end{equation}

\subsection{Non-fixed distance and non-interacting dipoles}

Having established the result for a simple, well-known case, we next consider the case where the two dipoles are not at fixed distance. Hence, we must now account for the radial diffusion between an inner radius $r=r_i$ and an outer radius $r=r_f$ (Figure \ref{fig:Fig1}(b)). 
This accounts for the changes in the distance between the dipoles due to sampling of different molecular conformation (for the case of intramolecular relaxation) and thermal fluctuations, and thus compared to the BPP-model, adds an additional degree of freedom to the dipoles.  This model can be an approximate description 
of $^1$H$-^1$H interactions on the two extremes of a long hydrocarbon chain for the case of like-spins, or the relaxation of $^1$H from water molecules in the inner-shell of paramagnetic ions for the case of unlike-spins. It is important to clarify that our formalism only accounts for rotational diffusion in the polar and azimuthal directions and translational diffusion along the radial axis. We emphasize that we do not model the translational exchange between the inner-shell and the bulk with respect to the central dipole, as in the Hwang-Freed model \cite{Hwang1975}.

Therefore, now Eq.\ \eqref{eq:Gen_FP_2} can be written by
\begin{equation}
\begin{split}
    \frac{\partial}{\partial t} \rho(\textbf{r},t) =& D \left[ \frac{1}{r^2} \frac{\partial}{\partial r} \left( r^2 \frac{\partial}{\partial r} \right) + \right. \\ & \left. \frac{1}{r^2 \sin \theta} \frac{\partial}{\partial \theta} \left( \sin \theta \frac{\partial}{\partial \theta} \right) + \frac{1}{r^2 \sin^2 \theta} \frac{\partial^2}{\partial \phi^2} \right] \rho(\textbf{r},t). \label{eq:FP_NI}
\end{split}
\end{equation}
Assuming isotropy, we have periodicity in $\theta$ and $\phi$; further, we use Neumann (reflecting) boundary conditions
\begin{equation}
\begin{split}
    \left. \frac{\partial \rho (r,\theta,\phi,t)}{\partial r} \right|_{r=r_i} = 0 \, , \\
    \left. \frac{\partial \rho (r,\theta,\phi,t)}{\partial r} \right|_{r=r_f} = 0 \, .
    \label{eq:bcs}
\end{split}
\end{equation}
The initial condition of the mobile dipole is given by 
\begin{equation}
    \begin{split}
        \rho(\textbf{r},0) &= \frac{\delta(r - r_0) \delta(\theta - \theta_0) \delta(\phi-\phi_0)}{r^2 \sin \theta} p(\textbf{r}_{0}),
    \end{split}
\end{equation}
where $p(\textbf{r}_{0})$ is the probability of having that initial discrete position. 

For the above conditions, the time-dependent equilibrium distribution function is given by 
\begin{equation}
\begin{split}
    \rho(\textbf{r}, t) = & \frac{3}{4\pi (r_f^3 - r_i^3)}  Y_{2}^{0}(t) Y_2^0(0) \\ & \sum_{k=1}^{\infty} \frac{j_2 \left( \lambda_{2,k} r_0 \right) j_2 \left( \lambda_{2,k} r \right)}{N_{2,k}} \exp \left( - \lambda_{2,k}^2 D t \right), \label{eq:final_rho}
\end{split}
\end{equation}
where $j_2(\lambda_{2,k} r)$ is the $2^{nd}$-order spherical Bessel function of the first kind, and $N_{2,k}$ is the normalization constant that arises from orthogonality of spherical Bessel functions assuming Neumann boundary conditions, i.e., \cite{Arfken1985}
\begin{equation}
    \begin{split}
     & \int_{r_i}^{r_f} j_2 \left( \lambda_{2,i} r \right) j_2 \left( \lambda_{2,k} r \right) r^2 dr = N_{2,k} \delta_{ik}, \\
     & N_{2,k} = \frac{r^3}{2} \left( 1 - \frac{6}{(r \lambda_{2,k})^2} \right) j_2 \left( \lambda_{2,k} r \right) \Biggr|_{r=r_i}^{r=r_f}. \label{eq:ortho_j1}
    \end{split}
\end{equation}
In order to satisfy the Neumann boundary conditions and having in mind the result from Eq. \eqref{eq:final_rho}, the only possible values of $\lambda_{2,k}$ are those that satisfy the corresponding Sturm-Liouville eigenvalue problem (SL-EVP) \cite{Reid1980}, which is given by
\begin{equation}
\begin{split}
    j_n' \left( \lambda_{2,k} r_i \right)y_n' \left( \lambda_{2,k} r_f \right) - y_n' \left( \lambda_{2,k} r_i \right) j_n' \left( \lambda_{2,k} r_f \right) = 0 \, , \label{eq:eigenv}
\end{split}
\end{equation}
where the prime indicate derivatives with respect to $r$. The eigenvalues $\lambda_{2,k}$ will be the positive roots of equation above, allowing the solution in Eq.\ \eqref{eq:final_rho} to be expanded in terms of their corresponding eigenfunctions \cite{Hildebrand1952}. Here, we ignore the the trivial solution $\lambda_{2,k} = 0$. Notice that the eigenvalues depend on the functional form of the Eq.\ \eqref{eq:eigenv}, which is a consequence of the boundary condition, and on the values of $r_i$ and $r_f$.

For simplicity, because it is implied that only the second moments are of interest to NMR relaxation, we simplified the notation for the eigenvalues from $\lambda_{2,k}$ to $\lambda_{k}$.
Using the result in Eq. \eqref{eq:Gm_ACF} and integrating over all possible coordinates in the system, and
using the orthogonality of spherical harmonics, we find 
\begin{equation}
    \begin{split}
     G(t) =& \frac{3}{4 \pi (r_f^3 - r_i^3)} \\ & \sum_{k=1}^{\infty} \exp \left( - \lambda_{k}^2 D t \right) \int_{r_i}^{r_f} \int_{r_i}^{r_f} \frac{1}{rr_0} \frac{j_2 \left( \lambda_{k} r \right) j_2 \left( \lambda_{k} r_0 \right)}{N_{2,k}} dr dr_0. \label{eq:NI}
    \end{split}
\end{equation}

It is important to highlight a couple of points about the result in Eq. \eqref{eq:NI}. Firstly, we observe that now the NMR autocorrelation function has a multi-exponential decay, which comes as an infinite set of discrete characteristic decay times that depend on the corresponding eigenvalues. Secondly, we must recall that Eq. \eqref{eq:NI} should be computed under the definition of the inner-product (orthogonality) in Eq. \eqref{eq:ortho_j1} that arise from the boundary conditions, such that the calculated quantities are projected into the correct orthogonal eigenfunction set. The detailed derivation can be found in the Supplementary Information.

Eq. \eqref{eq:NI} shows that the characteristic time for the $k^{th}$-mode is 
\begin{equation}
	\tau_k = \frac{1}{\lambda_k^2 D} \qquad k=1,2,\ldots,\infty  \label{eq:tauk_theory_full}
\end{equation}
and, after restoring the constant factors from Eq.~\eqref{eq:Gm}, the corresponding amplitude (second moment contribution to the autocorrelation function at $t=0$) is 
\begin{equation}
    \begin{split}
     P(\tau_k) =& \frac{3}{4 \pi (r_f^3 - r_i^3)} \sum_{k=1}^{\infty} \int_{r_i}^{r_f} \int_{r_i}^{r_f} \frac{1}{rr_0} \frac{j_2 \left( \lambda_{k} r \right) j_2 \left( \lambda_{k} r_0 \right)}{N_{2,k}} dr dr_0. \label{eq:modek_theory_full}
    \end{split}
\end{equation}

\section{Simulation Methodology}

We performed molecular simulations using LAMMPS\cite{Plimpton1995} to address the systems covered in our theoretical formulation. We have adopted two types of reduced units in this work: Lennard-Jones (LJ) reduced units for the simulation parameters, and NMR reduced units for the calculation of relaxation rates.

Consider first the LJ reduced units. The relationships between real (denoted by $^{\dagger}$) and LJ-reduced quantities are defined for quantities like mass ($M$), distance ($r$), time ($t$), frequency ($\omega$), temperature ($T$), and molecular friction constant ($f$) by
\begin{flalign}
 \begin{split}   
    M &= \frac{M^{\dagger}}{m}, \\
    r &= \frac{r^{\dagger}}{\sigma}, \\  
    t &= t^{\dagger} \sqrt{\frac{\epsilon}{m\sigma^2}}, \\  
    \omega &= \omega^{\dagger} \sqrt{\frac{m\sigma^2}{\epsilon}}, \\  
    T &= \frac{T^{\dagger} k_B}{\epsilon},\\  
    f &= f^{\dagger} \sqrt{\frac{\sigma^2}{m\epsilon}}.
 \end{split}      
\end{flalign}
In our simulation, LJ-particles with $m=1.0$ (a.u.), $\epsilon=0.1$ (a.u.), and $\sigma=0.1$ (a.u.) carry the dipoles. Usually LJ-reduced quantities are represented by ($^*$), but we omit this for simplicity.

Next consider NMR reduced units. The dimensionless (i.e., reduced) NMR relaxation rates ($1/T_{1,2}$) are related to real units $1/T_{1,2}^{\dagger}$ ($time^{-1}$) by
\begin{align} 
    \frac{1}{T_{1,2}} &= \frac{1}{T_{1,2}^{\dagger}} \frac{\sigma^6}{\alpha}\sqrt{\frac{\epsilon}{m\sigma^2}},   \label{eq:T1reduced}
\end{align}
where the constant $\alpha$ ($time^{-2} \times distance^{6}$) is generally given by
\begin{align}
	\alpha = \frac{4\pi}{5} \left( \frac{\mu_0}{4 \pi} \right)^2 \hbar^2 \gamma_I^2 \gamma_S^2 S(S+1) \, , \label{eq:alpha}
\end{align}
wherein $\mu_0$ is the vacuum permittivity, $\hbar$ is the modified Planck's constant $(\hbar=h/2\pi)$, and $\gamma_I$ and $\gamma_S$ are the gyromagnetic ratios for particles with spin $I$ and $S$, respectively. For like spins, $\gamma_I = \gamma_S$ and $I = S$. In this equation, the ``measured'' dipole has spin $I$ and gyromagnetic ratio $\gamma_I$, while the ``influencing'' dipole has spin $S$ and gyromagnetic ratio $\gamma_S$. Alternatively, $\alpha$ can be included in the definition of $G^m(t)$ in Eq.~\ref{eq:Gm} as a multiplicative factor.  Without loss of generality, we set $\alpha=1$ for NMR reduced units.

In our simulations, we use a cut-off distance $r_c=0.1$ always less than the minimum distance between the dipoles in the simulation box to ensure they do not interact.
For simplicity, the first dipole was kept at the center of the simulation box, while the second dipole was constrained in the annulus (Figure \ref{fig:Fig1}(b)) by LJ-WCA walls\cite{Weeks1971}. We use Langevin dynamics with different friction constants $f$ to propagate the dynamics, and a timestep of $\delta t=0.0001$ while sampling the configurations every 100 timesteps (the configuration sampling rate is $\delta t_{samp} = 0.01$).
The friction constant $f$ (in Langevin dynamics) is related to the diffusion coefficient $D$ of the dipoles by
\begin{align}
	D = \frac{k_B T}{f}, \label{eq:D_langevin}
\end{align}
where $k_B=1$ is Boltzmann's constant in LJ-reduced units. Note that we define $D$ as the diffusion coefficient for the bulk fluid (i.e., without confinement), which is time independent and normally defined as ``$D_0$'' in porous media.

It is important to highlight that LJ-reduced units imply the use of the corresponding states law, i.e., the solution in LJ-reduced units can be converted to any physical systems of interest given that appropriate LJ-parameters for the system are known ($m$, $\epsilon$, and $\sigma$). Overall, the results obtained with such LJ-reduced units are broad and general.

We calculate the autocorrelation function $G(t)$ from the MD trajectory  using in-house codes 
after rewriting Eq.\ \eqref{eq:Gm} as 
\begin{flalign}
    G(t) &= \frac{5}{16 \pi} \left< \frac{ \left( 3 \cos^2 \theta(t+t') - 1 \right)}{r^3(t+t')} \frac{ \left( 3 \cos^2 \theta(t') - 1 \right) }{r^3(t')} \right>_{t'}, \label{eq:Gt_MD}
\end{flalign}
where $t'$ is the lag-time, and the indexes $I$ and $S$ refer to the measured and influencing dipole in the dipole pair, respectively. Recall from Eq. \eqref{eq:Gm} that $G(t)$ is the normalized auto-correlation function. 

\subsection{Pad\'e-Laplace inversion recovery}

As shown in the theoretical development, given that dipoles undergoing NMR relaxation under normal diffusion yield a multi-exponential decay, we can write the $G(t)$ from molecular simulations as 
\begin{align}
	G(t) &= \sum_{j=1}^n P(\mu_j) \exp(-\mu_j t),
\end{align}
where $n$ is the number of exponential decays, and $P(\mu_j)$ is the amplitude (probability) for a mode with decay-rate $\mu_j$. The Laplace transform of $G(t)$ is given by \cite{Hellen2005}
\begin{align}
	L(p) &= \int_{0}^{\infty} \exp(-pt) G(t) dt  \label{eq:Laplace}, \\
	L(p) &= \sum_{j=1}^n \frac{P(\mu_j)}{p-\mu_j} \label{eq:Laplace2}.
\end{align}
Eq.~\ref{eq:Laplace2} makes it clear that the number of exponential decays is the same as the number of poles of the Laplace transform. We seek the Padé approximant $R_{n-1,n}(p)$ of $L(p)$ \cite{Yeramian1987}. That is, we seek 
\begin{align}
		L(p) &= R_{n-1,n}(p) = \frac{A_{n-1}(p)}{B_{n}(p)}, \label{PL}
\end{align}
in which $A_{n-1}(p)$ is a polynomial of order $n-1$ and $B_{n}(p)$ is a polynomial of order $n$.
The left-hand side of equation \eqref{PL} can be expanded in a Taylor series
about a point $p_0$, where a good choice for $p_0$ is the inverse of the time necessary for $G(t)$ to decay to half of its initial value. Therefore, 
\begin{align}
    \begin{split}
	   d_0 + d_1 (p-p_0) + \cdots + d_{2n-1} (p-p_0)^{2n-1} = \\ \frac{a_0 + a_1(p-p_0) + \cdots + a_{n-1}(p-p_0)^{n-1}}{1 + b_1(p-p_0) + \cdots + b_{n}(p-p_0)^{n}},	
    \end{split}
\end{align}
where the Taylor series coefficients $d_0,d_1,\cdots,d_{2n-1}$ are given by
\begin{align}
		d_i = \frac{1}{i!} \left( \frac{d^{(i)}L}{dp^{(i)}} \right)_{p=p_0}\, .
\end{align}
These coefficients can be calculated using numerical integration of the derivatives of equation \eqref{eq:Laplace} by
\begin{equation}
    \begin{split}
        d_i \frac{ i!}{\Delta t} = &  \frac{1}{2} \left[ (-t_0)^{i} \exp(-p_0 t_0)G(0) \right. \\ & \left. + (-t_{M-1})^{i} \exp(-p_0 t_{M-1})G(M-1) \right] \\ & + \sum_{j=1}^{M-2} (-t_{j})^{i} \exp(-p_0 t_{j})G(j) .
    \end{split}
\end{equation}

Finally, we can get the coefficients of the Pad\'e approximants $A_{n-1}(p)$ and $B_n(p)$ from the Taylor expansion coefficients. We then perform a partial fraction decomposition of $R_{n-1,n}(p) = \frac{A_{n-1}(p)}{B_{n}(p)}$ using MATLAB \cite{MATLAB}, returning a sum of fractions as in equation \eqref{eq:Laplace2}. By analyzing the magnitude of the numerator of these partial fractions, we can determine the amplitude and also the number of poles $N$ that are physical \cite{Hellen2005,Yeramian1987}. In practice, the number of poles is determined by trying different orders of Pad\'e approximants starting from $n=1$ up to the point $n=N+1$ when no more new physical poles are identified. As we add more physical poles, we are in effect constructing better models of $P(\tau)$. Thus, one could envision using a maximum entropy (MaxEnt) approach to determine when to cutoff the process of adding new physical poles. We plan to investigate this approach in future studies. Here we include all the physical poles.

If needed, the final physical poles and corresponding amplitudes from this inversion were refined using a standard Particle Swarm Optimization (PSO) procedure to improve the adjustment to the simulation data \cite{Gad2022}. We have presented the agreement between the original NMR dipole-dipole autocorrelation from MD simulations and its corresponding reconstructed signal through the Pad\'e-Laplace inversion in the Supplementary Information.

\section{Results and Discussion}

\subsection{Fixed distance and non-interacting dipoles}

\begin{figure}[!t]
    \centering
        \includegraphics[width=0.5\textwidth]{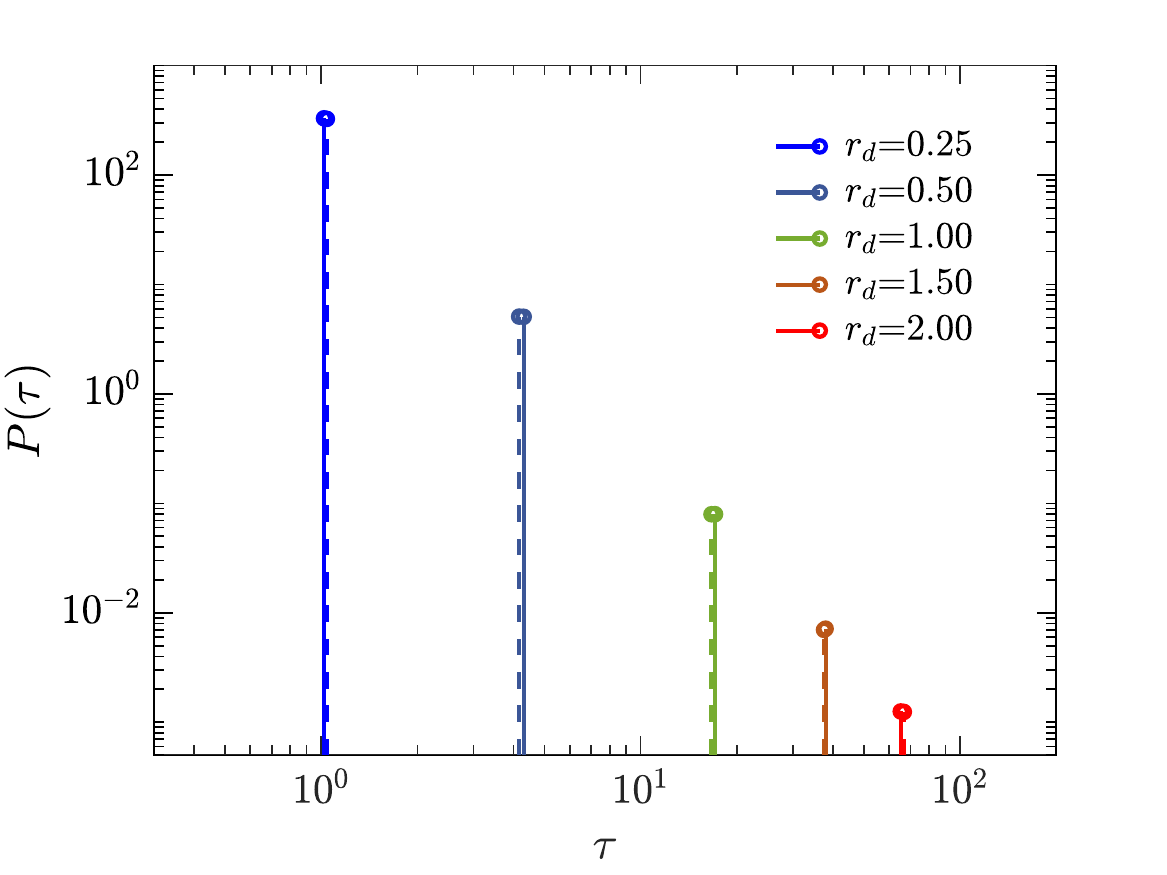}
    \caption{Molecular modes of NMR relaxation in LJ-reduced units for dipole pairs at different fixed distances $r_d$, with constant friction $f = 100$ and $T=1.00$. The solid lines represent the theoretical predictions, while the dashed lines represent the MD simulation results.}
    \label{fig:Fig2}
\end{figure}

For the case of fixed distance and non-interacting dipoles, we have seen from Eq. \eqref{eq:BPP} that the NMR autocorrelation function $G(t)$ has a single molecular mode (mono-exponential decay) with characteristic time $\tau_d$ given by Eq. \eqref{eq:tau_BPP} and the corresponding amplitude $P(\tau_d)$ given by Eq. \eqref{eq:Ptau_BPP}.
Since this result holds for isotropic systems, this mode is known for being a simple rotational diffusion component.
Figure \ref{fig:Fig2} shows the comparison of the MD simulation NMR relaxation modes recovered with the Pad\'e-Laplace method, and the theoretical results predicted from equations Eqs. \eqref{eq:tau_BPP} and \eqref{eq:Ptau_BPP}. We observe excellent agreement between simulations and theory across different LJ-reduced distances $r_d$ at constant friction $f = 100$ and $T=1.00$. To contextualize these parameters, we can give the example of liquid water \cite{Lin2004} ($m=3.0103\times 10^{-26}$ kg, $\sigma=2.725\times 10^{-10}$ m, $\epsilon=4.9115\times 10^{-21}$ J). Thus, the temperature in real units would correspond to $T^{\dagger}=358.38$ K and $f^{\dagger} \propto 10^{-12}$ kg/s with $D^{\dagger} \propto 10^{-9}$m$^2$/s, which is within the typical range of diffusion coefficients for liquids.

\begin{figure}[!t]
    \centering
        \includegraphics[width=0.5\textwidth]{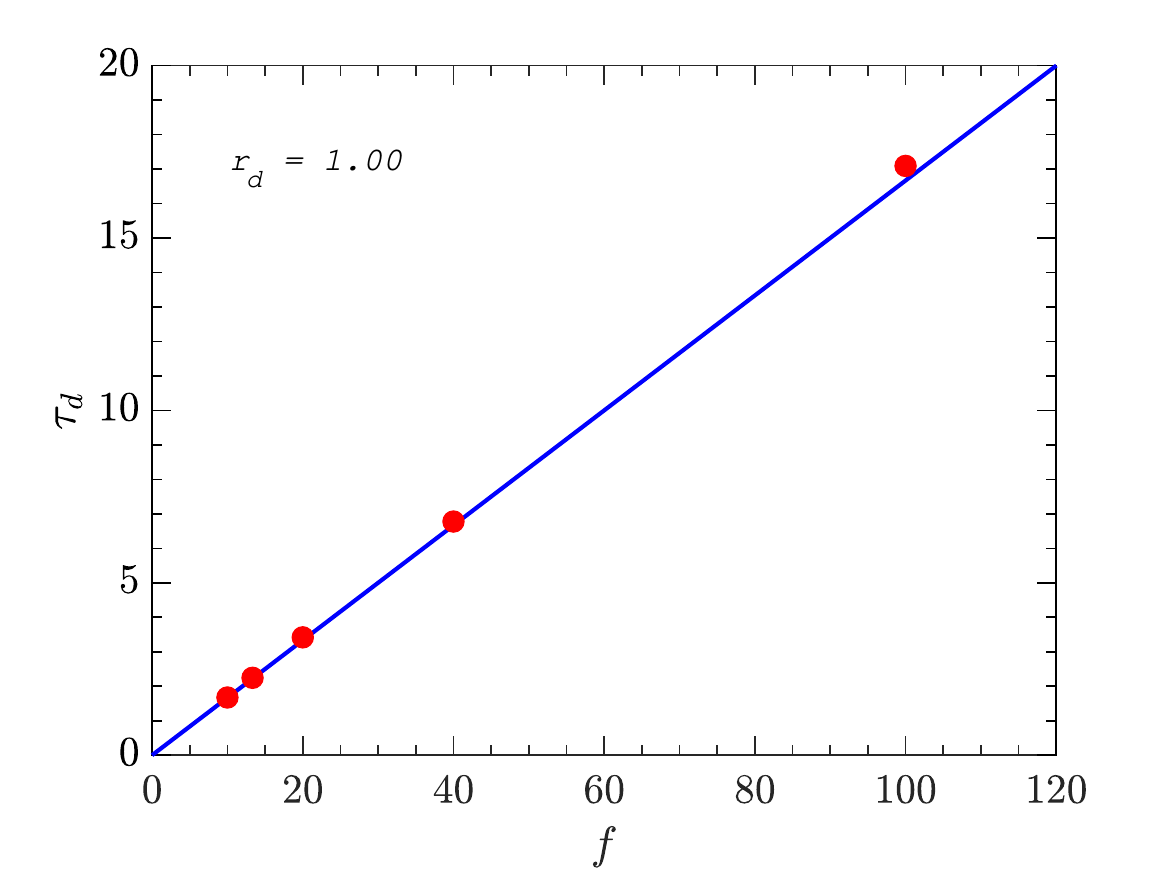}
    \caption{LJ-reduced characteristic time $\tau_d$ of NMR relaxation for dipole pairs with different friction constants $f$, at $r_d=1.00$ and $T=1.00$. The solid line represent the theoretical predictions, while the solid points represent the MD simulation results.} 
    \label{fig:Fig3}
\end{figure}

Figure \ref{fig:Fig3} shows a comparison of (LJ-reduced) characteristic relaxation times from MD simulations and the theoretical predictions across different LJ-reduced friction constants $f$ at constant distance $r_d = 1.00$ and $T=1.00$. For water as an example, this would correspond to $r^{\dagger}=2.725$ \r{A} or about one water molecular diameter, with diffusion coefficients $D^{\dagger} \propto 10^{-8}-10^{-9}$m$^2$/s, as is typical for 
liquids. We again observe excellent agreement, showing that the characteristic relaxation time increases with the increment of the molecular friction in the system, but its amplitude remains the same. Physically, this confirms that the magnitude of the NMR relaxation rate depends solely on the distance between the fixed dipoles, but the characteristic time of the rotational diffusion mode increases as the viscous forces increase in the system, i.e., when the dipole diffusion decreases.

Finally, we calculate the spin-lattice $1/T_1(\omega)$ and spin-spin $1/T_2(\omega)$ relation rates in the low frequency limit ($\omega_0 = 0$) for the case of like-spins. Given that $1/T_1(0) = 1/T_2(0) = 1/T_{1,2}(0)$ in the zero frequency limit, we show that
\begin{align}
	\frac{1}{T_{1,2}(0)} = 5 \,J(0) = 10 \,G(0)\,\tau_d \label{eq:T12_BPP}
\end{align}
in which $J(0)$ is the spectral density in Eq. \ref{eq:Jomega}. The real units $1/T_{1,2}^{\dagger}$ can be deduced from Eq. \ref{eq:T1reduced} and Eq. \ref{eq:alpha} where $\alpha$ can be calculated with $\gamma_I = \gamma_S$ and $I = S$ for the case of like spins.

The results of $T_{1,2}$ for both MD simulations and theory are summarized in Table \ref{Tab1} for the two dipoles at different fixed distances, and in Table \ref{Tab2} for the two dipoles at fixed distance but different friction constants.
Overall, the agreement obtained for the data serves as a validation for the MD simulations.

\begin{table}[!ht]
\centering
\begin{tblr}{
  cells = {c},
  cell{1}{1} = {r=2}{},
  cell{1}{2} = {c=2}{},
  hline{1,3,8} = {-}{},
  hline{2} = {2-3}{},
}
$r_d$ & $1/T_{1,2}(0)$        &        \\
     & MD simulations     & Theory \\
0.25 & 3367.8 $\pm$ 191.1 & 3395.3 \\
0.50 & 218.3 $\pm$ 14.2  & 212.2  \\
1.00 & 13.6 $\pm$ 1.2  & 13.3   \\
1.50 & 2.7 $\pm$ 0.3  & 2.6   \\
2.00 & 0.8 $\pm$ 0.1  & 0.8   
\end{tblr}
\caption{Reduced relaxation rate $1/T_{1,2}(0)$ at zero frequency for two dipoles at different $r_d$, with $f=100$ and $T=1.00$.}
\label{Tab1}
\end{table}

\begin{table}[!ht]
\centering
\begin{tblr}{
  cells = {c},
  cell{1}{1} = {r=2}{},
  cell{1}{2} = {c=2}{},
  hline{1,3,8} = {-}{},
  hline{2} = {2-3}{},
}
$f$ & $1/T_{1,2}(0)$        &        \\
     & MD simulations   & Theory \\
10.0 & 1.4 $\pm$ 0.1  & 1.3   \\
13.3 & 1.8 $\pm$ 0.1  & 1.8   \\
20.0 & 2.7 $\pm$ 0.1  & 2.7   \\
40.0 & 5.4 $\pm$ 0.4  & 5.3  \\
100.0 & 14.3 $\pm$ 1.2  & 13.3 \\
\end{tblr}
\caption{Reduced relaxation rate $1/T_{1,2}(0)$ at zero frequency for two dipoles at different $f$, with $r_d=1.00$ and $T=1.00$.}
\label{Tab2}
\end{table}

\subsection{Non-fixed distance and non-interacting dipoles}

For the case when one of the dipoles is free to move in the annulus, as noted above
we must find a multi-exponential (hence, multi-mode) decay. From Eq.\ \eqref{eq:NI}, we find that each characteristic time $\tau_k$ is given by Eq. \eqref{eq:tauk_theory_full}, and the corresponding amplitude $P(\tau_k)$ is given by Eq. \eqref{eq:modek_theory_full}.
Notice that the amplitude $P(\tau_k)$ of each mode $\tau_k$ depends on the eigenvalue $\lambda_{k}$ and the geometry of the system defined by the inner radius $r_i$ and the outer radius $r_f$. Further, the integral needs to calculated whilst observing the definition of the inner-product (orthogonality) in Eq.\ \eqref{eq:ortho_j1} that depends on the boundary conditions.

\begin{figure}[!t]
\includegraphics[width=0.50\textwidth]{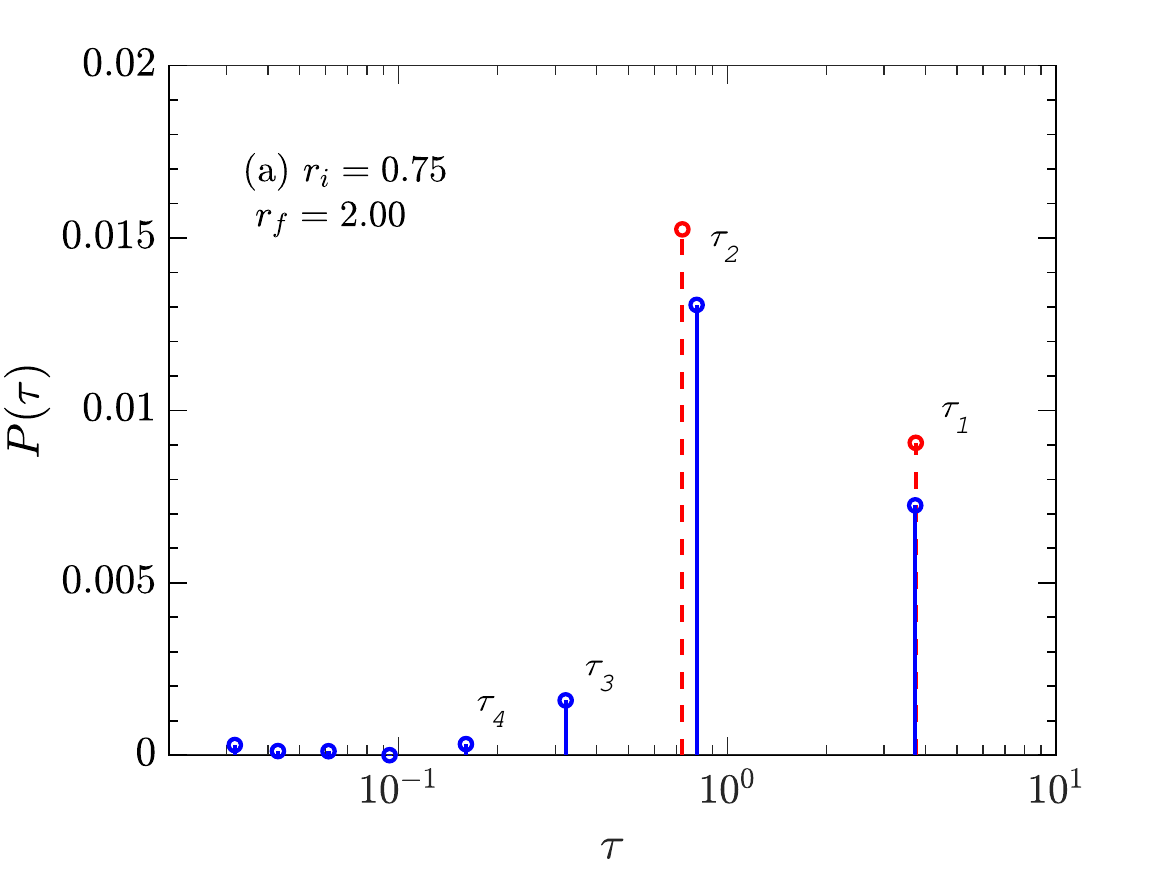}
\includegraphics[width=0.50\textwidth]{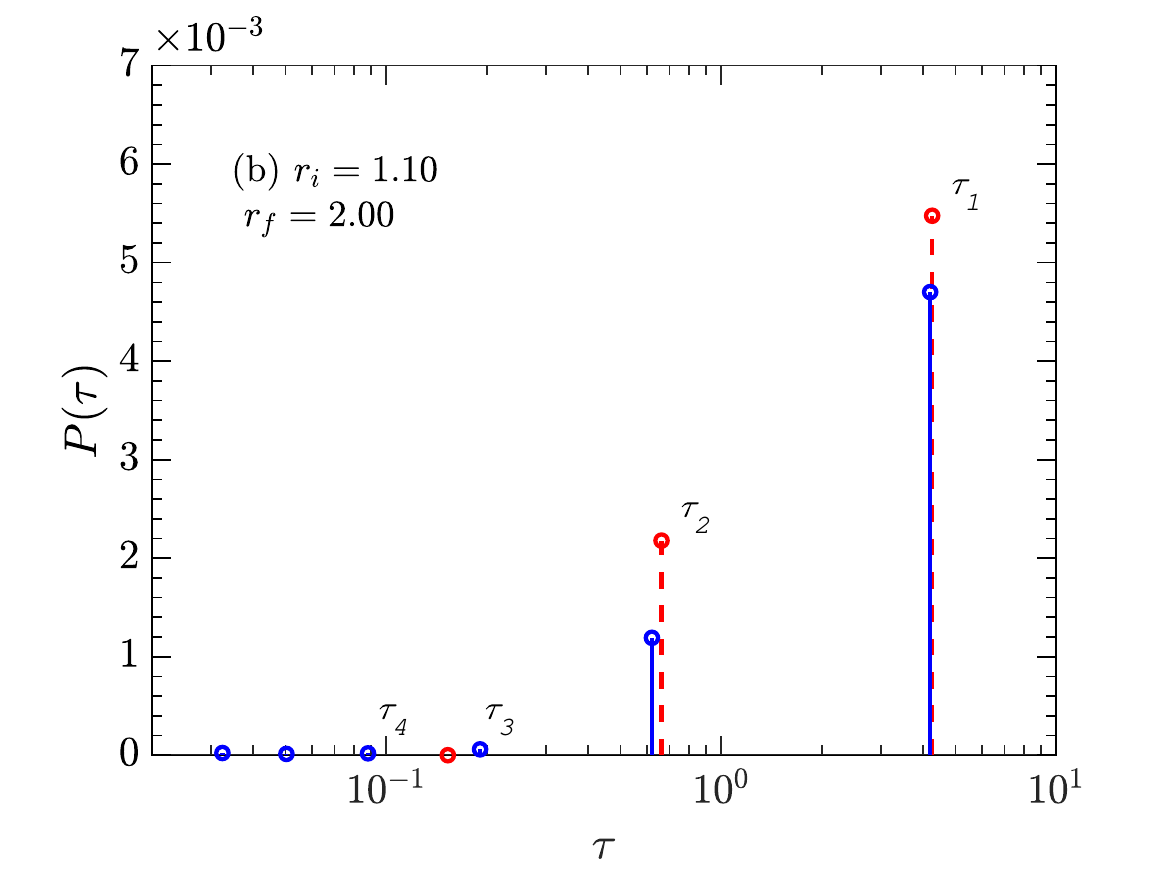}
    \caption{Molecular modes of NMR relaxation in LJ-reduced units for dipole pairs at non-fixed distances for the case of (a) $r_i=0.75$ and $r_f=2.00$ and (b) $r_i=1.10$ and $r_f=2.00$, with constant friction $f=10$ and $T=1.00$. The solid lines represent the theoretical predictions, while the dashed line represents the MD simulation results.}
    \label{fig:Fig4}
\end{figure}

Figures \ref{fig:Fig4}(a) and \ref{fig:Fig4}(b) show the $P(\tau_k)$ distribution for two systems with spherical shell thickness limited by $r_f=2.00$ and (a) with $r_i=0.75$ and (b) with $r_i=1.10$. In both cases, we employed $f=10$ and $T=1.0$. For water as an example, these distances correspond to $r_i^{\dagger}\approx 2.0 - 5.5$ \r{A} (from a fraction of of the size of water to two water molecule diameters), the physical temperature is roughly $T^{\dagger}=358.38$ K, and $f^{\dagger} \propto 10^{-11}$~kg/s with $D^{\dagger} \propto 10^{-8}$~m$^2$/s, as expected for liquids. Two key features stand out. First, it is worth noticing that MD simulations do capture the multi-exponential behavior, but usually the shorter modes \emph{are not} captured because (i) $\tau_k \leq \delta t_{samp} = 0.01$, the sampling time for configurations, hindering the Pad\'e-Laplace method from detecting such short modes, and/or (ii) 
the amplitudes of these modes small enough to be swamped by simulation noise (low signal to noise ratio). Recalling that the summation of all the amplitudes $P(\tau_k)$ should equal $G(0)$, it is interesting to notice that the simulation results \textit{fold in} the effect of the un-captured shorter modes within the first two main captured modes, increasing their amplitudes to satisfy the correct $G(0)$.
Second, we notice from Figures \ref{fig:Fig4}(a) and \ref{fig:Fig4}(b) that there is an inversion in the amplitude between the first and second largest modes. By keeping $r_f=2.00$ constant, the second mode $\tau_2$ presents a higher amplitude than $\tau_1$ in the case where $r_i=0.75$, while an opposite behavior is observed for $r_i=1.10$. The physical significance of these modes will be further explored later.

\begin{figure}[!t]
    \centering
        \includegraphics[width=0.525\textwidth]{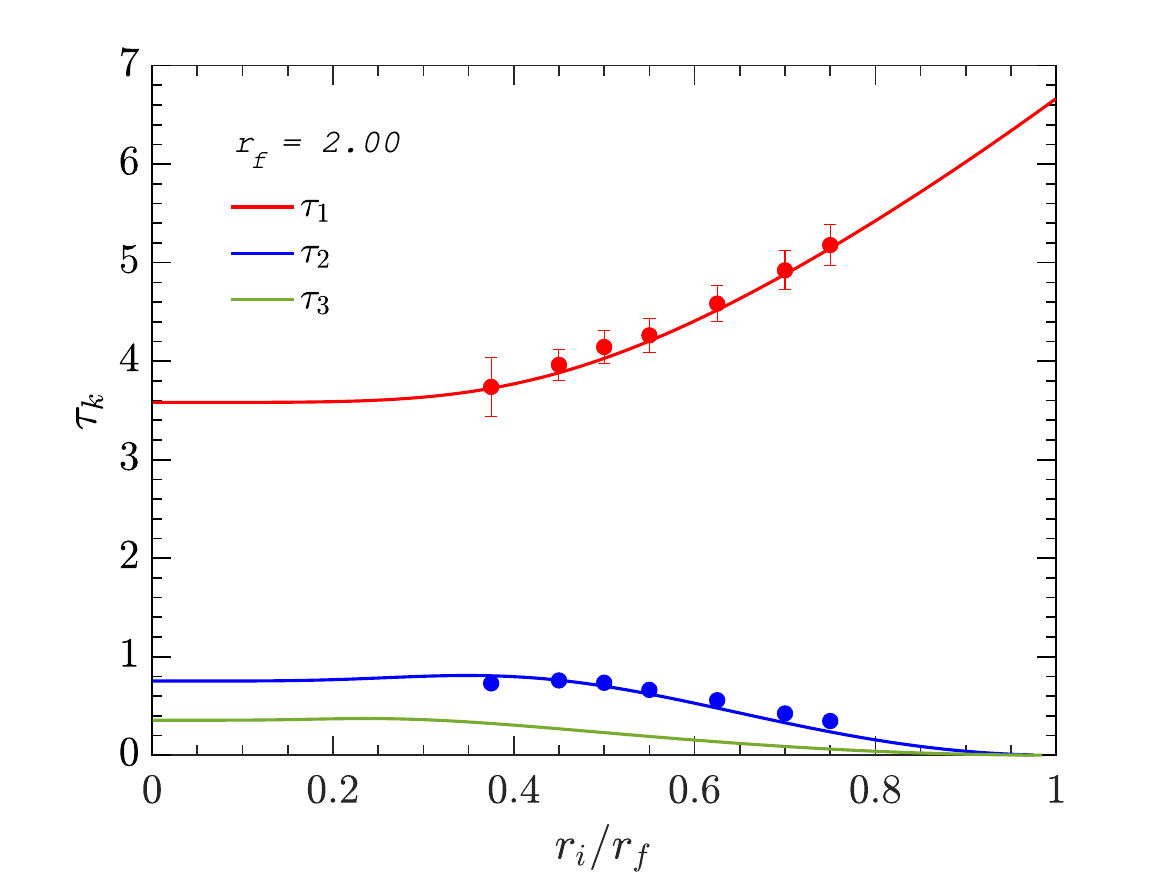}
    \caption{Molecular modes LJ-reduced characteristic times $\tau_k$ up to the third largest mode as a function of $r_i/r_f$, for the case of $r_f=2.00$ with $f=10$ and $T=1.0$. The solid lines represent the theoretical predictions, while the solid points represent the MD simulation results.}
    \label{fig:Fig5}
\end{figure}

\begin{figure}[!t]
    \centering
        \includegraphics[width=0.525\textwidth]{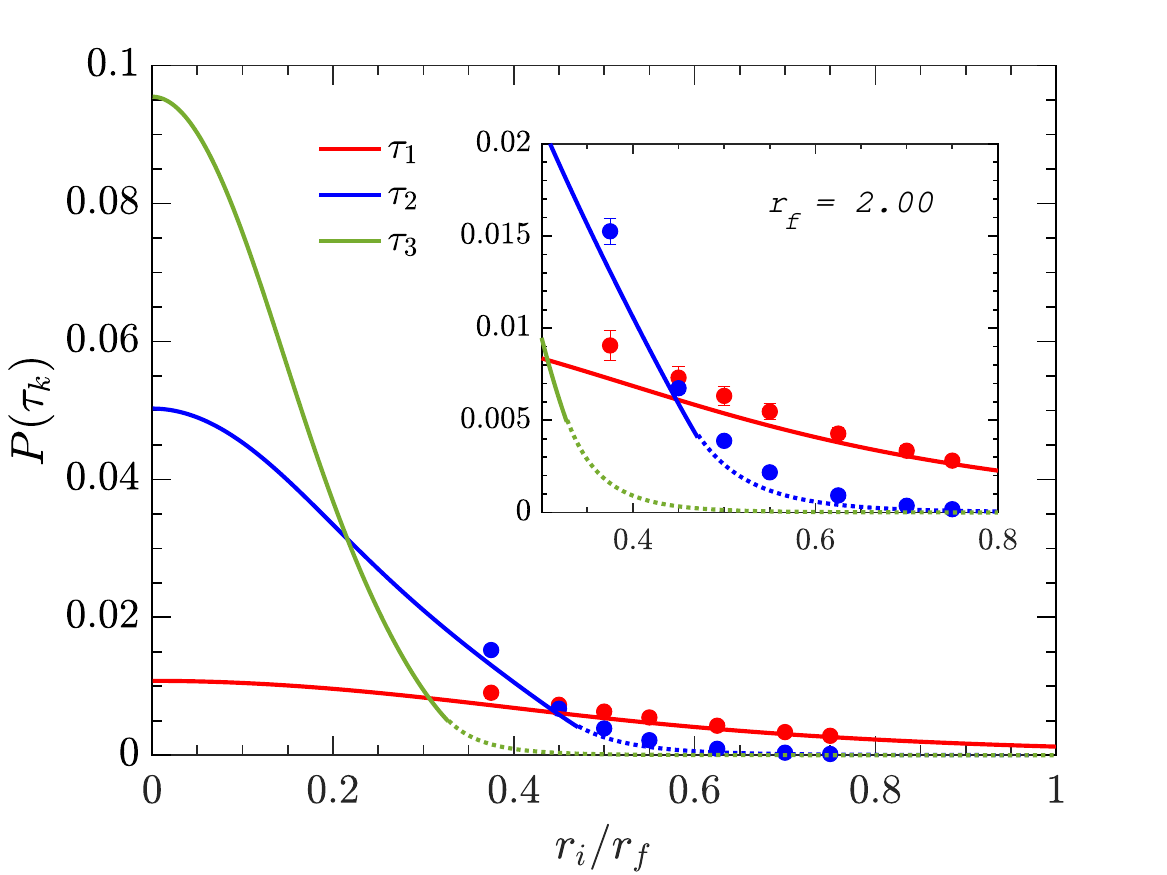}
    \caption{Molecular modes amplitudes $P(\tau_k)$ up to the third largest mode as a function of $r_i/r_f$, for the case of $r_f=2.00$ with $f=10$ and $T=1.0$. The solid lines represent the theoretical predictions, while the solid points represent the MD simulation results. The dotted lines are the interpolation of the theoretical predictions between points of higher numerical integration accuracy.}
    \label{fig:Fig6}
\end{figure}

Figures \ref{fig:Fig5} and \ref{fig:Fig6} show respectively the $\tau_k$ values and their corresponding $P(\tau_k)$ across different spherical thickness as a function of the ratio $r_i/r_f$, keeping $r_f=2.00$ constant. Again, $f=10$ and $T=1.0$ in all cases. For simplicity, only the theoretical results up to $k=3$ were shown.
Observe from Figure \ref{fig:Fig5} that the curves for $\tau_k$ never cross each other, signifying that these characteristic relaxation times are never degenerate, as expected.
In Figure \ref{fig:Fig6}, the dotted line encompass a region where not all the points can be calculated with high numerical accuracy given our approximated numerical approach that uses the regular definition of inner-product (Euclidean space). This dotted region represent a monotonic curved based on interpolation of points computed within acceptable numerical accuracy in that region given the deviations from the regular Euclidean space and the non-euclidean space defined by the inner-product and orthogonality condition in Eq. \eqref{eq:ortho_j1}. More details about the numerical criteria and the interpolation can be found in the Supplementary Information.
Overall, the agreement between MD simulation results and the theoretical predictions is satisfactory and supports our formalism. Moreover, it is important to analyze how the theory behaves in the limiting cases where $r_i/r_f \rightarrow 1$ and $r_i/r_f \rightarrow 0$.

When $r_i/r_f \rightarrow 1$, the thickness of the spherical shell is zero and the dipoles are at fixed distances, and a single exponential decay (single mode) will be expected in the system according to the traditional BPP theory. Observe from Figures \ref{fig:Fig5} and \ref{fig:Fig6} that when $r_i/r_f \rightarrow 1$, the theory predicts that $\tau_1$ will have a non-zero finite value, with also a non-zero finite corresponding amplitude $P(\tau_1)$. On the other hand, all the remaining modes ($k=2,3,...,\infty$) shorten to zero in this limit, with a corresponding amplitude that also converges to zero. Moreover, the values of $\tau_1$ and  $P(\tau_1)$ predicted by our solutions match the expected BPP values in this limit.

When $r_i/r_f \rightarrow 0$, the two dipoles are so close that any small relative motion will affect the angular terms that drive NMR relaxation (Eq. \eqref{eq:Gt_MD}) within a very short time, meaning that shorter characteristic time modes will play a bigger role than longer ones.
The actual limit $r_i/r_f = 0$ is non-physical, since the two dipoles cannot occupy the same point in space. Observe from Figures \ref{fig:Fig5} and \ref{fig:Fig6} that, as we approach $r_i/r_f \rightarrow 0$, the theory predicts that all $\tau_k$ will have a non-zero finite value, with a non-zero finite amplitude $P(\tau_k)$ that grows as $k$ increases, i.e., $P(\tau_{k+1}) > P(\tau_k)$. 
Overall, the fact that our analytical results make physical sense in the limiting cases of $r_i/r_f$, together with the agreement with MD simulations for intermediate $r_i/r_f$ cases, supports our theoretical solution.

By combining Eqs. \eqref{eq:tauk_theory_full} and \eqref{eq:D_langevin}, and recognizing that the Langevin friction parameter $f$ has an Arrhenius-like temperature dependence \cite{Johnson2023}, we can write that
\begin{align}
	\tau_k = \frac{f}{\lambda_k^2 k_B T} \propto  \frac{A}{\lambda_k^2 k_B T} \exp \left( \frac{E_A}{k_B T} \right), \label{eq:tauk_temp}
\end{align}
where $A$ is a pre-exponential Arrhenius parameter, and $E_A$ is the associated activation energy in LJ-reduced units. 
Notice that Eq. \eqref{eq:tauk_temp} clarifies the temperature dependence of each of these characteristic times $\tau_k$. Firstly, we observe that this temperature dependence does not strictly follow an Arrhenius behavior, but for high enough temperatures, the temperature dependence of $\tau_k$ will follow an Arrhenius-like behavior.
A similar conclusion has been made empirically for the case of $^1$H of water relaxing on the inner-shell hydration around a paramagnetic ion, in which an Arrhenius-like behavior was observed for the temperature dependence of the molecular modes of the system \cite{Pinheiro2022}.

Similarly to the constant radius case, we also estimate $1/T_1(0) = 1/T_2(0) = 1/T_{1,2}(0)$ in the zero frequency limit ($\omega_0 = 0$) for like spins, such that
\begin{align}
	\frac{1}{T_{1,2}(0)} = 10 \, G(0) \left< \tau \right> = 10 \sum_k P(\tau_k) \,\tau_k \, . \label{eq:T12_full}
\end{align}
Note that even though a distribution in correlation times $P(\tau_k)$ exists (which leads to a multi-exponential decay of the autocorrelation function $G(t)$), the observed (i.e., measured) relaxation rate $1/T_{1,2}$ in Eq. \ref{eq:T12_full} is single valued, resulting in  a single-exponential decay of the observed magnetization $M(t)$. 

\begin{figure}[!ht]
    \centering
        \includegraphics[width=0.525\textwidth]{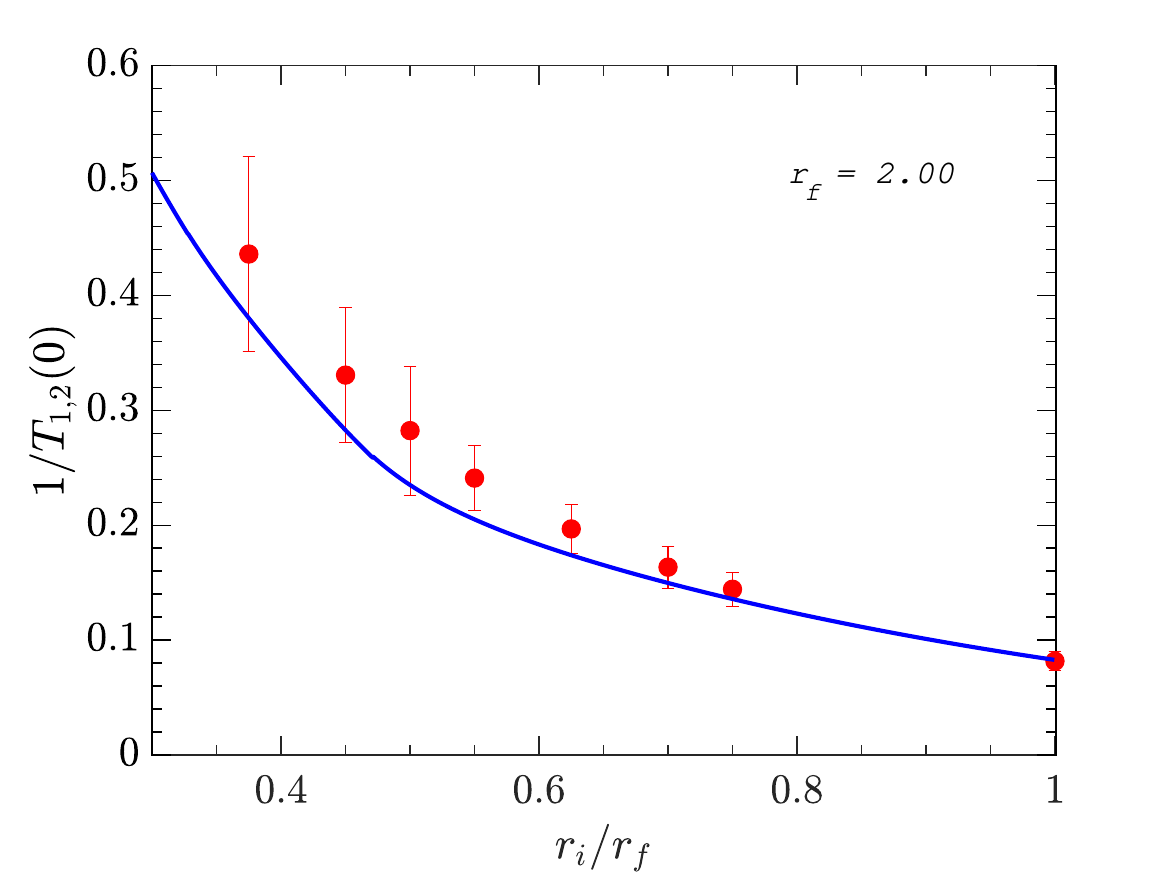}
    \caption{Reduced relaxation rate $1/T_{1,2}$ at $\omega_0 = 0$ (i.e., the zero frequency limit) for different $r_i/r_f$, with $r_f=2.00$, $f=10$ and $T=1.0$. The solid line represents the theoretical prediction, while the solid points represent the MD simulation results. We have also included the MD simulation result in the limit of $r_i/r_f \rightarrow 1$, which is the limit of fixed distance dipoles.}
    \label{fig:Fig7}
\end{figure}

Figure \ref{fig:Fig7} illustrates good agreement between the theoretical predictions for $1/T_{1,2}(0)$ and the results from MD simulations. We have provided in the Supplementary Information a percentage breakdown of the contribution from different molecular modes to the relaxation rate $1/T_{1,2}(0)$ over different ranges of $r_i/r_f$. We find that the first mode contributes to 100\% of $1/T_{1,2}(0)$ at $r_i/r_f=1$, as expected in the BPP limit. However, the contribution from the first mode to $1/T_{1,2}(0)$ decreases to $\simeq 71\%$ at $r_i/r_f=0.375$ (for example), with the remaining $\simeq 29\%$ coming from higher order modes ($k \geq 2$). 

\subsection{Analysis of the NMR dispersion}

The $1/T_{1}(\omega_0)$ dispersion can also be computed from the distribution in modes $P(\tau_k)$, for either like spins \cite{valiyaparambathu:jpcl2023} or unlike spins \cite{Pinheiro2022}. 
For the case of like spins, one can show that \cite{valiyaparambathu:jpcl2023}
\begin{align}
	\frac{1}{T_{1}(\omega_0)} = 2 \sum_k \,P(\tau_k) \left[\frac{\tau_k}{1+(\omega_0\tau_k)^2 } + \frac{4\tau_k}{1+(2\omega_0\tau_k)^2 }\right] \, , \label{eq:T1_dispersion}
\end{align}
In the Supplementary Information we provide the expression for the case of unlike spins.

\begin{figure}[!t]
    \centering
        \includegraphics[width=0.525\textwidth]{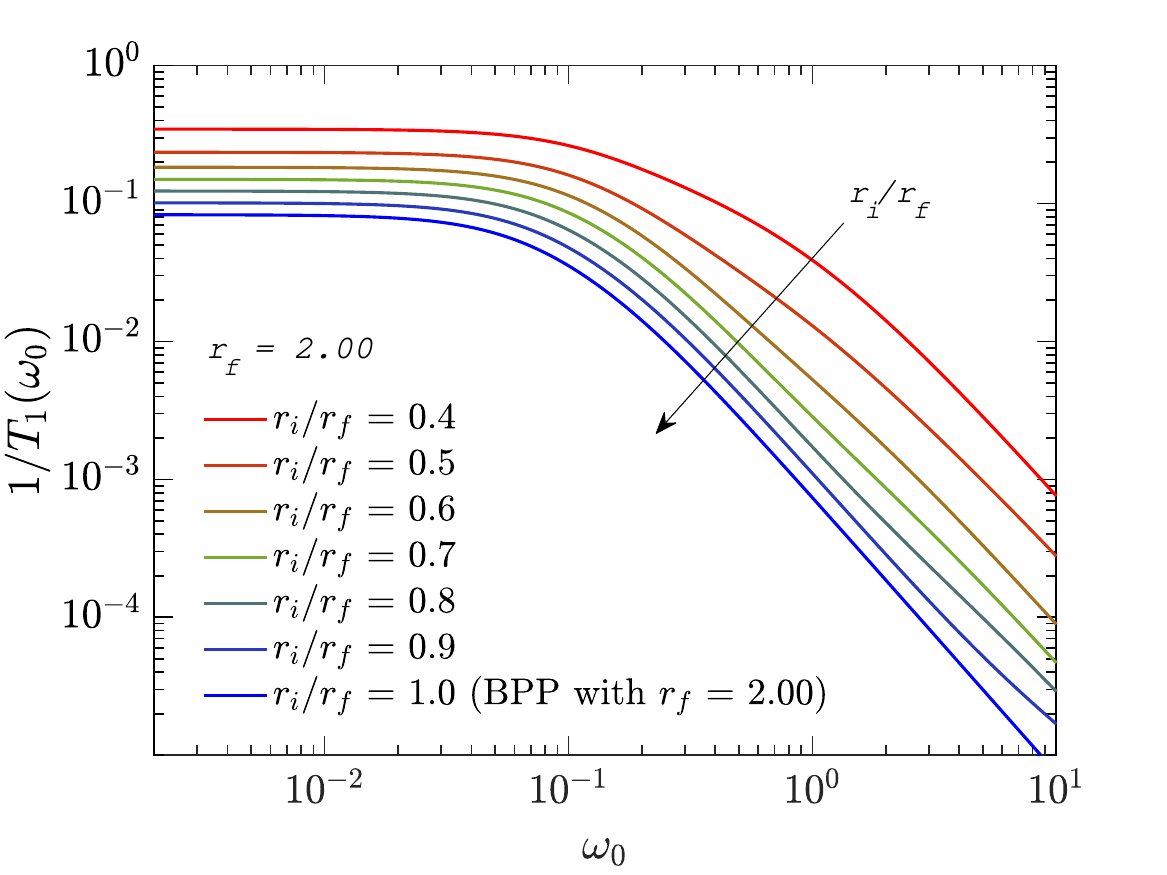}
    \caption{Reduced relaxation rate $1/T_{1}$ dispersion as a function of the frequency $\omega_0$ for different $r_i/r_f$, with $r_f=2.00$, $f=10$ and $T=1.0$.}
    \label{fig:Fig8}
\end{figure} 

\begin{figure}[!t]
    \centering
        \includegraphics[width=0.525\textwidth]{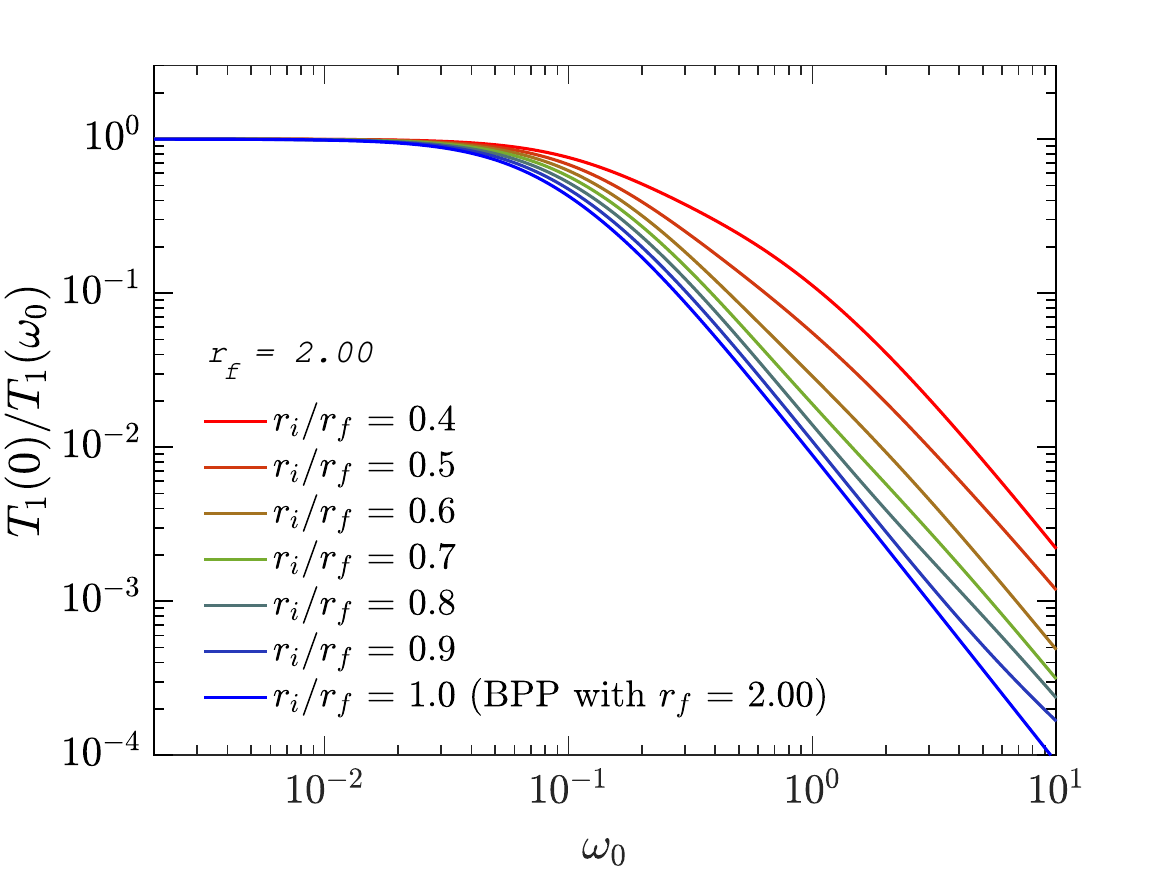}
    \caption{Reduced relaxation rate $1/T_{1}$ dispersion normalized by the corresponding zero frequency limit $1/T_{1}(0)$ for different $r_i/r_f$, with $r_f=2.00$, $f=10$ and $T=1.0$.}
    \label{fig:Fig9}
\end{figure} 

We know that, in general, for any $r_i/r_f (< 1)$, the contribution of the first mode to $1/T_{1}(\omega_0)$ will monotonically decrease with increasing frequency when $\omega_0 \tau_1 \gtrsim 1$.
Thus, as the frequency $\omega_0$ increases, the importance of higher order modes also increase.
Figure \ref{fig:Fig8} shows the NMR dispersion for $1/T_1(\omega_0)$ for different values of $r_i/r_f$.
This plot includes modes up to $k=4$, which for cases studied here 
allow the determination of $1/T_1$ up to $\omega_0=10$. For higher $\omega_0$, more modes will be required. This result confirms that, as the thickness of the annulus increases ($r_i/r_f$ decreases), higher order modes play a bigger role and increase $1/T_1$. Meanwhile, $1/T_{2}(\omega_0)$ shows only minor dispersion.

Figure \ref{fig:Fig9} presents the NMR dispersion for $1/T_1(\omega_0)$ normalized by the corresponding zero frequency limit, i.e., $T_1(0)/T_{1}(\omega_0)$. The plot includes the case $r_i/r_f=1$, which is equivalent to the BPP limit where $1/T_1(\omega_0) \propto \omega_0^{-2}$ at high frequencies. 
For $r_i/r_f < 1$, however, deviations from BPP arise at high frequencies including modified power-law and non power-law dispersion. This signifies that the BPP model for $G(t)$ is not sufficient for $r_i/r_f<1$ at high frequencies when higher order modes start to play a bigger role. It is natural to wonder what is the exact frequency dependence for $1/T_1(\omega_0)$ at high frequencies when $r_i/r_f<1$; an exhaustive study of this point is necessarily left for future investigations.

In the Supplementary Information we show that at $r_i/r_f=0.375$ (for example), the percentage contribution from the first mode to $1/T_{1}(\omega_0)$ decreases as \{71\%,   57\%,  39\%\} with increasing frequency $\omega_0 = \{0,1/\tau_d,2/\tau_d\} = \{0,0.15,0.30\}$, respectively, where $\tau_d$ is the BPP limit at $r_i/r_f=1$ defined in Eq. \eqref{eq:tau_BPP}. This highlights the importance of higher order modes ($k \geq 2$) in $1/T_{1}(\omega_0)$ with increasing  $\omega_0$ and decreasing $r_i/r_f$. Furthermore, for similar physical systems our formalism can determine how many modes are required to describe $1/T_{1}(\omega_0)$ within a specified accuracy, where an increasing number of modes ($k \geq 2$) will be required with increasing $\omega_0$ and decreasing $r_i/r_f$.

\begin{figure*}[!ht]
    \centering
        \includegraphics[width=1.0\textwidth]{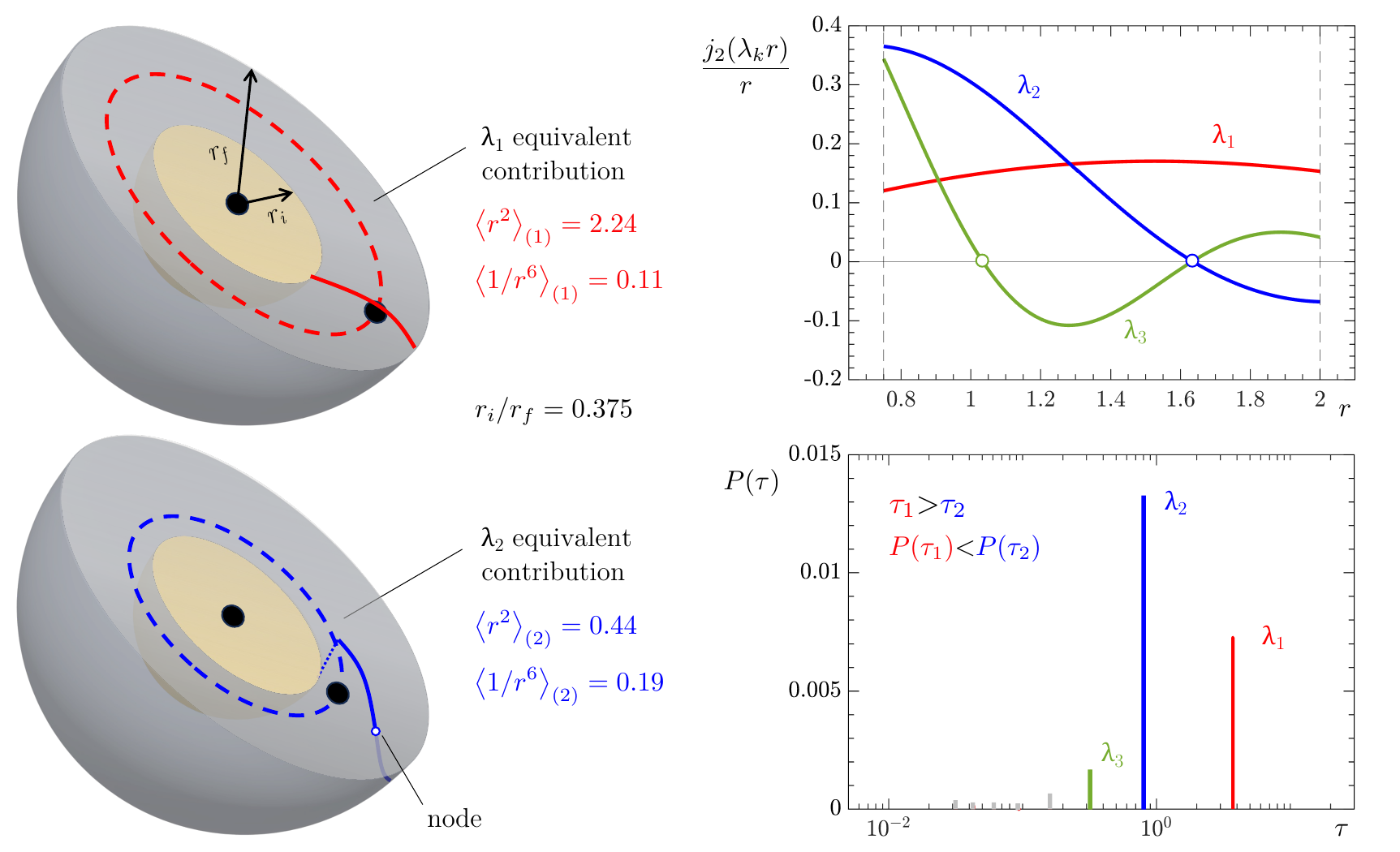}
    \caption{Schematic representation of the molecular modes of NMR relaxation. The multi-modal (multi-exponential) decay arise from the different eigenvalues $\lambda_k$ and their corresponding eigenfunctions. We observe the $k^{th}$ contribution to the total radial probability distribution will present $k-1$ nodes (zero probability points), and hence regions where the eigenfunction goes negative will present a depletion on the probability distribution. Each mode $k$ has a corresponding $\left< r^{2} \right>_{(k)} \propto \tau_k$ and $\left< 1/r^{6} \right>_{(k)} \propto P(\tau_k)$ given by a weighted average through that mode's contribution to the (total) phase distribution function.} 
    \label{fig:Fig10} 
\end{figure*}

\begin{figure*}[!ht]
    \centering
        \includegraphics[width=0.49\textwidth]{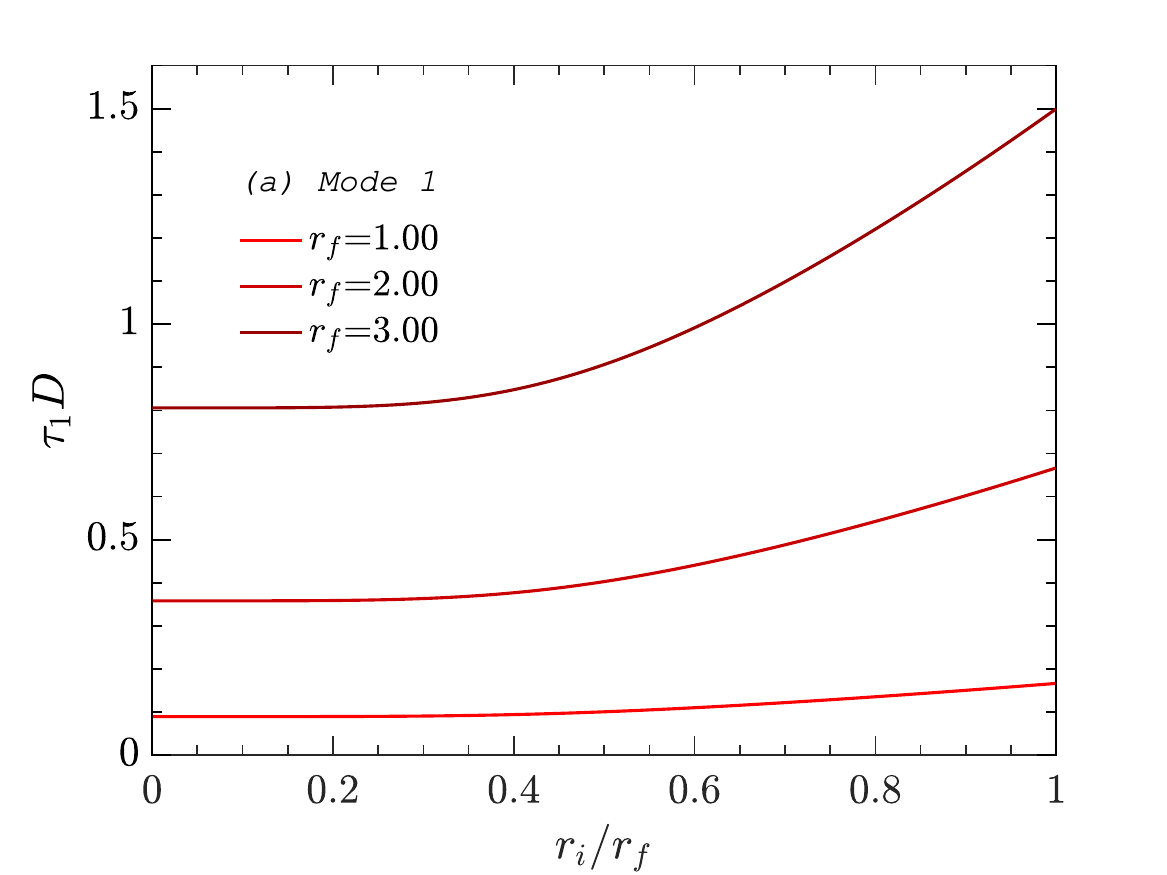}
        \includegraphics[width=0.49\textwidth]{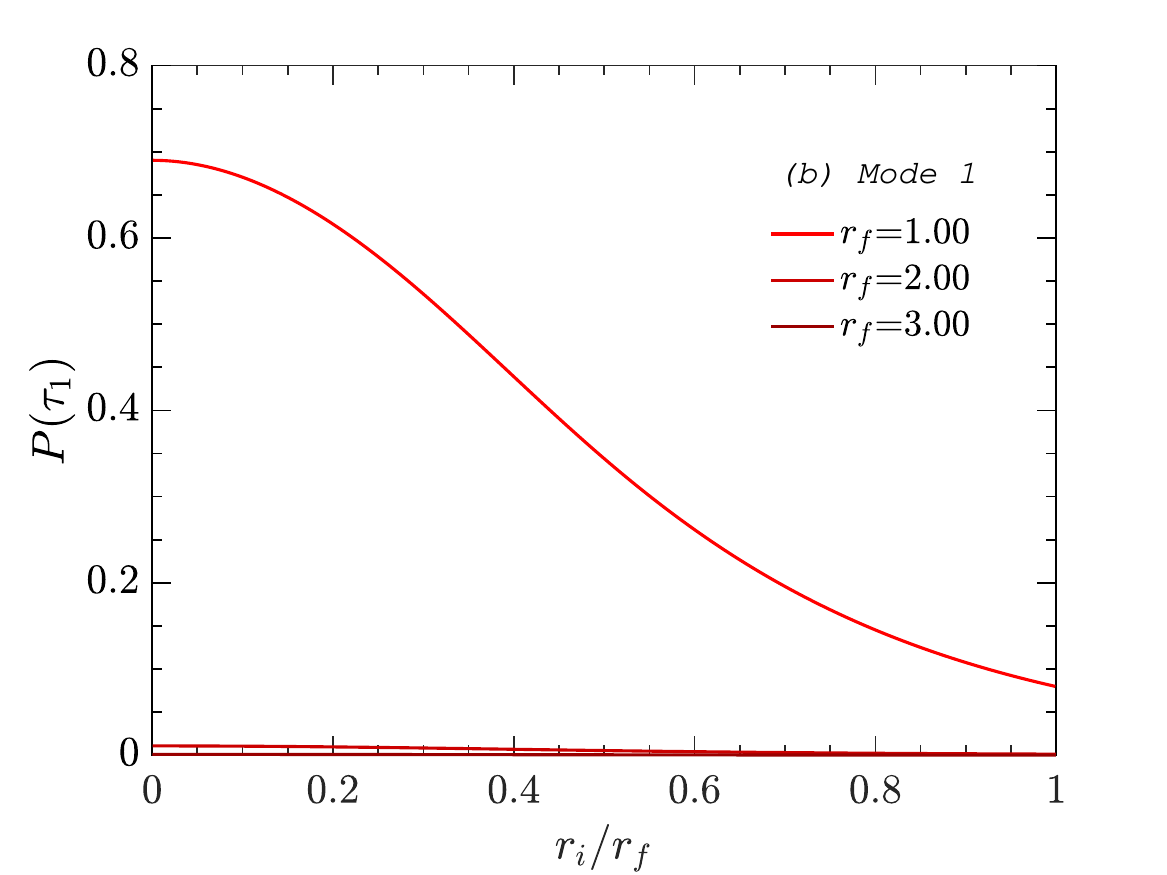}
        \includegraphics[width=0.49\textwidth]{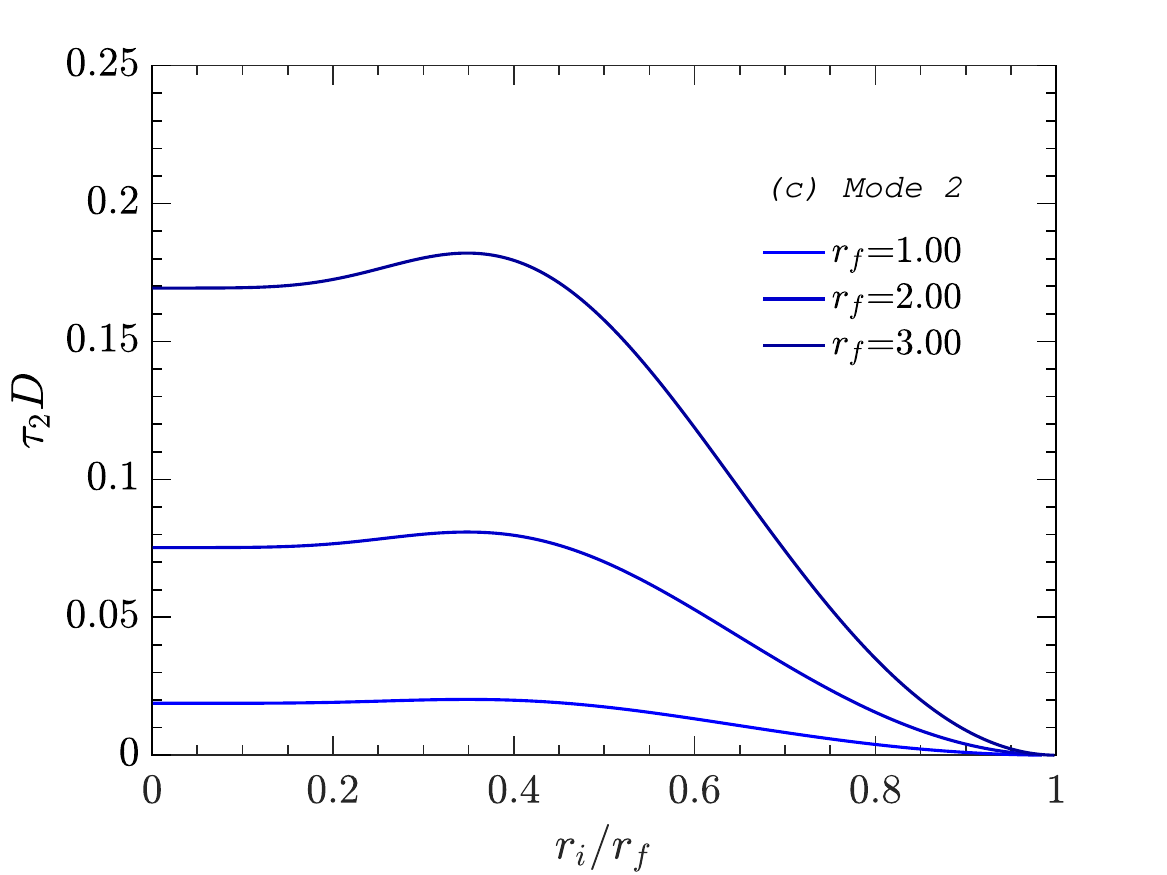}
        \includegraphics[width=0.49\textwidth]{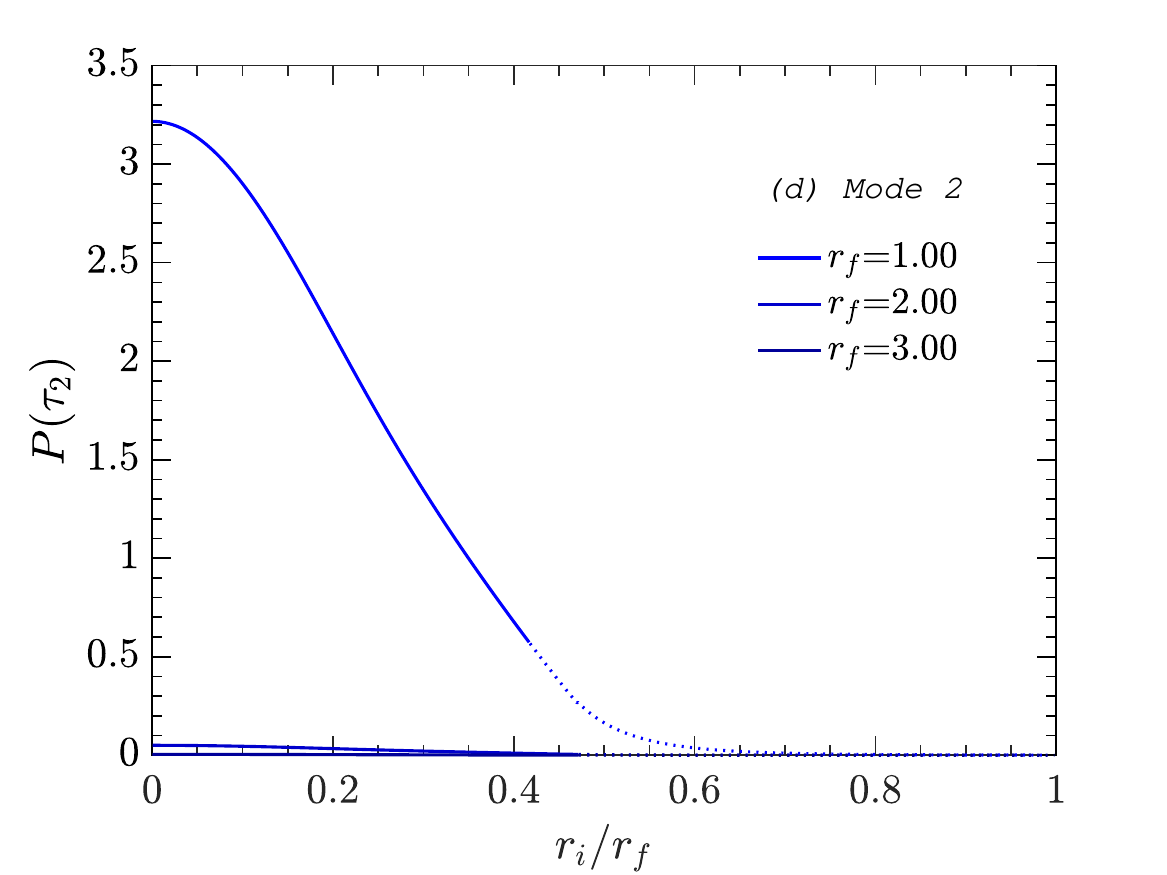}
    \caption{Theoretical prediction of molecular modes characteristic times $\tau_k$ and corresponding amplitudes $P(\tau_k)$ for different $r_f$ values, with $f=10$ and $T=1.0$. Plot (a) shows $\tau_1D$ for the first molecular mode and plot (b) shows its corresponding amplitude $P(\tau_1)$ across $r_i/r_f$. Plot (c) shows $\tau_2D$ for the second molecular mode and plot (d) shows its corresponding amplitude $P(\tau_2)$ across $r_i/r_f$.}
    \label{fig:Fig11}
\end{figure*}

\subsection{Physical interpretation of modes}
The physical interpretation of the molecular modes of NMR relaxation leads to the conclusion that they depend both on the dynamics and the structure of the system. The dynamic contribution can be understood as the effect of the diffusion constant $D$ on the characteristic times $\tau_{k}$, given by Eq. \eqref{eq:tauk_theory_full} for the case under study. This implies that low frictional (viscous) and highly diffusional systems will present shorter relaxation times. The structural effect on the molecular modes of relaxation arises from the individual contributions to the phase distribution function $\rho(\textbf{r},t)$ of the system between the limits $r_i$ and $r_f$. This is analogous to how the quantum harmonic oscillator solution splits into different eigenstates, with discrete energies. In our case, the summation in Eq. \eqref{eq:final_rho} teaches us how the overall (physical) probability distribution partitions over the underlying subsets (states) defined by the eigenmodes, with discrete characteristic times. Figure \ref{fig:Fig10} illustrates how the $k^{th}$ mode, i.e.\ the $k^{th}$-eigenfunction, contributes to the probability distribution $\rho(\textbf{r},t)$. Notice that the mode corresponding to $\lambda_k$ has $k-1$ nodes in $r$, i.e., points where $j_2(\lambda_{k}r)/r=0$ and thus $\rho(\textbf{r},t)=0$. For each mode, the zones where the eigenfunction goes negative will usually involve a depletion of probability and conversely when the eigenfunction is positive. This implies that each $k^{th}$ mode will contribute with a unique and inherent structural probability density in $r$, reminiscent of the idea of inherent structures in liquids \cite{Stillinger1983}.

\begin{figure*}[!ht]
    \centering
        \includegraphics[width=0.49\textwidth]{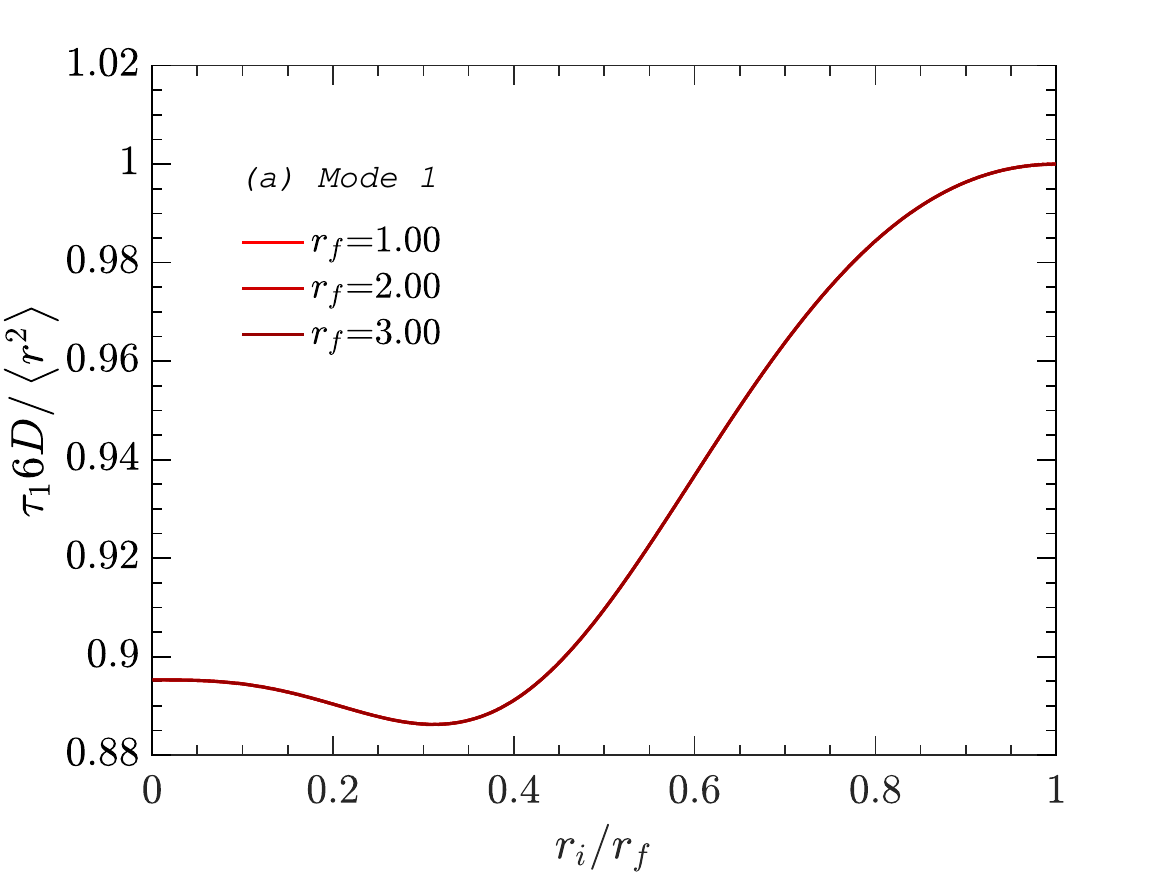}
        \includegraphics[width=0.49\textwidth]{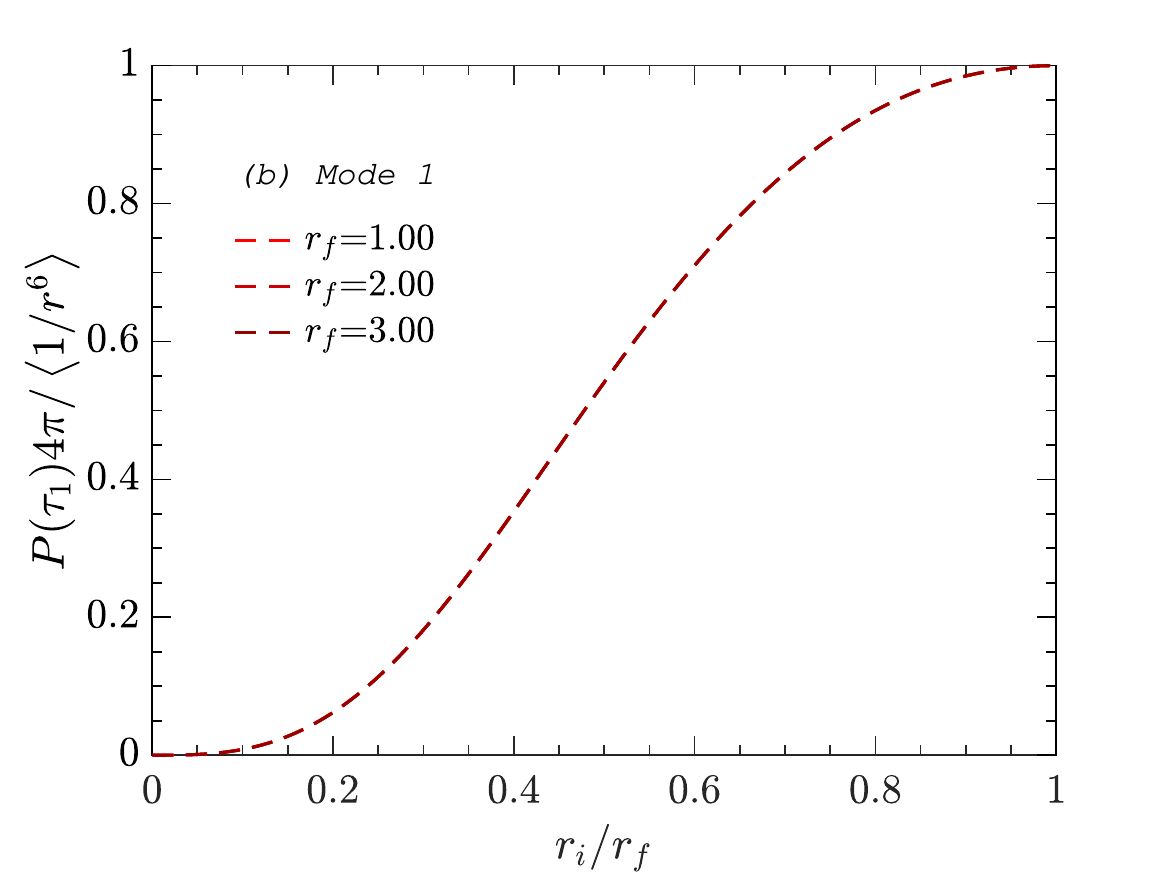}
        \includegraphics[width=0.49\textwidth]{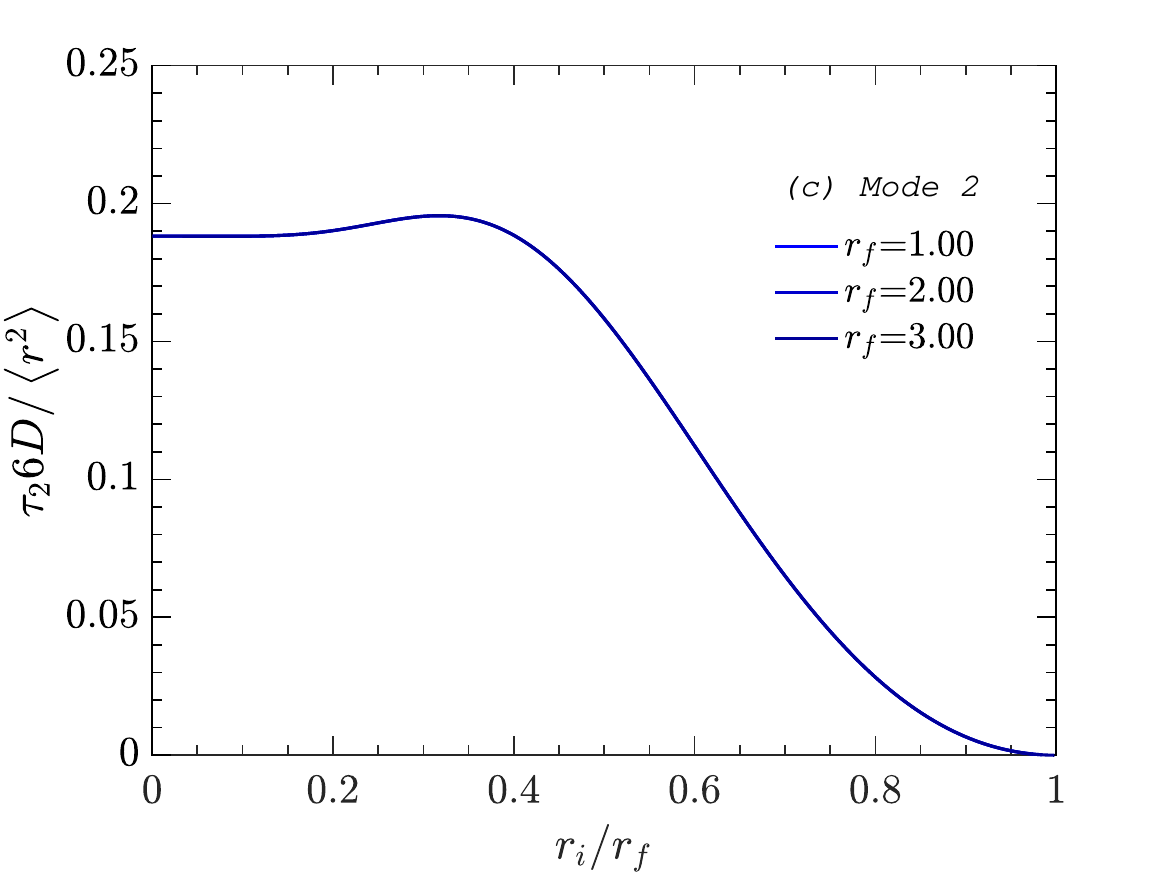}
        \includegraphics[width=0.49\textwidth]{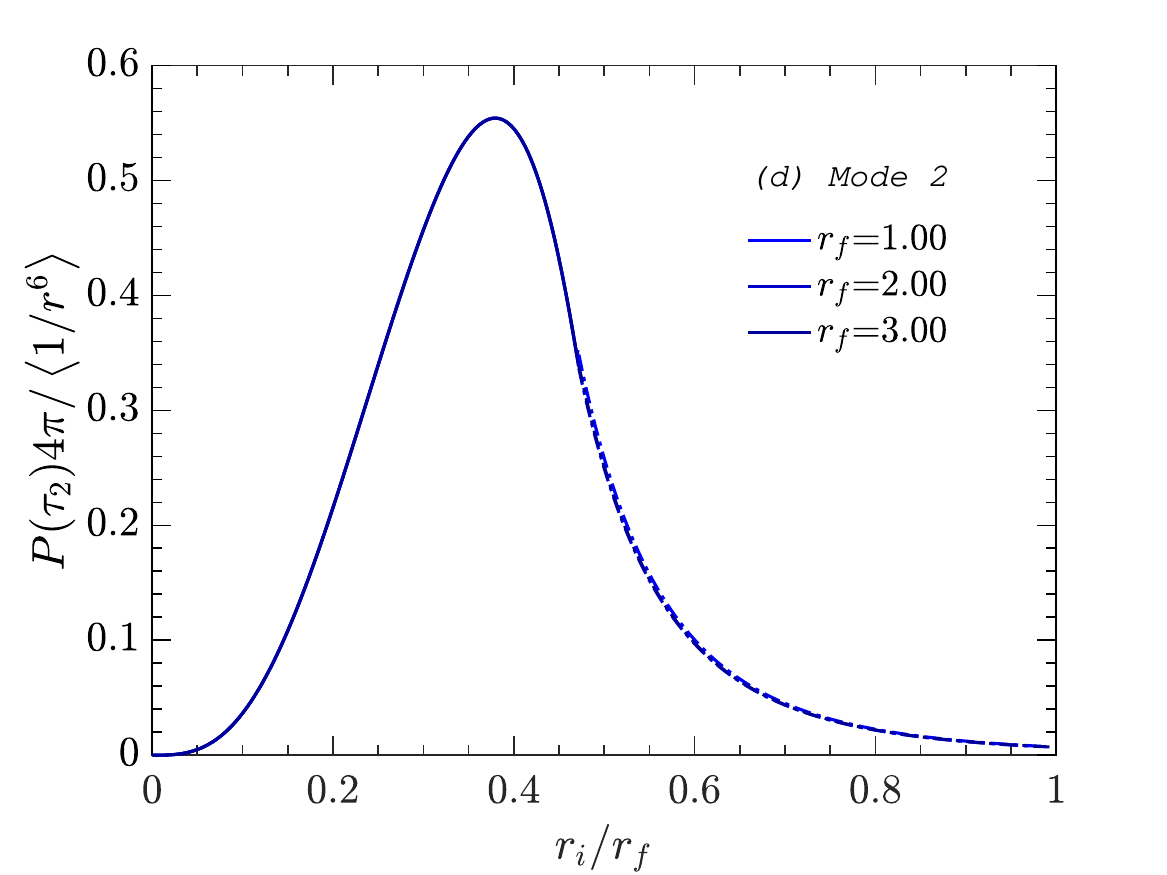}
    \caption{Theoretical prediction of scaled molecular modes characteristic times $6D\tau_k/\left< r^{2} \right>$ and scaled amplitudes $4\pi P(\tau_k)/\left< 1/r^{6} \right>$ for different $r_f$ values, with $f=10$ and $T=1.0$. Plot (a) shows $6D\tau_1/\left< r^{2} \right>$ for the first molecular mode and plot (b) shows its corresponding scaled amplitude $4\pi P(\tau_1)/\left< 1/r^{6} \right>$ across $r_i/r_f$. Plot (c) shows $6D\tau_2/\left< r^{2} \right>$ for the second molecular mode and plot (d) shows its corresponding scaled amplitude $4\pi P(\tau_2)/\left< 1/r^{6} \right>$ across $r_i/r_f$.}
    \label{fig:Fig12}
\end{figure*}

Another way of interpreting these molecular modes of relaxation is by attributing an equivalent BPP-like rotational diffusion contribution to each of them. For any mode $k$, there will exist a weighted average leading to $\left< r^{2} \right>_{(k)}$ that is proportional to $\tau_k$, and similarly a weighted average leading to $\left< 1/r^{6} \right>_{(k)}$ that is proportional to $P(\tau_k)$. The weight in these averages is the mode's contribution to the total phase distribution function. By increasing the order of the mode $k$, $\left<r^{2} \right>_{(k)}$ will always decrease and get closer to the central dipole, signifying shorter characteristic times $\tau_{k}$. On the other hand, $\left<1/r^{6} \right>_{(k)}$ does not always increase with order of the mode $k$, and hence the corresponding amplitude $P(\tau_{k})$ of each mode also depends on the radial spacing of the annulus.
As implied in Eq. \ref{eq:T12_full}, the net effect of these two contributions on the relaxation rate $1/T_{1,2}$ from mode $k$ alone is the product $P(\tau_{k})\, \tau_{k} \propto \left< r^{2} \right>_{(k)}\left< 1/r^{6} \right>_{(k)}$.

Consider the first mode corresponding to $\lambda_{1}$. This mode has no nodes in the radial direction. On average, this is equivalent to a rotational diffusion at $\left< r^{2} \right>_{(1)}$ close to the geometrical average $\left< r^{2} \right>$ between $r_i$ and $r_f$, depending on how flat the corresponding probability density is in that spherical shell. Importantly, in the limit of $r_i/r_f \rightarrow 1$, the diffusion happens exactly at $\left< r^{2} \right>$, which is the BPP limit. 

Next consider the second mode $\lambda_{2}$. The corresponding eigenfunction has one node in the radial direction. Thus, the range of $r$ where the density function is positive contributes positively to the average rotational diffusion of that mode, and conversely the negative part of the density function depletes the probability for its corresponding range of $r$. This gives rise to a unique averaged quantity $\left< r^{2} \right>_{(2)}$ (from the inherent structure of mode 2), which is associated with an equivalent rotational diffusion for the mode.
In a similar fashion, one can interpret the contribution of the higher modes

We can gain further insights into above ideas. Since the magnitude of the NMR relaxation rate ($1/T_{1,2}$) is sensitive to the distance between the two dipoles, it proves helpful to investigate how the modes behave as we change the outer radius $r_f$. Physically, we expect that the larger the distance between the dipoles, the smaller the NMR relaxation rate. Figure \ref{fig:Fig11} shows the characteristic times and corresponding amplitudes for the first two molecular modes across $r_i/r_f$, for different values of $r_f$. 

Figures \ref{fig:Fig11}(a) and \ref{fig:Fig11}(c) show that the characteristic times $\tau_1$ and $\tau_2$ increase in magnitude as we increase $r_f$ at constant diffusivity. This makes physical sense, given that a larger distance between the dipoles means that the second dipole will need to take a longer path to present a certain angular deviation with respect to the central dipole and affect the angular terms that drive NMR relaxation. On the other hand, Figures \ref{fig:Fig11}(b) and \ref{fig:Fig11}(d) show that the amplitudes $P(\tau_1)$ and $P(\tau_2)$ decrease in magnitude as we increase $r_f$ at constant diffusivity. This also agrees with the expected physical results, since the farther away the dipoles, the smaller the strength of NMR relaxation. In the limit that the two particles are infinitely distant (in this case, around $r_f \sim 3$ or greater), the amplitudes of such modes will decay to zero.

Interestingly, our results reveal a scaling behavior. For each characteristic time $\tau_k$, the profiles for different $r_i$ and $r_f$ across $r_i/r_f$ collapse to the same curve if we evaluate the normalized quantity $6D\tau_k/\left<r^{2}\right>$.
Figures \ref{fig:Fig12}(a) and \ref{fig:Fig12}(c) illustrate that all curves match for the normalized $\tau_1$ and $\tau_2$ regardless of the actual spherical thickness for translational diffusion. Similarly, for each amplitude $P(\tau_k)$, the profiles for different $r_i$ and $r_f$ across $r_i/r_f$ collapse to the same curve if we evaluate the normalized quantity $4\pi P(\tau_k)/\left< 1/r^{6}\right>$ (Figures \ref{fig:Fig12}(b) and \ref{fig:Fig12}(d)), regardless of the annulus thickness. 

In the limit $r_i/r_f \rightarrow 1$ both the normalized characteristic time $\tau_1$ and the amplitude $P(\tau_1)$ for the first mode converge to unity, while characteristic times and amplitudes for the higher modes converge to zero, as it should in the 
limit where BPP is valid (cf.\ Eqs.\ \eqref{eq:tau_BPP} and \eqref{eq:Ptau_BPP}). The fact that we are able to describe the entire profile of $\tau_k$'s and $P(\tau_k)$'s over $r_i/r_f$ is an advantage: one can use the normalized results to predict the characteristic time and amplitude for a mode for different spherical thicknesses, and for any absolute distance of interest between the dipoles.

Note that in the (non-physical) limit $r_i/r_f \rightarrow 0$, the amplitudes $P(\tau_k)$ remain finite for $k = \{1,2\}$ (Figures \ref{fig:Fig11}(b,d)), while the normalized $4\pi P(\tau_k)/\left< 1/r^{6}\right>$ decay to zero for $k = \{1,2\}$ due to the divergence $\left< 1/r^{6}\right> \rightarrow \infty$ (Figures \ref{fig:Fig12}(b,d)). 
In this limit, the amplitudes $P(\tau_k)$ grow with $k$, and $P(\tau_k) \rightarrow \infty$ diverge when $\tau_k \rightarrow 0$. Thus, in the non-physical limit of $r_i/r_f \rightarrow 0$, the contributions from modes $k = \{1,2\}$ become increasingly negligible.

\section{Extensions of the model}

 We discuss some possible extensions of our formalism to other cases of physical interest. In the case of a system with $m$ dipoles interacting with a central dipole $c$, the total equilibrium phase distribution function of the system is given by
\begin{equation}
    \rho(\textbf{r}, t) = \sum_{i=1}^{m} \rho_{c,i}(\textbf{r}, t), \label{eq:sum_rho}
\end{equation}
where $\rho_{c,i}(\textbf{r}, t)$ is the probability function of finding dipole $i$ around dipole $c$, for $i=1,2,3,...m$, and subject to the presence of all other dipoles in the system. If we again assume an isotropic system and that all the dipoles are non-interacting and constrained to lie in defined shells around the central dipole, we can get $\rho_{c,i}(\textbf{r}, t)$ through the result derived in Eq. \eqref{eq:final_rho}.

\begin{figure*}[!ht]
    \centering
        \includegraphics[width=0.9\textwidth]{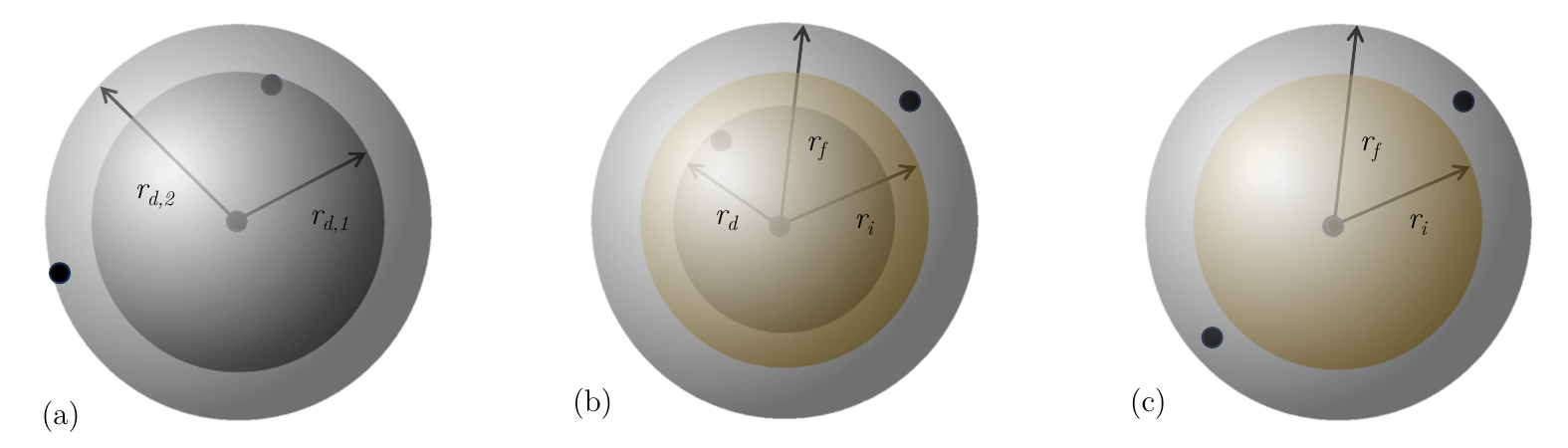}
        \includegraphics[width=0.3\textwidth]{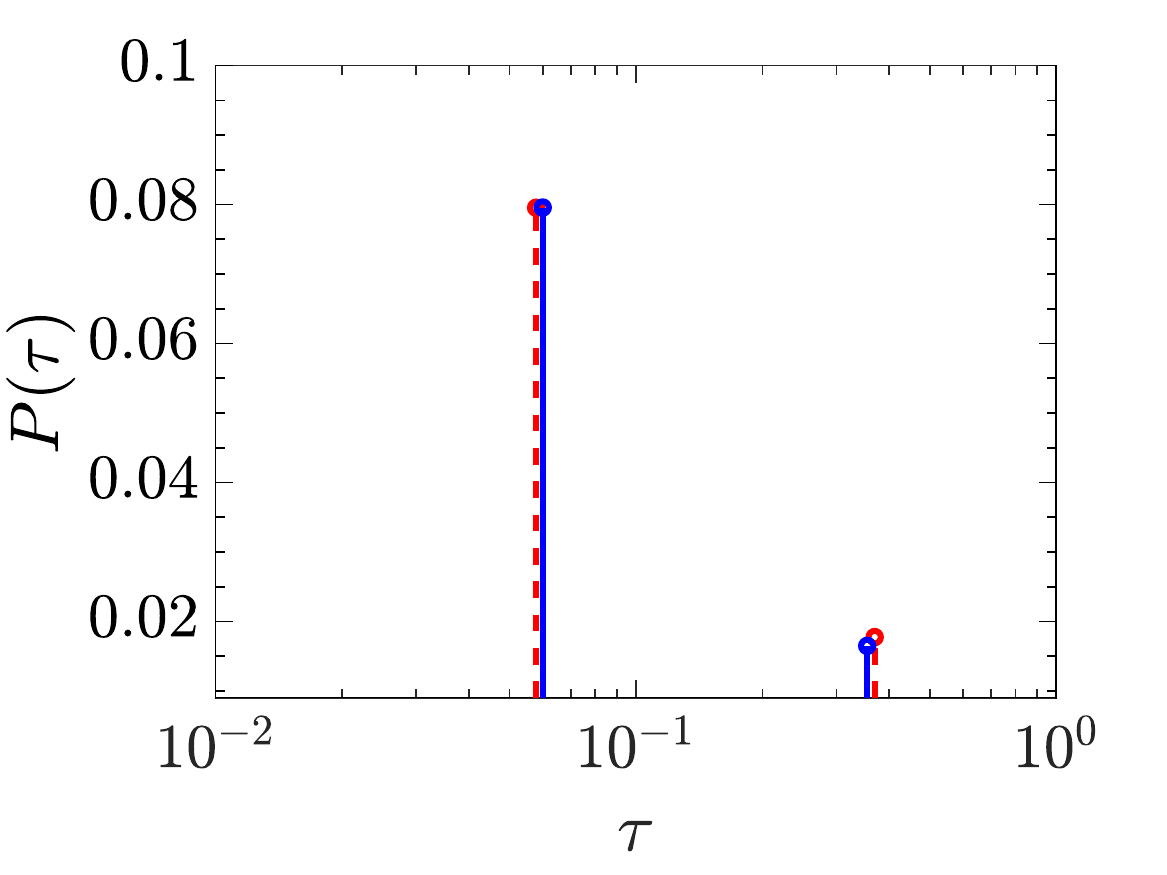}
        \includegraphics[width=0.3\textwidth]{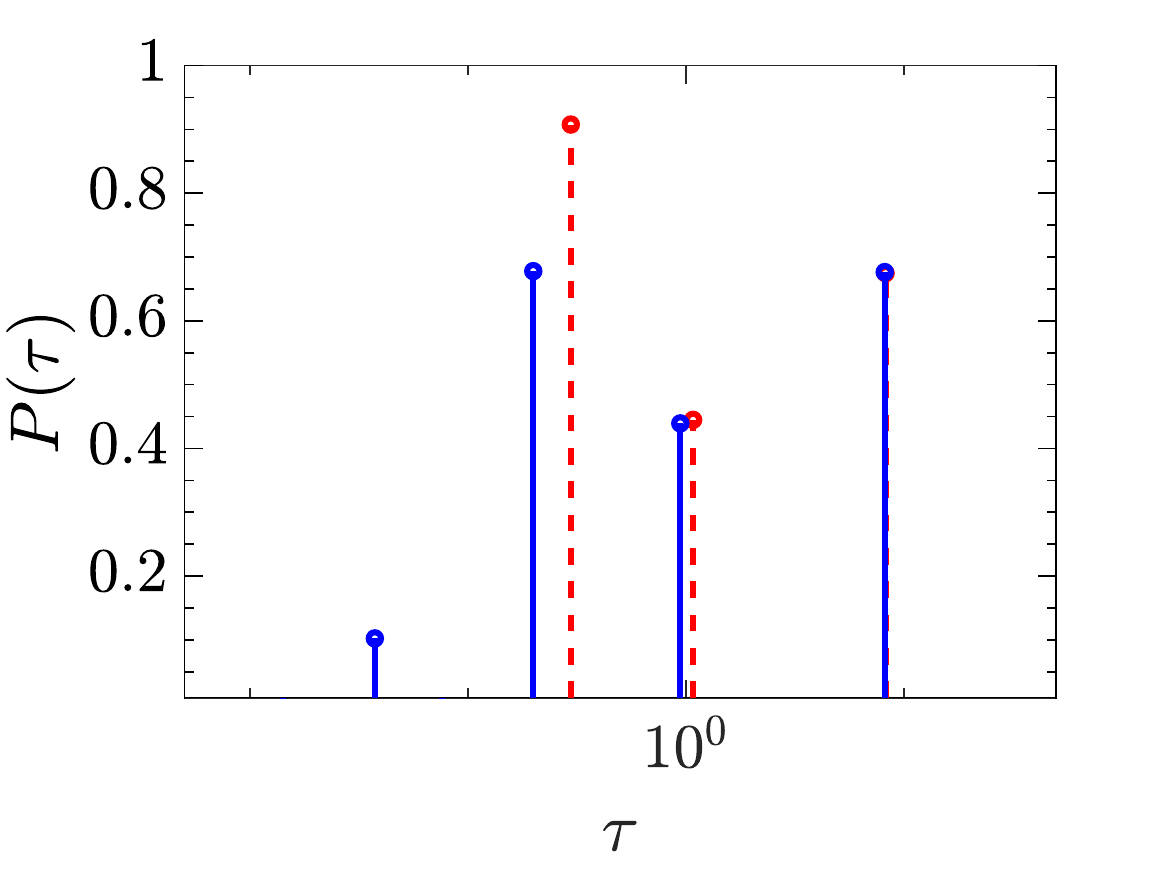}
        \includegraphics[width=0.3\textwidth]{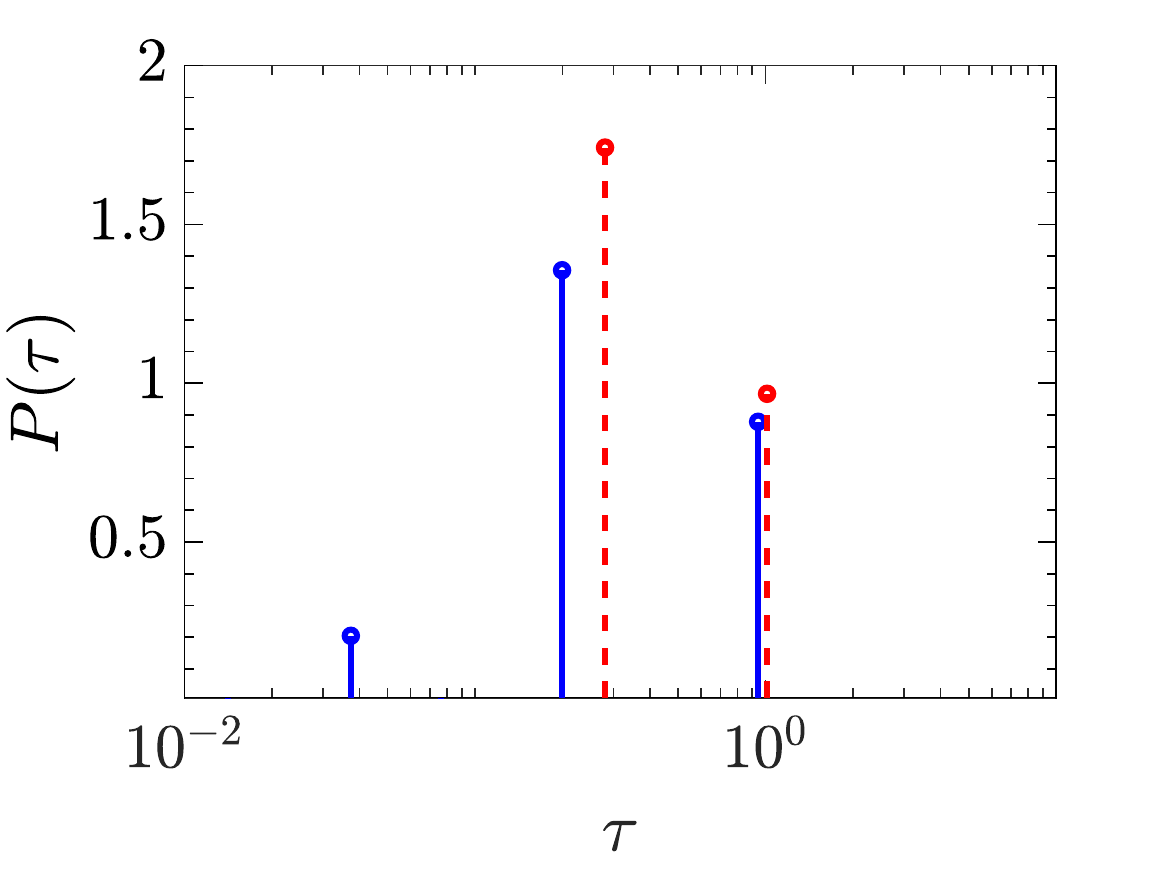}
    \caption{Extended toy-cases of multiple non-interacting dipoles undergoing NMR relaxation with respect to a central dipole and their corresponding molecular modes distribution at $T=1.00$. The solid lines represent the theoretical predictions, while the dashed lines represent the MD simulation results. Case (a) shows two dipoles around the central one at fixed distances $r_{d,1}=1.00$ with $f_1=0.01$, and $r_{d,2}=1.30$ and $f_2=0.10$. Case (b) shows one dipole at a fixed distance $r_{d}=0.70$ with $f_1=0.01$, and another dipole at non-fixed distance diffusing between $r_i=0.40$ and $r_f=1.00$ with $f=10$. Case (c) shows two dipoles at non-fixed distances diffusing between $r_i=0.40$ and $r_f=1.00$ with $f=10$.}
    \label{fig:Fig13}
\end{figure*}

Consider, for example, two non-interacting dipoles around a central one. Figure \ref{fig:Fig13} shows three different systems, having a central dipole and (a) two fixed dipoles around it, (b) one dipole at a fixed distance and another one diffusing on a thick spherical shell, and (c) two dipoles diffusing on a thick spherical shell.
Observe that the NMR autocorrelation function for dipole pairs from  MD simulations is a single averaged function that subsumes all the modes present in the system. The task for the Pad\'e-Laplace inversion is to decode the underlying exponential decays. Hence, if the modes are above the frequency of sampling from the simulation and their amplitudes present a high signal-to-noise ratio, we find that the Pad\'e-Laplace inversion will capture these modes. For this example, we find that the theory can also predict relatively well the multi-modal behavior.

Another possible extensions is the inclusion of a potential of interaction $U(r)$ between the dipoles in Eq.\ \eqref{eq:Gen_FP_2}, which would affect the corresponding $\rho(\textbf{r}, t)$ and, ultimately, the molecular modes and the spin-spin and spin-lattice relaxation times. (Studies in this direction are underway and the results will be communicated later.) 
Finally, a natural extension of the model would be to investigate intra-molecular relaxation between dipoles in a chain (e.g., alkanes), by assuming that $r_f$ is the maximum distance between the two dipoles when the chain is fully stretched, and $r_i$ is the minimum distance between the dipoles when the chain is coiled at its maximum. In this case, however, relaxing boundary conditions would be needed instead of reflective ones, implying that SL-EVP for the calculation of the characteristic time will have a different mathematical form, and the definition of orthogonality of the eigenfunctions will change accordingly.

\section{Conclusion}

We investigate the emergence of a multi-exponential NMR autocorrelation function for dipole pairs undergoing regular rotational and translational diffusion, and propose a theoretical framework to model such multi-exponential behavior, here coined as molecular modes of NMR relaxation. This multi-exponential behavior in the autocorrelation function leads to anomalous NMR relaxation dispersion compared to the traditional BPP model. A key feature is that there are no adjustable parameters in interpreting the NMR relaxation behavior. We examine the case of two non-interacting dipoles in a viscous and isotropic system, where distance between the dipoles can freely vary between defined lower and upper limits. Even this simple case leads to a multi-mode (multi-exponential) behavior, which our theory is able to capture. Such multi-exponential behavior of the NMR autocorrelation function essentially arises from a corresponding eigenvalue problem, that depends on the boundary conditions of the problem and the geometry of the systems. Molecular simulations associated with the Pad\'e-Laplace inversion validate the theoretical predictions with good agreement.

Our investigations clarify the physical meaning of such molecular modes of NMR relaxation. The dynamics of the system directly impacts these characteristic times $\tau_k$, which are inversely proportional to the diffusion constant $D$ of the dipoles in the viscous fluid. However, structural effects also play a major role in these molecular modes, and we observe that different modes $\tau_k$ contribute to the solution with a corresponding eigenfunction with $k-1$ nodes in the radial direction (zero probability density). The first mode (largest characteristic time) has no nodes in the radial distribution, and then the averaged angular diffusion given by this mode happens on a sphere with radius close to the geometrical average $\left<r^2\right>$ over the full radial range. The higher-order modes present nodes in the radial distribution, and hence the averaged angular diffusion for such nodes happens in different equivalent radii.

A key result is that our theoretical formalism agrees with the traditional BPP theory in the limit where the distance between the two dipoles becomes constant ($r_i/r_f \rightarrow 1$). The model also agrees with the expected response in the limit where the two dipoles come close to each other ($r_i/r_f \rightarrow 0$), and it also supports the empirical observations that the temperature dependence of the modes' characteristic times $\tau_k$ follows an Arrhenius-like model for high temperatures.
For the cases studied in this work, the characteristic times of the molecular modes $\tau_k$ and their corresponding amplitudes $P(\tau_k)$ can be scaled by $6D\tau_k/\left< r^2 \right>$ and $4\pi P(\tau_k)/\left< 1/r^6 \right>$, respectively. This allows us to predict the molecular modes and the corresponding amplitudes for any $r_i$ and $r_f$ combination between the dipoles undergoing relaxation.

We have also presented the NMR dispersion for $1/T_1(\omega_0)$ for the case of like-spins, showing that the importance of higher order modes (and hence, the deviation from the BPP model) increases with increasing $\omega_0$ and decreasing $r_i/r_f$.
Finally, we discussed some extensions of our formalism and future work to cover more cases of interest, like the case with potential of interaction or the intramolecular relaxation of dipole pairs in a molecular chain. Being able to model and understand the individual modes that lead to NMR dipole-dipole relaxation can provide crucial physical insights into phenomena on the molecular level.


\section*{Acknowledgements}

The authors thank Arjun Valya Parambathu for the insightful discussions. We also thank the Ken Kennedy Institute, the Rice University Creative Ventures Fund (Faculty Initiatives Fund), and the Robert A. Welch Foundation for the financial support. 
We gratefully acknowledge the National Energy Research Scientific Computing Center, which is supported by the Office of Science of the U.S. Department of Energy (No.\ DE-AC02-05CH11231) and the Texas Advanced Computing Center (TACC) at The University of Texas at Austin for high-performance computer time and support. 
Research at Oak Ridge National Laboratory is supported under contract DE-AC05-00OR22725 from the U.S. Department of Energy to UT-Battelle, LLC. This research used resources of National Energy Research Scientific Computing Center, which is supported by the Office of Science of the U.S. Department of Energy under Contract \# DE-AC02-05CH11231.

\section*{Supplementary Information}

The Supplementary Information provides the detailed analytical solution of the NMR dipole-dipole autocorrelation function for the cases of fixed and non-fixed distance between the dipoles, as well as the comparison between our theoretical predictions and MD simulations raw data for the modes' amplitudes and characteristics times for all studied cases. We have also included a comparison of the NMR dipole-dipole autocorrelation obtained using MD simulations and the corresponding reconstructed signal using the Pad\'e-Laplace method. Moreover, we provide an analysis of the accuracy of the numerical approximations made to compute integrals using the regular Euclidean definition of inner-product. Finally, we provide the theoretical breakdown contribution from different molecular modes to NMR relaxation dispersion.

\section*{Data Availability}

The data that support the findings of this study are available from the corresponding authors upon reasonable request.

\section*{References}

%

\end{document}


\preprint{}

\title[NMR modes]{Supplementary Information: Theory and modeling of molecular modes in the NMR relaxation of fluids}

\author{Thiago J. Pinheiro dos Santos}
\affiliation{Department of Chemical and Biomolecular Engineering, Rice University, Houston, Texas 77005, USA.}
\author{Betul Orcan-Ekmekci}
\affiliation{Department of Mathematics, Rice University, Houston, Texas 77005, USA.}
\author{Walter G. Chapman}
\affiliation{Department of Chemical and Biomolecular Engineering, Rice University, Houston, Texas 77005, USA.}
\author{Philip M. Singer}
\affiliation{Department of Chemical and Biomolecular Engineering, Rice University, Houston, Texas 77005, USA.}
\email{ps41@rice.edu}
\author{Dilipkumar N. Asthagiri}
\affiliation{Oak Ridge National Laboratory, Oak Ridge, Tennessee 37830, USA.}
\email{asthagiridn@ornl.gov}

\date{\today}
             
\setcounter{figure}{0}
\setcounter{equation}{0}

\renewcommand{\thefigure}{S\arabic{figure}}
\renewcommand{\theequation}{S\arabic{equation}}

\begin{absolutelynopagebreak}

\maketitle

\noindent \scriptsize{Notice: This manuscript has been authored by UT-Battelle, LLC, under contract DE-AC05-00OR22725 with the US Department of Energy (DOE). The US government retains and the publisher, by accepting the article for publication, acknowledges that the US government retains a nonexclusive, paid-up, irrevocable, worldwide license to publish or reproduce the published form of this manuscript, or allow others to do so, for US government purposes. DOE will provide public access to these results of federally sponsored research in accordance with the DOE Public Access Plan (http://energy.gov/downloads/doe-public-access-plan).}

\end{absolutelynopagebreak}

\tableofcontents

\clearpage
\subsection{Fixed distance and non-interacting dipoles}

Let us consider the case where two, non-interacting dipoles are at fixed distance $r_d$.
For convenience, we fix one of the dipoles at the center of coordinates, and 
solve for the probability density of the second dipole around the central one. In spherical coordinates, we have $\textbf{r}_d = \{ r_d, \theta, \phi \}$ and $U(\textbf{r}_d)=0$. Hence, Eq. (8) can be written as
\begin{equation}
    \frac{\partial}{\partial t} \rho(\textbf{r}_d,t) = D \nabla^2_r \rho(\textbf{r}_d,t). \label{eqSI:FP_BPP}
\end{equation}
where the Laplacian operator at constant $r=r_d$ is given by
\begin{equation}
         \nabla^2_r = - \frac{\hat{I}^2}{r_d^2} = \frac{1}{r_d^2 \sin \theta} \frac{\partial}{\partial \theta} \left( \sin \theta \frac{\partial}{\partial \theta} \right) + \frac{1}{r_d^2 \sin^2 \theta} \frac{\partial^2}{\partial \phi^2}, \label{eqSI:mom_op_r}
\end{equation}
where $\hat{I}^2$ is the angular momentum operator from quantum mechanics. This problem should be solved observing periodic boundary condition for the angular terms, i.e., $\rho (r_d,\theta,\phi,t)=\rho (r_d,\theta+\pi,\phi,t)$ and $\rho (r_d,\theta,\phi,t)=\rho (r_d,\theta,\phi+2\pi,t)$.

Assuming separation of variables, the solution $\rho(\textbf{r}_d,t)$ can be written as 
\begin{equation}
    \rho(\textbf{r}_d, t) = T(t) \Phi(\theta,\phi) \label{eqSI:expeq_BPP}
\end{equation}
where $T(t)$ is the part of the solution that depends only on time, and $\Phi(\theta,\phi)$ contains the angular dependence of the solution. Substituting Eq. \eqref{eqSI:expeq_BPP} into Eq. \eqref{eqSI:FP_BPP}, we find that
\begin{equation}
    \frac{r_d^2}{D T(t)}\frac{\partial}{\partial t} T(t) = \frac{1}{\Phi(\theta,\phi)} \left( \frac{1}{ \sin \theta} \frac{\partial}{\partial \theta} \left( \sin \theta \frac{\partial}{\partial \theta} \right) \Phi(\theta,\phi) + \frac{1}{  \sin^2 \theta} \frac{\partial^2}{\partial \phi^2} Y(\theta,\phi) \right) \Phi(\theta,\phi) = - n(n+1),
\end{equation}
in which the first term of the equality only depends on time, the second one only depends on the angular components, and then both should be equal to a constant $-n(n+1)$ chosen for convenience. Therefore, the time-dependent part will be given by
\begin{equation}
    \begin{split}
        \frac{1}{T(t)}\frac{\partial}{\partial t} T(t) = - \frac{n(n+1) D}{r_d^2} \\
        T(t) = C_1 \exp \left( - \frac{n(n+1) D}{r_d^2} t \right)
    \end{split}
\end{equation}
where $C_1$ is a constant of integration. On the other hand, the angular-dependent part will be given by
\begin{equation}
    \begin{split}
        \left( \frac{1}{ \sin \theta} \frac{\partial}{\partial \theta} \left( \sin \theta \frac{\partial}{\partial \theta} \right) Y(\theta,\phi) + \frac{1}{  \sin^2 \theta} \frac{\partial^2}{\partial \phi^2} Y(\theta,\phi) \right) Y(\theta,\phi) &= - n(n+1) Y(\theta,\phi) \\
        \hat{I}^2 \Phi(\theta,\phi) &= n(n+1) \Phi (\theta,\phi).
    \end{split}
\end{equation}

At this point, it is important to recognize that the eigenfunctions and eigenvalues of the angular momentum operator follow
\begin{equation}
    \hat{I}^2 Y_n^m(\theta, \phi) = n(n+1) Y_n^m(\theta, \phi).
\end{equation}
where $Y_n^m(\theta, \phi)$ are the spherical harmonic functions associated with the corresponding Legendre polynomials of order $n$ and $m$, i.e., $P_n^m(x)$. Hence, the angular-dependent solution $\Phi(\theta,\phi)$ is given by the linear combination of spherical harmonic functions, i.e., 
\begin{equation}
    \Phi(\theta,\phi) = \sum_{n=0}^{\infty} \sum_{m=-n}^{n} C_{n,m} Y_n^m(\theta, \phi).
\end{equation}
Finally, the solution for $\rho(r_d, \theta, \phi,t)$ is given by
\begin{equation}
    \rho(\textbf{r}_d, t) = \sum_{n=0}^{\infty} \sum_{m=-n}^{n} A_{n,m} Y_n^m(\theta, \phi) \exp \left( - \frac{n(n+1) D}{r_d^2} t \right)
\end{equation}
where the coefficients $A_{n,m}$ depend on the initial condition $\rho(\textbf{r}_d,0)$ as
\begin{equation}
    A_{n,m} = \frac{ \int_0^{2\pi} \int_0^{\pi} \rho(\textbf{r}_d,0) Y_n^m(\theta, \phi) \sin \theta d\theta d\phi }{\int_0^{2\pi} \int_0^{\pi} | Y_n^m(\theta, \phi) |^2 \sin \theta d\theta d\phi}.
\end{equation}

The initial distribution of the system is given by the delta functions at an initial configuration $\{ r_d, \theta_0, \phi_0 \}$, which is proportional to the spherical harmonic closure relationship\footnote{
From the closure relationship of spherical harmonics, we know that 
\begin{equation}
    \begin{split}
    \frac{\delta (\theta-\theta_0) \delta (\phi-\phi_0)}{\sin \theta} = \sum_{n=0}^{\infty} \sum_{m=-n}^{n} Y_{n}^{m*}(\theta,\phi) Y_{n}^{m}(\theta_0,\phi_0)
\end{split}
\end{equation}
}, i.e.,
\begin{equation}
    \begin{split}
        \rho(\textbf{r}_d,0) &= \frac{\delta(\theta - \theta_0) \delta(\phi-\phi_0)}{\sin \theta} p(\textbf{r}_{d,0}) \\
        &= p(\textbf{r}_{d,0})  \sum_{n=0}^{\infty} \sum_{m=-n}^{n} Y_{n}^{m*}(\theta,\phi) Y_{n}^{m}(\theta_0,\phi_0)
    \end{split}
\end{equation}
where the probability $p(\textbf{r}_{d,0})$ of having the initial condition configuration at equilibrium is given by
\begin{equation}
    \begin{split}
        p(\textbf{r}_{d,0}) &= \frac{p(\theta_0) p(\phi_0)}{\sin \theta_0}\\
        &= \frac{1}{\sin \theta_0}\frac{\sin \theta_0}{\int_0^{\pi} \sin \theta d \theta} \frac{1}{\int_0^{2\pi} d\phi} \\
        & = \frac{1}{4\pi},
    \end{split}
\end{equation}
such that
\begin{equation}
    \begin{split}
        \rho(\textbf{r}_d,0) &= \frac{1}{4\pi}  \sum_{n=0}^{\infty} \sum_{m=-n}^{n} Y_{n}^{m*}(\theta,\phi) Y_{n}^{m} (\theta_0,\phi_0).
    \end{split}
\end{equation}

In NMR relaxation, we are only interested in the second moment ($n=2$), and we will also assume that the system is isotropic ($m=0$). Thus, the general solution to $\rho(\textbf{r}_d,t)$ simplifies to
\begin{equation}
    \rho(\textbf{r}_d, t) = A_{2,0} Y_n^m(\theta, \phi) \exp \left( - \frac{6D}{r_d^2} t \right),
\end{equation}
with the corresponding initial condition
\begin{equation}
    \begin{split}
        \rho(\textbf{r}_d,0) &= \frac{1}{4\pi}  Y_{2}^{0*}(\theta,\phi) Y_{2}^{0}(\theta_0,\phi_0).
    \end{split}
\end{equation}
Thus, the coefficient $A_{2,0}$ is given by
\begin{equation}
    \begin{split}
        A_{2,0} & = \frac{1}{4\pi} \frac{ \int_0^{2\pi} \int_0^{\pi} Y_{2}^{0*}(\theta,\phi) Y_{2}^{0}(\theta_0,\phi_0) Y_2^0(\theta, \phi) \sin \theta d\theta d\phi }{\int_0^{2\pi} \int_0^{\pi} | Y_2^0(\theta, \phi) |^2 \sin \theta d\theta d\phi} \\
        & = \frac{1}{4\pi} Y_{2}^{0}(\theta_0,\phi_0),
    \end{split}
\end{equation}
in which we consider the orthogonality of spherical harmonics\footnote{
From the orthogonality relationship of spherical harmonics, we know that 
\begin{equation}
    \int_{0}^{2\pi} \int_{0}^{\pi} Y_n^m(\theta,\phi) Y_n^{m*}(\theta,\phi) \sin \theta d\theta d\phi = \delta_{n,n} \delta_{m,m}.
\end{equation}
}.

The final solution to the time-dependent equilibrium distribution probability between the two dipoles undergoing NMR relaxation at constant distance is given by
\begin{equation}
    \begin{split}
        \rho(\textbf{r}_d,t) &= \frac{1}{4\pi}  Y_{2}^{0*}(\theta,\phi) Y_{2}^{0}(\theta_0,\phi_0) \exp \left( -\frac{6Dt}{r_d^2} \right).
    \end{split}
\end{equation}

Using this result in Eq. (5) without integrating over the constant $r_d$, we find that
\begin{equation}
    \begin{split}
        G(t) & = \frac{4\pi}{5} \left( \frac{\mu_0}{4 \pi} \right)^2 \hbar^2 \gamma_I^2 \gamma_S^2 S(S+1) \frac{1}{4\pi} \frac{1}{r_d^6} \exp \left( -\frac{6Dt}{r_d^2} \right) \\ & \int_0^{2\pi} \int_0^{\pi} Y_2^0 \left( \theta_0, \phi_0 \right) Y_2^0 \left( \theta_0, \phi_0 \right) \sin \theta_0 d\theta_0 d\phi_0 \int_0^{2\pi} \int_0^{\pi} Y_{2}^{0*}(\theta,\phi) Y_{2}^{0}(\theta,\phi) \sin \theta d\theta d\phi, \label{eqSI:Gm_BPP}
    \end{split}
\end{equation}
and by the orthogonality of the spherical harmonics, we obtain that
\begin{equation}
    \begin{split}
        & G(t) = \frac{4\pi}{5} \left( \frac{\mu_0}{4 \pi} \right)^2 \hbar^2 \gamma_I^2 \gamma_S^2 S(S+1) \frac{1}{4\pi r_d^6} \exp \left( -\frac{6Dt}{r_d^2} \right). \label{eqSI:BPP}
    \end{split}
\end{equation}
This result is simply the mono-exponential decay from the traditional BPP theory that has been derived in different ways by others\cite{Bruce2000,Bloembergen1961}. \newline

\clearpage
\subsection{Non-fixed distance and non-interacting dipoles}

Let us next consider the case where the two, non-interacting dipoles are not at fixed distance. Again, we fix one of the dipoles at the center of coordinates for convenience. If we adopt spherical coordinates to solve the problem, we have that $\textbf{r} = \{ r, \theta, \phi \}$, $U(\textbf{r})=0$, and Eq. (8) becomes 
\begin{equation}
    \frac{\partial}{\partial t} \rho(\textbf{r},t) = D \nabla^2 \rho(\textbf{r},t), \label{eqSI:FP_NI}
\end{equation}
where the full Laplacian operator is given by
\begin{equation}
         \nabla^2 = - \frac{\hat{I}^2}{r^2} = \frac{1}{r^2} \frac{\partial}{\partial r} \left( r^2 \frac{\partial}{\partial r} \right) + \frac{1}{r^2 \sin \theta} \frac{\partial}{\partial \theta} \left( \sin \theta \frac{\partial}{\partial \theta} \right) + \frac{1}{r^2 \sin^2 \theta} \frac{\partial^2}{\partial \phi^2}. \label{eqSI:mom_op}
\end{equation}
Again, this problem should be solved observing periodic boundary condition for the angular terms as in $\rho (r,\theta,\phi,t)=\rho (r,\theta+\pi,\phi,t)$ and $\rho (r,\theta,\phi,t)=\rho (r,\theta,\phi+2\pi,t)$. We employed Neumann boundary conditions for the radial term (reflecting boundary conditions), given by
\begin{equation}
\begin{split}
    \left. \frac{\partial \rho (r,\theta,\phi,t)}{\partial r} \right|_{r=r_i} = 0, \\
    \left. \frac{\partial \rho (r,\theta,\phi,t)}{\partial r} \right|_{r=r_f} = 0.
    \label{eqSI:bcs}
\end{split}
\end{equation}

Assuming separation of variables, the solution $\rho(\textbf{r},t)$ can be written as 
\begin{equation}
    \rho(\textbf{r}, t) = T(t) R(r) \Phi(\theta,\phi), \label{eqSI:expeq_NI}
\end{equation}
where $T(t)$ is the part of the solution that depends only on time, $R(r)$ is the radial dependent solution, and $\Phi(\theta,\phi)$ contains the angular dependency of the solution. Hence, by substituting Eq. \eqref{eqSI:expeq_NI} into Eq. \eqref{eqSI:FP_NI}, we find that
\begin{equation}
    \begin{split}
    \frac{1}{D T(t)}\frac{\partial}{\partial t} T(t) &= \frac{1}{r^2 R(r)} \frac{\partial}{\partial r} \left( r^2 \frac{\partial}{\partial r} \right) R(r) + \frac{1}{\Phi(\theta,\phi)} \frac{1}{ r^2 \sin \theta} \frac{\partial}{\partial \theta} \left( \sin \theta \frac{\partial}{\partial \theta} \right) \Phi(\theta,\phi) \\ 
    &+ \frac{1}{\Phi(\theta,\phi)} \frac{1}{ r^2 \sin^2 \theta} \frac{\partial^2}{\partial \phi^2} Y(\theta,\phi) = - \lambda_{n,k}^2,
    \end{split}
\end{equation}
in which the first term of the equality only depends on time, the second term only depends on the radial and angular components, and then both should be equal to a constant $-\lambda_{n,k}^2$ chosen by convenience. Therefore, the time-dependent solution will be given by
\begin{equation}
    \begin{split}
        \frac{1}{DT(t)}\frac{\partial}{\partial t} T(t) = -\lambda_{n,k}^2 \\
        T(t) = C_1 \exp \left( -\lambda_{n,k}^2 D t \right),
    \end{split}
\end{equation}
where $C_1$ is a constant of integration. On the other hand, the radial and angular-dependent part leads to
\begin{equation}
    \begin{split}
        \frac{1}{R(r)} \frac{\partial}{\partial r} \left( r^2 \frac{\partial}{\partial r} \right) R(r) + \lambda_{n,k}^2 r^2 &= - \frac{1}{\Phi(\theta,\phi)} \frac{1}{ \sin \theta} \frac{\partial}{\partial \theta} \left( \sin \theta \frac{\partial}{\partial \theta} \right) \Phi(\theta,\phi) \\ & - \frac{1}{\Phi(\theta,\phi)} \frac{1}{\sin^2 \theta} \frac{\partial^2}{\partial \phi^2} \Phi(\theta,\phi) = n(n+1),
    \end{split}
\end{equation}
in which the first term of the equality only depends on the radius, the second term only depends on the angular components, and then both should be equal to a constant $n(n+1)$ chosen for convenience. Hence, we get two separate equations. The angular equation
\begin{equation}
    - \left( \frac{1}{ \sin \theta} \frac{\partial}{\partial \theta} \left( \sin \theta \frac{\partial}{\partial \theta} \right) - \frac{1}{\sin^2 \theta} \frac{\partial^2}{\partial \phi^2}  \right) \Phi(\theta,\phi) = n(n+1) \Phi(\theta,\phi),
\end{equation}
has a solution already shown to be
\begin{equation}
    \Phi(\theta,\phi) = \sum_{n=0}^{\infty} \sum_{m=-n}^{n} C_{n,m} Y_n^m(\theta, \phi).
\end{equation}
The radial component equation is given by
\begin{equation}
    \frac{\partial}{\partial r} \left( r^2 \frac{\partial}{\partial r} \right) R(r) = - \left[ \lambda_{n,k}^2 r^2 - n(n+1) \right] R(r). \label{eqSI:Bessel_eq}
\end{equation}
It is important to recognize that Eq. \eqref{eqSI:Bessel_eq} is a Sturm–Liouville equation with solution 
\begin{equation}
    R(r) = \sum_{n=0}^{\infty} \sum_{k=1}^{\infty} \left[ D_{n,k} j_n \left( \lambda_{n,k} r \right) + E_{n,k} y_n \left( \lambda_{n,k} r \right) \right], \label{eqSI:R_sol}
\end{equation}
where $j_n(\lambda_{n,k} r)$ and $y_n (\lambda_{n,k} r)$ are the $n^{th}$-order spherical Bessel function of the first and second kind, respectively.
In order to satisfy the Neumann boundary conditions, and having in mind the result from Eq. \eqref{eqSI:R_sol}, we should obey 
\begin{equation}
\begin{split}
    D_{n,k} \lambda_{n,k} j_n' \left( \lambda_{n,k} r_i \right) + E_{n,k} \lambda_{n,k} y_n' \left( \lambda_{n,k} r_i \right) = 0, \\
    D_{n,k} \lambda_{n,k} j_n' \left( \lambda_{n,k} r_f \right) + E_{n,k} \lambda_{n,k} y_n' \left( \lambda_{n,k} r_f \right) = 0,
\end{split}
\end{equation}
from where we conclude that the only possible values of $\lambda_{n,k}$ are those that satisfy the determinant
\begin{equation}
\begin{split}
    \begin{vmatrix}
     j_n' \left( \lambda_{n,k} r_i \right) & y_n' \left( \lambda_{n,k} r_i \right) \\
     j_n' \left( \lambda_{n,k} r_f \right) & y_n' \left( \lambda_{n,k} r_f \right)
    \end{vmatrix} = 0, \\
    j_n' \left( \lambda_{n,k} r_i \right)y_n' \left( \lambda_{n,k} r_f \right) - y_n' \left( \lambda_{n,k} r_i \right) j_n' \left( \lambda_{n,k} r_f \right) = 0. \label{eq:eigenprob}
\end{split}
\end{equation}
Thus, the eigenvalues $\lambda_{2,k}$ will be the positive roots of Eq. \eqref{eq:eigenprob}, such that the solution will be expanded in terms of their corresponding eigenfunctions. Here, we ignore the the trivial solution $\lambda_{n,k} = 0$.

Because of the impenetrability condition given by $R'(0) = 0$, all $E_{n,k}$ coefficients are zero since the spherical Bessel functions of the second-kind $y_n(r)$ does not obey that condition (the solution blows up when $r_i \rightarrow 0$). 
Finally, the solution for $\rho(r, \theta, \phi,t)$ is given by
\begin{equation}
    \rho(\textbf{r}, t) = \sum_{n=0}^{\infty} \sum_{m=-n}^{n} \sum_{k=1}^{\infty} F_{n,m,k} j_n \left( \lambda_{n,k} r \right)  Y_n^m(\theta, \phi) \exp \left( - \lambda_{n,k}^2 D t \right)
\end{equation}
in which the coefficients $F_{n,m,k}$ are the combination (product) of all the previous coefficients for each separable solution, and they depend on the initial condition $\rho(\textbf{r},0)$ as
\begin{equation}
    \begin{split}
        F_{n,m,k} &= \frac{ \int_0^{2\pi} \int_0^{\pi} \int_{r_i}^{r_f} \rho(\textbf{r},0) j_n \left( \lambda_{n,k} r \right) Y_n^m(\theta, \phi) r^2 \sin \theta d\theta d\phi }{\int_0^{2\pi} \int_0^{\pi} \int_{r_i}^{r_f} | j_n (\lambda_{n,k} r) Y_n^m(\theta, \phi) |^2 r^2 \sin \theta d\theta d\phi}.
    \end{split}
\end{equation}

The initial distribution of the system is given by the delta functions at a initial configuration $\{ r_0, \theta_0, \phi_0 \}$ and the corresponding closure relationships, i.e.,
\begin{equation}
    \begin{split}
        \rho(\textbf{r},0) &= \frac{\delta(r - r_0) \delta(\theta - \theta_0) \delta(\phi-\phi_0)}{r^2 \sin \theta} p(\textbf{r}_{0}) \\
        &= p(\textbf{r}_{0}) \sum_{n=0}^{\infty} \sum_{m=-n}^{n} \sum_{k=1}^{\infty} Y_{n}^{m*}(\theta,\phi) Y_{n}^{m}(\theta_0,\phi_0) \frac{j_n \left( \lambda_{n,k} r \right) j_n \left( \lambda_{n,k} r_0 \right)}{N_{n,k}},
    \end{split}
\end{equation}
in which the normalization constant (for Neumann boundary conditions) is given by
\begin{equation}
\begin{split}
    N_{n,k} = \int_{r_i}^{r_f} j_n \left( \lambda_{n,k} r \right) j_n \left( \lambda_{n,k} r \right) r^2 dr &= \\ \frac{r_f^3}{2} \left( 1 - \frac{n(n+1)}{(r_f \lambda_{n,k})^2} \right) j_n \left( \lambda_{n,k} r_f \right) - \frac{r_i^3}{2} \left( 1 - \frac{n(n+1)}{(r_i \lambda_{n,k})^2} \right) j_n \left( \lambda_{n,k} r_i \right) \label{eqSI:ortho_j1}.
\end{split}
\end{equation}
The probability $p(r_0,\theta_0,\phi_0)$ of having the initial condition configuration at equilibrium is given by
\begin{equation}
    \begin{split}
        p(\textbf{r}_{0}) &= \frac{p(r_0) p(\theta_0) p(\phi_0)}{r_0^2 \sin \theta_0}\\
        &= \frac{1}{r_0^2 \sin \theta_0} \frac{r_0^2}{\int_{r_i}^{r_f} r^2 dr} \frac{\sin \theta_0}{\int_0^{\pi} \sin \theta d \theta} \frac{1}{\int_0^{2\pi} d\phi} \\
        & = \frac{3}{4\pi (r_f^3-r_i^3)},
    \end{split}
\end{equation}
such that
\begin{equation}
    \rho(\textbf{r},0) = \frac{3}{4\pi (r_f^3-r_i^3)} \sum_{n=0}^{\infty} \sum_{m=-n}^{n} \sum_{k=1}^{\infty}  Y_{n}^{m*}(\theta,\phi) Y_{n}^{m}(\theta_0,\phi_0) \frac{j_n \left( \xi_{n,k} r \right) j_n \left( \xi_{n,k} r_0 \right)}{N_{n,k}}.
\end{equation}

In NMR relaxation, we are only interested in the second moment ($n=2$), and we will also assume that the system is isotropic ($m=0$). Thus, the general solution $\rho(\textbf{r},t)$ simplifies to
\begin{equation}
    \rho(\textbf{r}, t) = \sum_{k=1}^{\infty}F_{2,0,k} j_2 \left( \lambda_{2,k} r \right)  Y_2^0(\theta, \phi) \exp \left( - \lambda_{2,k}^2 D t \right),
\end{equation}
with the corresponding initial condition
\begin{equation}
    \rho(\textbf{r},0) = \frac{3}{4\pi (r_f^3-r_i^3)} \sum_{k=1}^{\infty}  Y_{2}^{0*}(\theta,\phi) Y_{2}^{0}(\theta_0,\phi_0) \frac{j_2 \left( \lambda_{2,k} r \right) j_2 \left( \lambda_{2,k} r_0 \right)}{N_{2,k}}.
\end{equation}
Thus, the coefficients are given by
\begin{equation}
    \begin{split}
        F_{2,0,k} &= \frac{ \int_0^{2\pi} \int_0^{\pi} \int_{r_i}^{r_f} \rho(\textbf{r},0) j_2 \left( \lambda_{2,k} r \right) Y_2^0(\theta, \phi) r^2 \sin \theta d\theta d\phi }{\int_0^{2\pi} \int_0^{\pi} \int_{r_i}^{r_f} | j_2 (\lambda_{2,k} r) Y_2^0(\theta, \phi) |^2 r^2 \sin \theta d\theta d\phi}.
    \end{split}
\end{equation}

Considering the orthogonality relationship of spherical harmonics and spherical Bessel functions, the coefficients of the solution will be given by
\begin{equation}
    \begin{split}
        F_{2,0,k} &= \frac{3 Y_{2}^{0}(\theta_0,\phi_0)}{2\pi r_f^6} \frac{\int_0^{2\pi} \int_0^{\pi} Y_2^0(\theta, \phi) Y_{2}^{0*}(\theta,\phi) \sin \theta d\theta d\phi  \int_{r_i}^{r_f} r^2 j_2 \left( \lambda_{2,k} r \right) \sum_{i=1}^{\infty}  \frac{j_2 \left( \xi_{2,i} r \right) j_2 \left( \xi_{2,i} r_0 \right)}{N_{2,i}} dr}{\int_0^{2\pi} \int_0^{\pi}  |Y_2^0(\theta, \phi) |^2 \sin \theta d\theta d\phi \int_{r_i}^{r_f} | j_2 \left( \lambda_{2,k} \right)|^2 r^2 dr} \\
        &= \frac{3 Y_{2}^{0}(\theta_0,\phi_0)}{2\pi r_f^6} \frac{\sum_{i=1}^{\infty} \frac{ j_2 \left( \xi_{2,i} r_0 \right) }{N_{2,i}} \int_{r_i}^{r_f} j_2 \left( \lambda_{2,k} r \right) j_2 \left( \xi_{2,i} r \right) r^2 dr}{\int_{r_i}^{r_f} | j_2 \left( \lambda_{2,k} r \right)|^2 r^2 dr} \\
        &= \frac{3}{4\pi (r_f^3-r_i^3)}  \frac{Y_{2}^{0}(\theta_0,\phi_0) j_2 \left( \lambda_{2,k} r_0 \right)}{N_{2,k}}.
    \end{split}
\end{equation}

The final solution to the time-dependent equilibrium distribution probability between the two dipoles undergoing NMR relaxation at constant distance is given by
\begin{equation}
    \rho(\textbf{r}, t) = \frac{3}{4\pi (r_f^3 - r_i^3)}  Y_{2}^{0}(\theta_0,\phi_0) Y_2^0(\theta, \phi)  \sum_{k=1}^{\infty} \frac{j_2 \left( \lambda_{2,k} r_0 \right)}{N_{2,k}}  j_2 \left( \lambda_{2,k} r \right) \exp \left( - \lambda_{2,k}^2 D t \right). \label{eqSI:final_rho}
\end{equation}

Using this result in Eq. (5) and integrating over all possible coordinates in the system, we find that
\begin{equation}
    \begin{split}
        G(t) & = \frac{4\pi}{5} \left( \frac{\mu_0}{4 \pi} \right)^2 \hbar^2 \gamma_I^2 \gamma_S^2 S(S+1) \frac{3}{4\pi (r_f^3-r_i^3)}  \\ & \int_0^{2\pi} \int_0^{\pi} Y_2^0 \left( \theta_0, \phi_0 \right) Y_2^0 \left( \theta_0, \phi_0 \right) \sin \theta_0 d\theta_0 d\phi_0 \int_0^{2\pi} \int_0^{\pi} Y_{2}^{0*}(\theta,\phi) Y_{2}^{0}(\theta,\phi) \sin \theta d\theta d\phi \\ & \int_{r_i}^{r_f} \int_{r_i}^{r_f} \sum_{k=1}^{\infty}  r^2 \frac{1}{r^3} r_0^2 \frac{1}{r_0^3} \exp \left( - \lambda_{2,k}^2 D t \right) \frac{j_2 \left( \lambda_{2,k} r \right) j_2 \left( \lambda_{2,k} r_0 \right)}{N_{2,k}} dr dr_0 , \label{eqSI:Gm_NI}
    \end{split}
\end{equation}
and by the orthogonality of the spherical harmonics, we obtain that
\begin{equation}
    \begin{split}
        & G(t) = \left( \frac{\mu_0}{4 \pi} \right)^2 \hbar^2 \gamma_I^2 \gamma_S^2 S(S+1) \frac{3}{5 (r_f^3 - r_i^3)} \sum_{k=1}^{\infty} \exp \left( - \lambda_{2,k}^2 D t \right) \int_{r_i}^{r_f} \int_{r_i}^{r_f} \frac{1}{rr_0} \frac{j_2 \left( \lambda_{2,k} r \right) j_2 \left( \lambda_{2,k} r_0 \right)}{N_{2,k}} dr dr_0. \label{eqSI:NI}
    \end{split}
\end{equation}
Observe that now the NMR autocorrelation function has a multi-exponential decay over
an infinite set of discrete characteristic decay times. It is important to emphasize that the quantity in Eq. \eqref{eqSI:NI} should be computed under the definition of the inner-product (orthogonality) in Eq. \eqref{eqSI:ortho_j1} that arise from the boundary conditions, such that the calculated quantities are projected into the correct orthogonal eigenvector space.

\clearpage
\subsection{Pad\'e-Laplace inversion of $G(t)$}

\begin{figure}[!ht]
    \centering
        \includegraphics[width=0.49\textwidth]{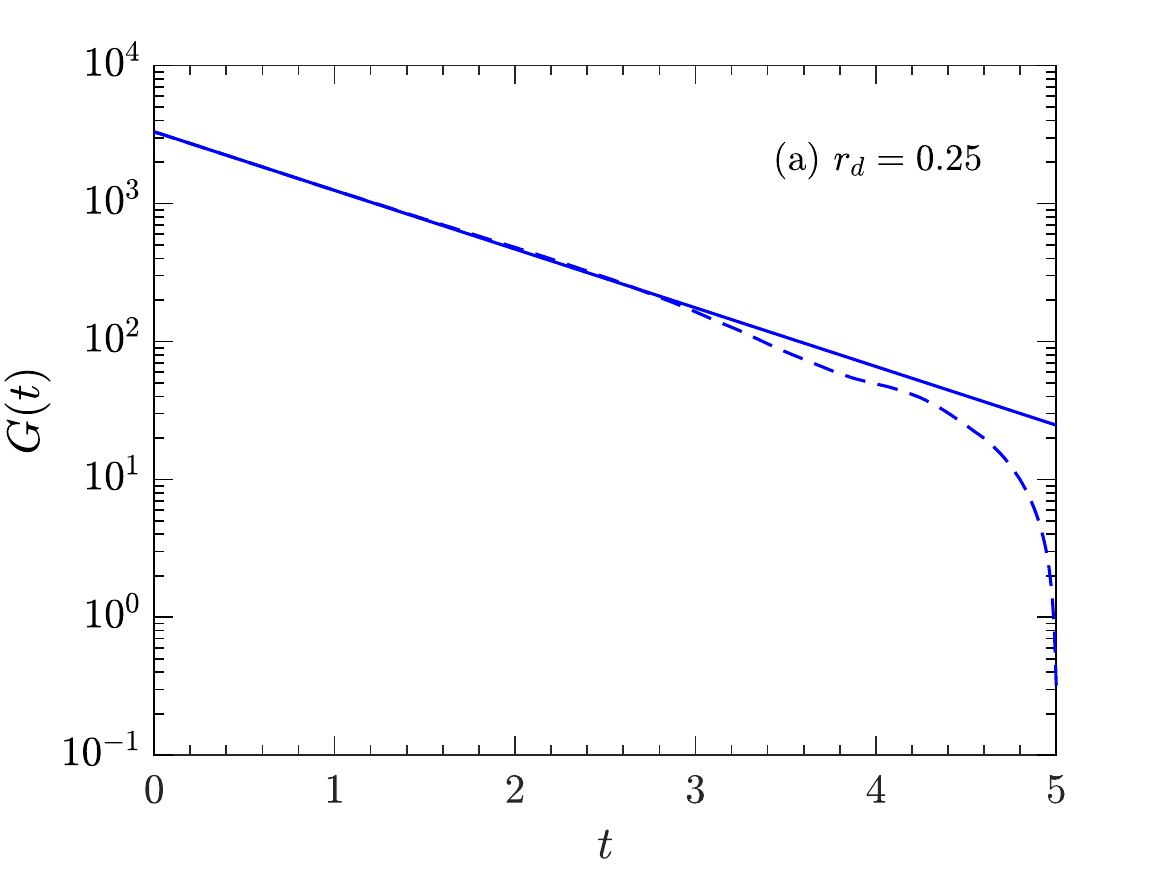}
        \includegraphics[width=0.49\textwidth]{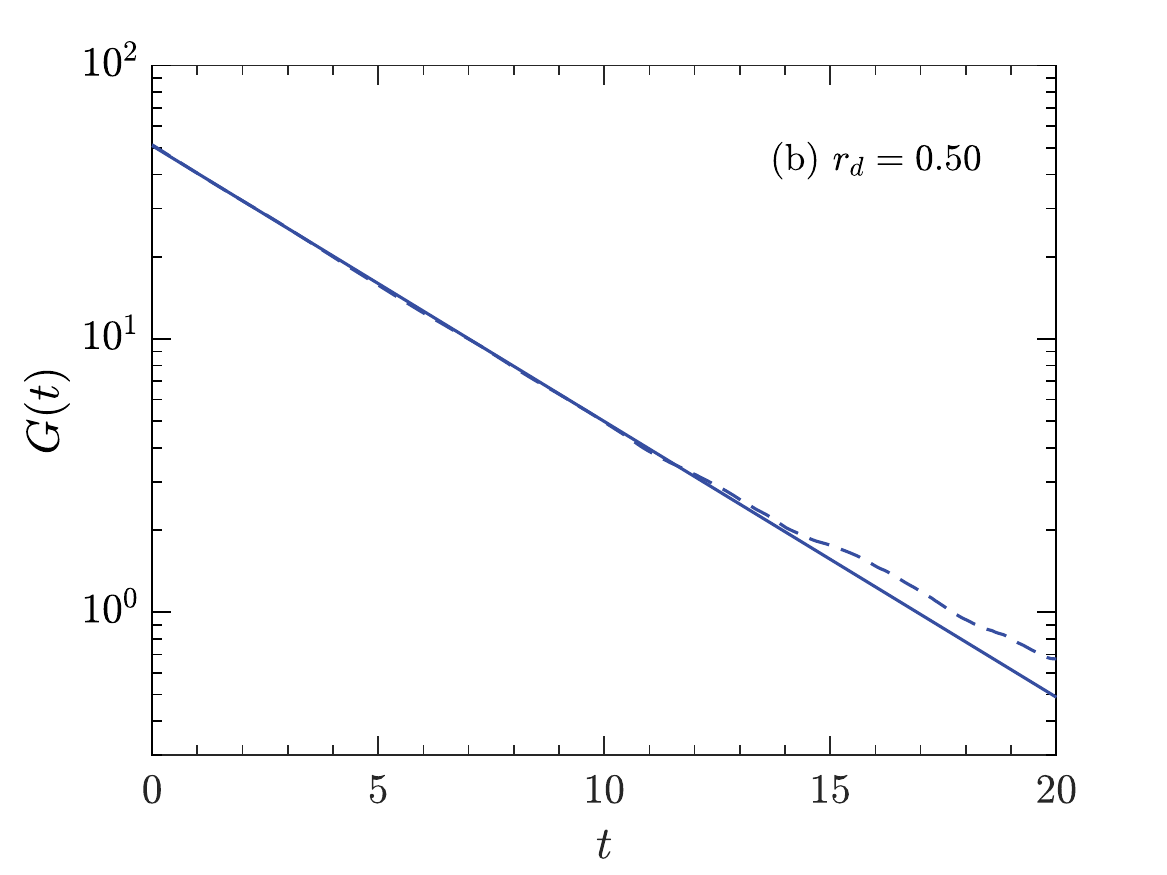}
        \includegraphics[width=0.49\textwidth]{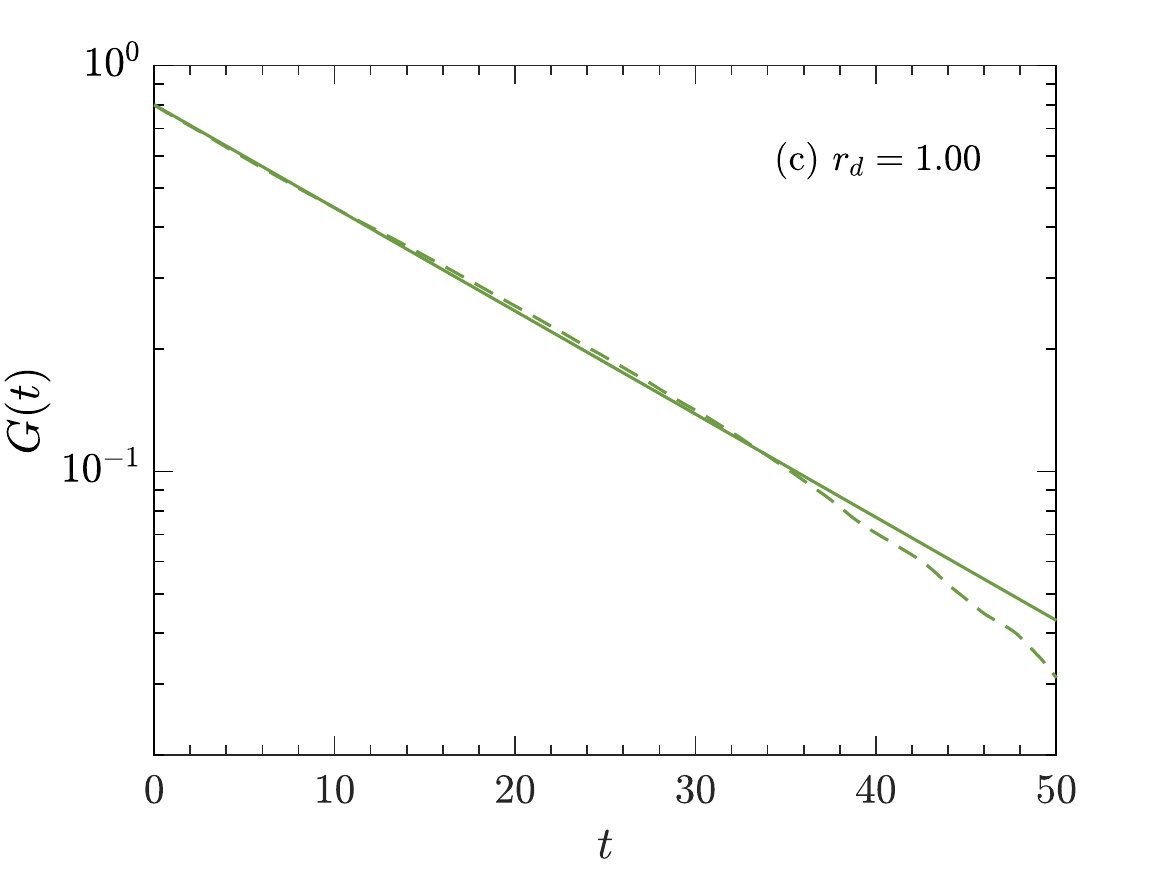}
        \includegraphics[width=0.49\textwidth]{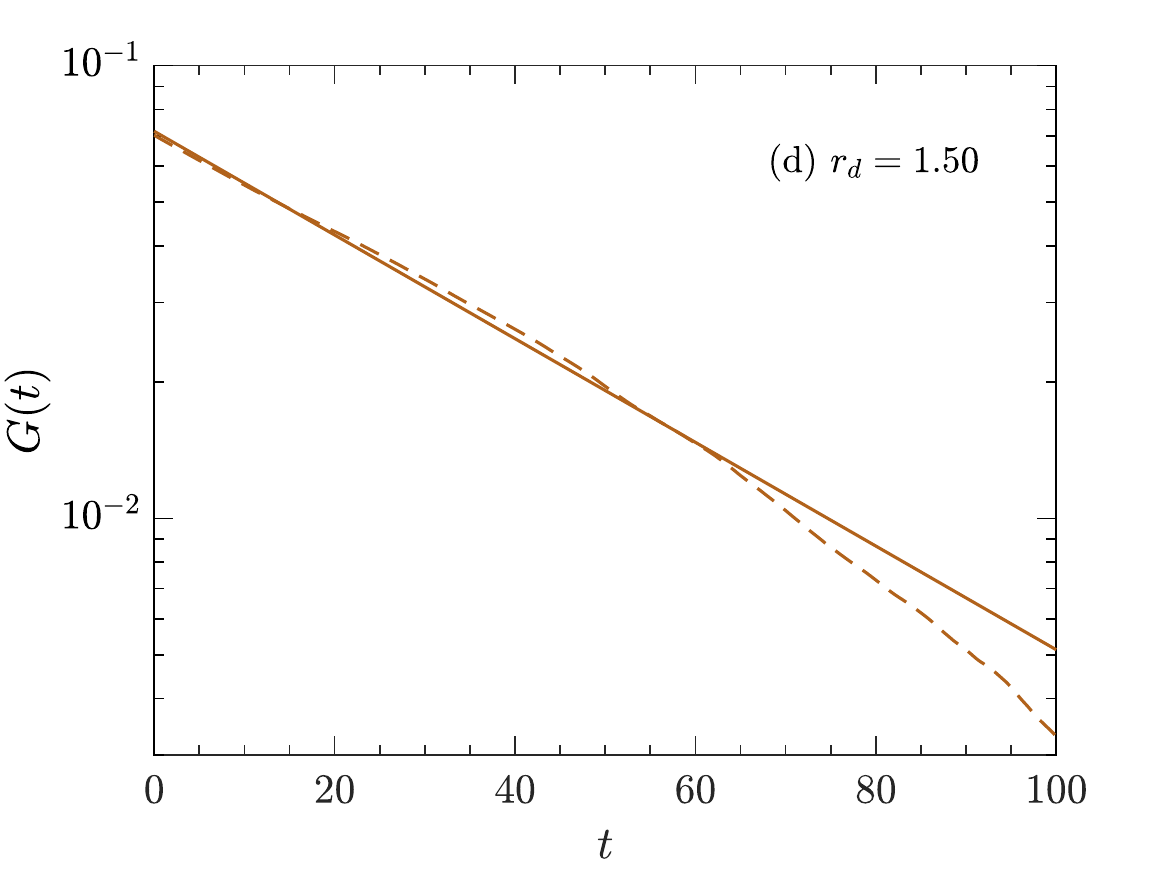}
        \includegraphics[width=0.49\textwidth]{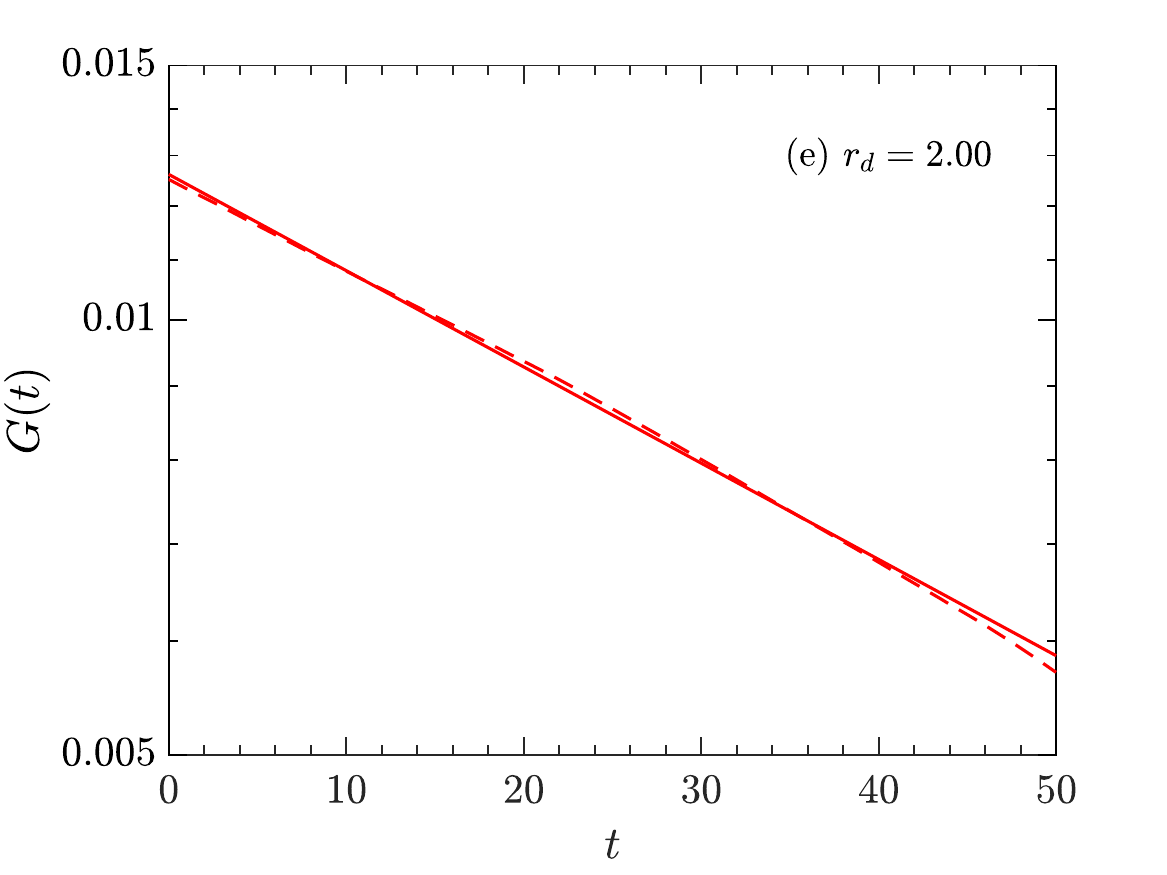}
    \caption{Reduced NMR dipole-dipole autocorrelation function between two dipoles for difference fixed distances $r_d$, with $f=10$ and $T=1.00$. The dashed lines represent the MD simulation results, while the solid lines represent the reconstruction of the signal given by the Pad\'e-Laplace inversion method.}
    \label{fig:FigS1}
\end{figure}

\begin{figure}[!ht]
    \centering
        \includegraphics[width=0.60\textwidth]{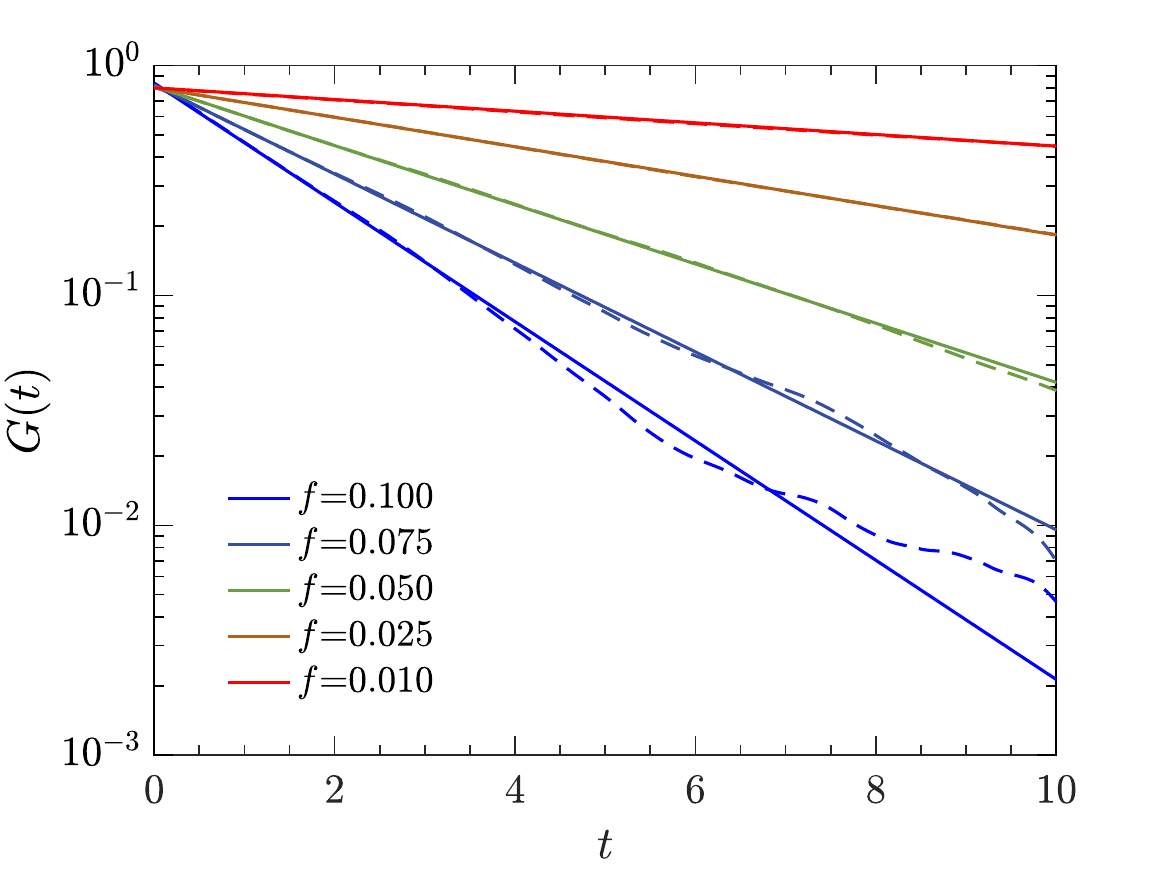}
    \caption{Reduced NMR dipole-dipole autocorrelation function between two dipoles at a fixed distance $r_d=1.00$ and with different friction constants $f$, with $T=1.00$. The dashed lines represent the MD simulation results, while the solid lines represent the reconstruction of the signal given by the Pad\'e-Laplace inversion method.}
    \label{fig:FigS2}
\end{figure}

\begin{figure}[!ht]
    \centering
        \includegraphics[width=0.60\textwidth]{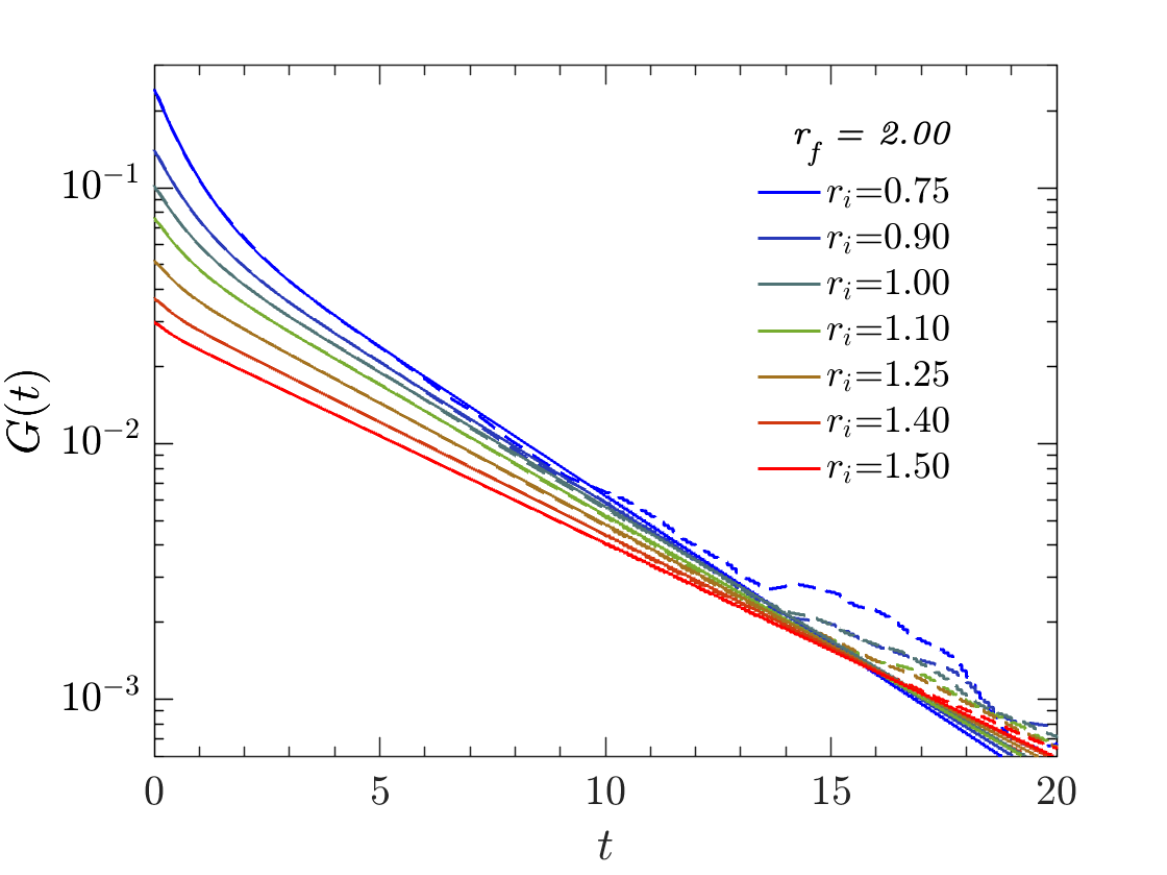}
    \caption{Reduced NMR dipole-dipole autocorrelation function between two dipoles with the second one diffusing on a thick spherical shell, for different values of inner radius $r_i$ and with $r_f=2.00$, $f=10$, and $T=1.00$. The dashed lines represent the MD simulation results, while the solid lines represent the reconstruction of the signal given by the Pad\'e-Laplace inversion method.}
    \label{fig:FigS3}
\end{figure}

In this section, we present the NMR dipole-dipole autocorrelation function $G(t)$ from MD simulations and the corresponding reconstruction of the signal using the Pad\'e-Laplace inversion method through the captured modes. Figure \ref{fig:FigS1} and \ref{fig:FigS2} compile the reduced NMR dipole-dipole autocorrelation functions for two dipoles at different fixed distances (and constant friction) and for different frictions rates (and constant distance), respectively. Figure \ref{fig:FigS3} shows the reduced NMR dipole-dipole autocorrelation functions for the case of a dipole diffusing on a thick spherical shell around the first dipole, for different reduced inner radius $r_i$. Overall, we observe very good agreement between the original autocorrelation functions from MD simulations and the corresponding reconstructed signal using the Pad\'e-Laplace inversion method. Notice that the autocorrelation functions from MD simulations present noise, which increases with time. On the other hand, the reconstructed signal from the captured modes is smooth and only contains the exponential contributions obtained through the Pad\'e-Laplace inversion (no noise).

\clearpage
\subsection{Numerical approximations for integral computations}

When computing the result in Eq. \eqref{eqSI:NI}, we have to observe that the inner-product definition is not the regular Euclidean definition commonly used in algebraic calculations and software packages. Observe that such integrals should be strictly evaluated in a non-Euclidean space respecting the inner-product definition (orthogonality) given by Eq. \eqref{eqSI:ortho_j1}, which arises from the reflective boundary conditions.

\begin{figure}[!t]
    \centering
        \includegraphics[width=0.65\textwidth]{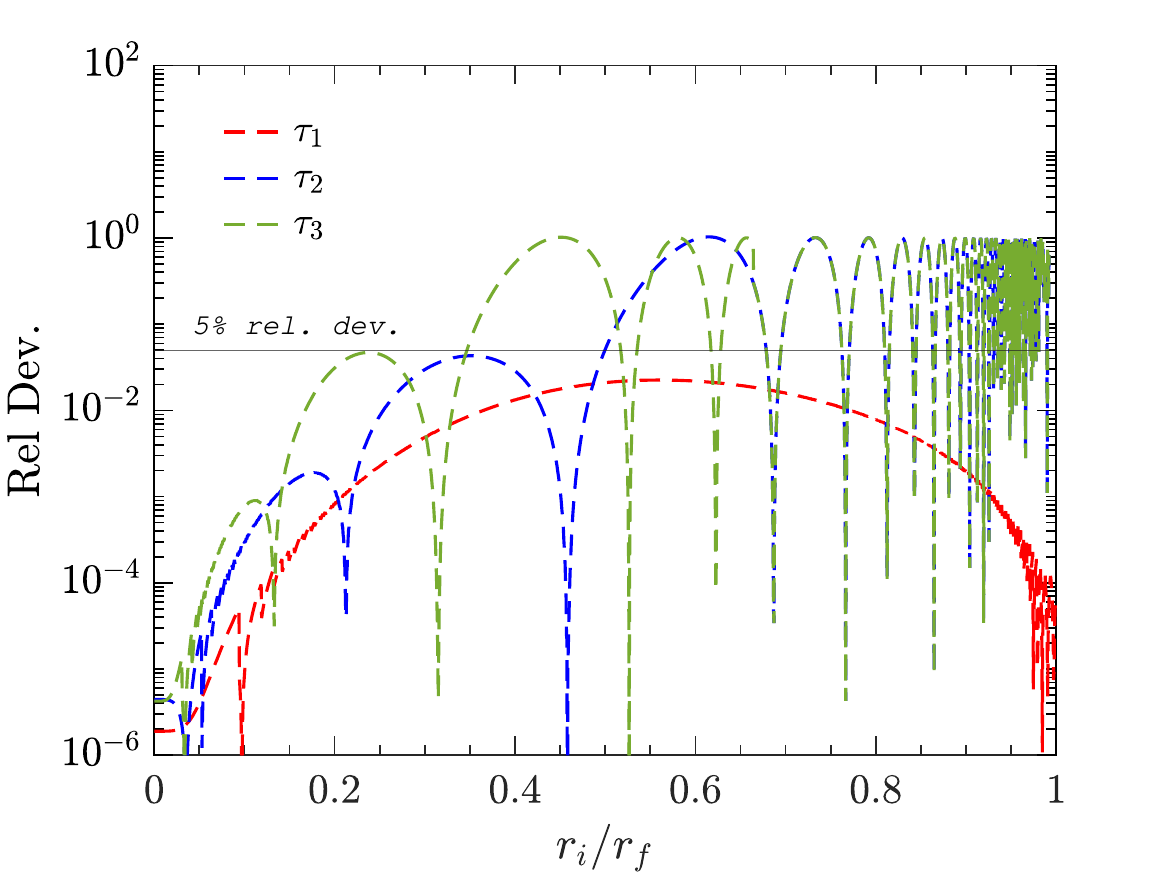}
    \caption{Relative deviation on the computation of the orthogonality integral (Eq. \eqref{eqSI:ortho_j1}) using the regular Euclidean inner-product and the non-Euclidean inner-product defined by the boundary conditions, across $r_i/r_f$ and with $r_f=2.00$.}
    \label{fig:FiS4}
\end{figure}

For the particular problem under investigation, we observed that for the first molecular mode ($k=1$) the orthogonality integral evaluated with the Euclidean inner-product definition is approximately the same as the analytical evaluation in the corresponding non-Euclidean space, within less than 5\% numerical deviation across $0<r_i/r_f<1$. Hence, it is expected that numerical errors for the computation of the integral in Eq. \eqref{eqSI:NI} (to which no analytical solution is available) due to deviations from the non-Euclidean inner-product will be negligible. However, for higher order molecular modes ($k\geq2$), the deviations of the inner-product integral from the Euclidean definition with respect to the exact non-Euclidean computation is not negligible (above 5\% in relative error) for certain values of $r_i/r_f$, and hence so will the quantity in Eq. \eqref{eqSI:NI} if computed with the regular Euclidean inner-product. 
Figure \ref{fig:FiS4} shows the relative error in the orthogonality condition integrals between using the Euclidean inner-product and the analytical solution (non-Euclidean solution), for different molecular modes and across the entire range of $r_i/r_f$ and with $r_f=2.00$.

\begin{figure}[!t]
    \centering
        \includegraphics[width=0.49\textwidth]{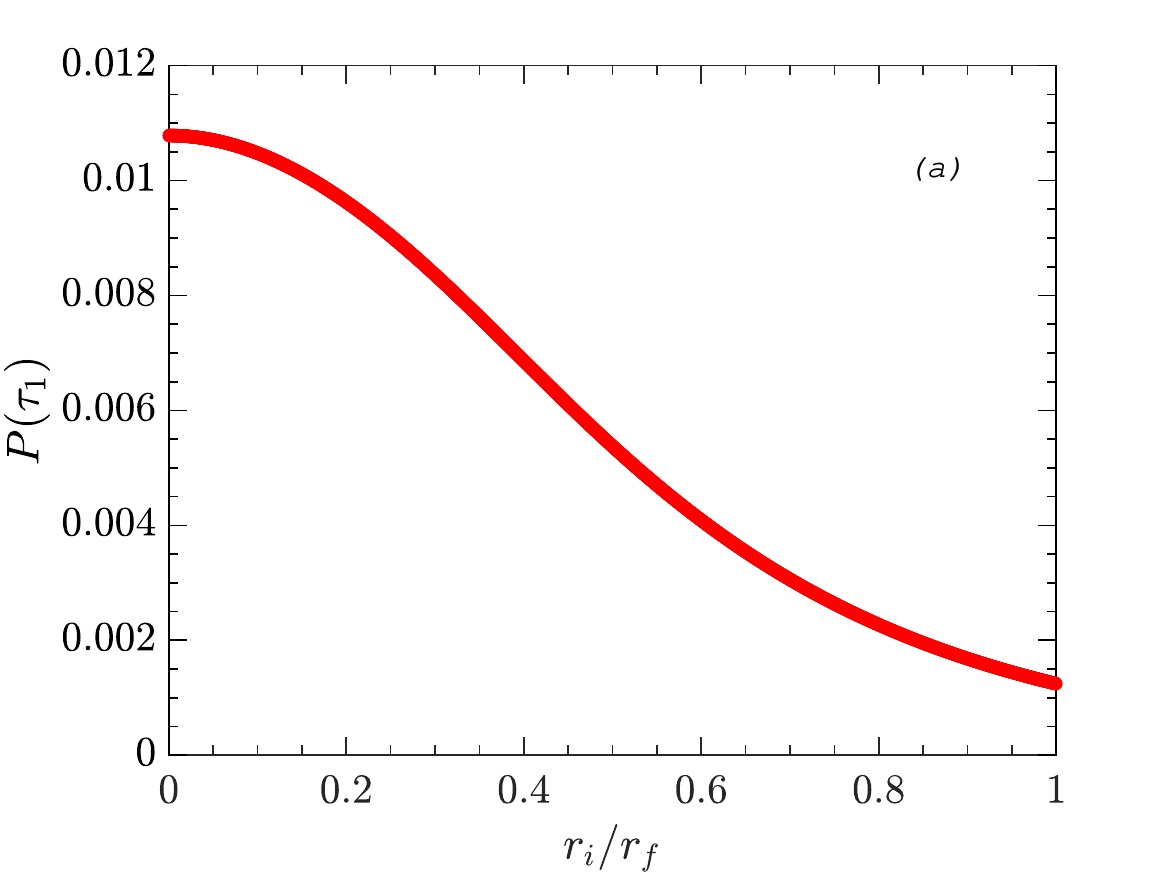}
        \includegraphics[width=0.49\textwidth]{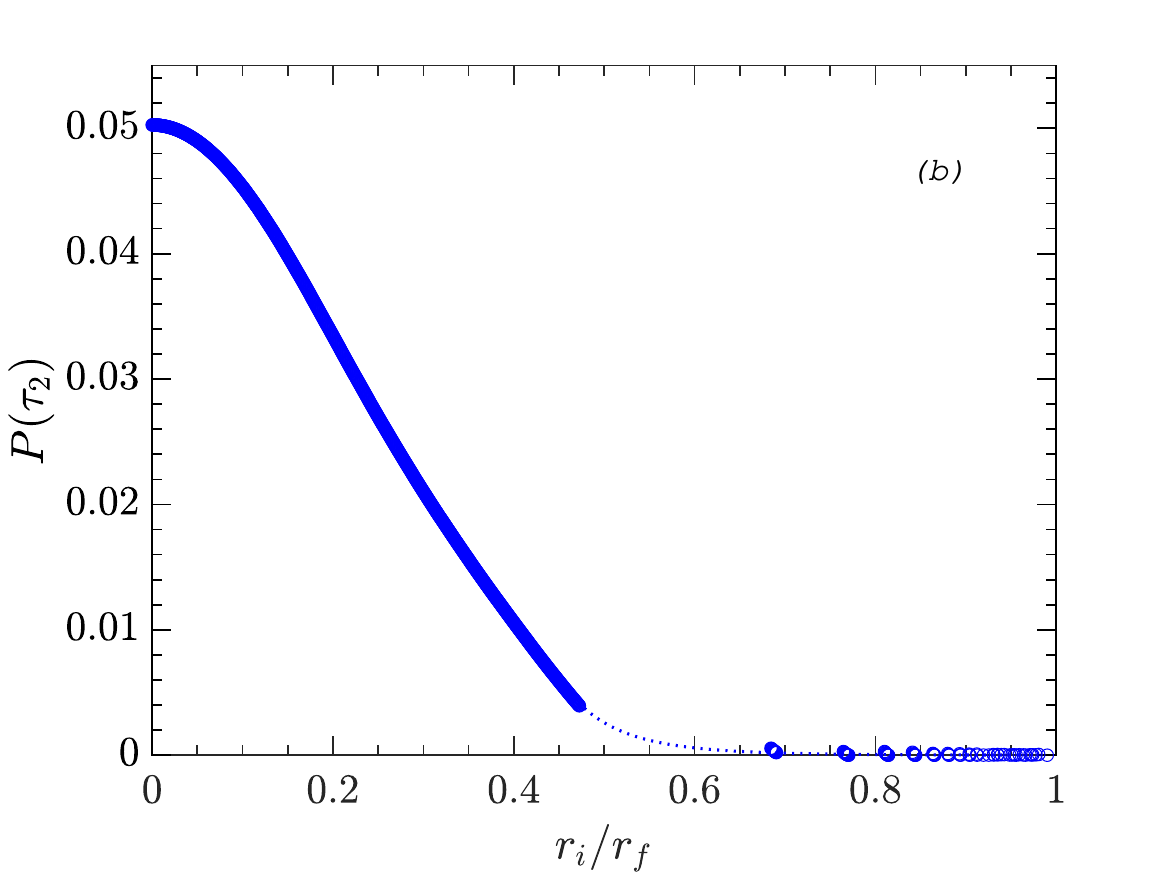}
        \includegraphics[width=0.49\textwidth]{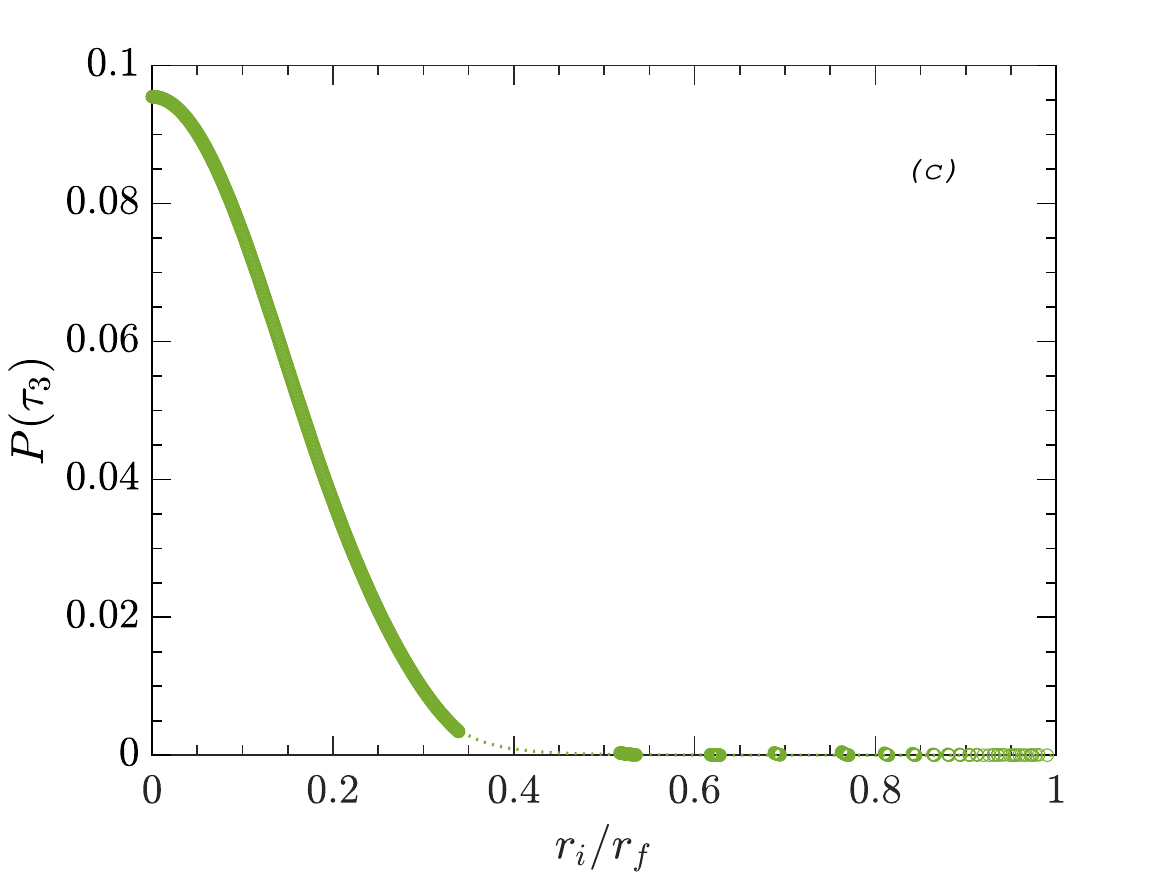}
    \caption{Molecular modes reduced amplitudes $P(\tau_k)$ for the (a) first, (b) second, and (c) third modes across $r_i/r_f$, for the case of $r_f=2.00$ with $f=10$ and $T=1.0$. The circle symbols represent the theoretical predictions using the regular Euclidean inner-product but with relative error in the orthogonality integral below 5\%, while the dotted lines are the smooth interpolation of the theoretical predictions.}
    \label{fig:FigS5}
\end{figure}

To by-pass this limitation, and having in mind that no analytical solution to the integration to Eq. \eqref{eqSI:NI} in the corresponding non-Euclidean space is available, we have computed the integrals using the regular inner product definition for the cases where the relative deviation on the orthogonality condition was less than 5\%. When this error becomes greater than that, we did not compute the integrations numerically.
However, given that the deviation is oscillatory and we are able to compute the integrals for points before and after high error regions, we can interpolate the curve with a smooth function, represented through dotted lines in our plots. This is a valid argument since the actual solution is expected to be monotonic.
Figure \ref{fig:FigS5} shows the points where the integration is numerically accurate (deviation below 5\%), and the corresponding interpolation curve. For simplicity, after a certain point, we only included the interpolation curve in the paper plots. We highlight, however, that there are accurate data points in the dotted interpolation region to support the proposed trend.

\clearpage
\subsection{Molecular modes via theory and MD simulations}

The raw data for the molecular modes obtained by both our theory and MD simulations are presented in Tables \ref{Tb:Table1}, \ref{Tb:Table2}, and \ref{Tb:Table3} for the cases of different constant radii $r_d$, $r_i/r_f = 0.375$, and $r_i/r_f = 0.550$, respectively.

\begin{table}[h!]
\centering
\caption{Molecular modes of NMR relaxation in LJ-reduced units for dipole pairs at different fixed distances $r_d$, with constant friction $f = 100$ and $T=1.00$ \newline}
\begin{tblr}{
  cells = {c},
  cell{1}{1} = {r=2}{},
  cell{1}{2} = {c=2}{},
  cell{1}{4} = {c=2}{},
  hline{1,3,8} = {-}{},
  hline{2} = {2-5}{},
}
$r_d$ & $\tau$    &                & $P(\tau)$   &                \\
     & Theory & MD simulations & Theory    & MD simulations \\
0.25 & 1.04   & 1.02           & 3.259E+02 & 3.300E+02      \\
0.50 & 4.17   & 4.30           & 5.093E+00 & 5.070E+00      \\
1.00 & 16.67  & 17.09          & 7.958E-02 & 7.969E-02      \\
1.50 & 37.50  & 37.93          & 6.986E-03 & 7.162E-03      \\
2.00 & 66.67  & 65.19          & 1.243E-03 & 1.255E-03      
\end{tblr} \label{Tb:Table1}
\end{table}

\begin{table}
\centering
\caption{Molecular modes of NMR relaxation in LJ-reduced units for dipole pairs at non-fixed distances for the case of $r_i=0.75$ and $r_f=2.00$, with constant friction $f=10$ and $T=1.00$ \newline}
\begin{tblr}{
  cells = {c},
  cell{1}{1} = {r=2}{},
  cell{1}{2} = {c=2}{},
  cell{1}{4} = {c=2}{},
  vline{3} = {1}{},
  vline{4} = {2-7}{},
  hline{1,3,8} = {-}{},
  hline{2} = {2-5}{},
}
$k$ & $\tau_k$ &                & $P(\tau_k$) &                \\
  & Theory & MD simulations & Theory    & MD simulations \\
1 & 3.72   & 3.74           & 7.246E-03 & 9.06E-03       \\
2 & 0.81   & 0.73           & 1.306E-02 & 1.53E-02       \\
3 & 0.32   & --             & 1.590E-03 & --             \\
4 & 0.16   & --             & 3.219E-04 & --             \\
5 & 0.09   & --             & 7.147E-07 & --             
\end{tblr} \label{Tb:Table2}
\end{table}

\begin{table}
\centering
\caption{Molecular modes of NMR relaxation in LJ-reduced units for dipole pairs at non-fixed distances for the case of $r_i=1.10$ and $r_f=2.00$, with constant friction $f=10$ and $T=1.00$ \newline}
\begin{tblr}{
  cells = {c},
  cell{1}{1} = {r=2}{},
  cell{1}{2} = {c=2}{},
  cell{1}{4} = {c=2}{},
  vline{3} = {1}{},
  vline{4} = {2-7}{},
  hline{1,3,8} = {-}{},
  hline{2} = {2-5}{},
}
$k$ & $\tau_k$ &                & $P(\tau_k)$ &                \\
  & Theory & MD simulations & Theory    & MD simulations \\
1 & 4.20   & 4.26           & 4.701E-03 & 5.48E-03       \\
2 & 0.62   & 0.66           & 1.192E-03 & 2.18E-03       \\
3 & 0.19   & 0.15           & 5.946E-05 & 1.37E-07       \\
4 & 0.09   & --             & 1.879E-05 & --             \\
5 & 0.05   & --             & 1.296E-05 & --             
\end{tblr} \label{Tb:Table3}
\end{table}

\clearpage
\subsection{Mode contributions to relaxation dispersion}

\begin{figure}[!ht]
    \centering
        \includegraphics[width=0.60\textwidth]{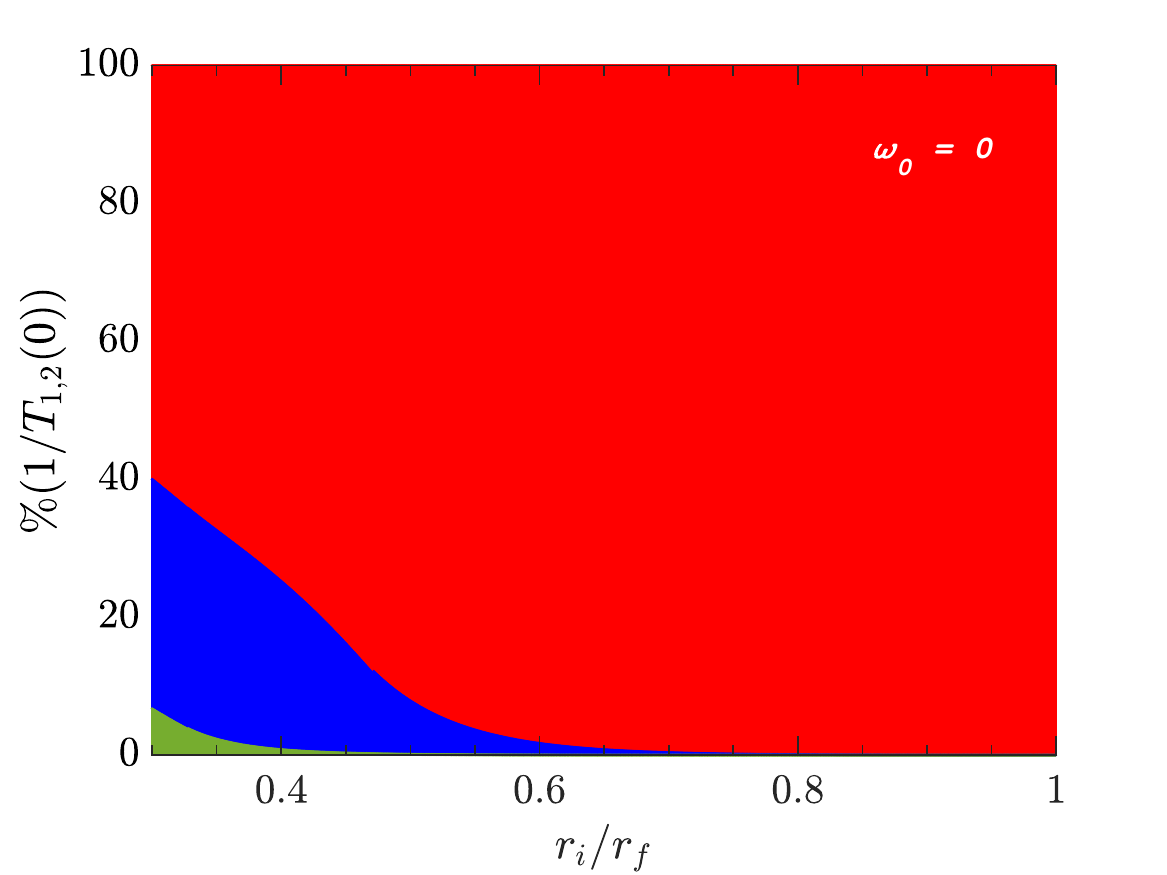}
        \caption{Percentage theoretical contribution to $1/T_{1,2}$ at $\omega_0 = 0$ from the first molecular mode (red), second mode (blue), and third mode (green).}
    \label{fig:FigS6}
\end{figure}

Figure \ref{fig:FigS6} shows the theoretical percentage contribution to the relaxation rate $1/T_{1,2}$ at zero frequency ($\omega_0 = 0$) from different molecular modes of relaxation, over a certain range of $r_i/r_f$. The contribution to $1/T_{1,2}$ at $\omega_0 = 0$ from the $k^{th}$-mode is proportional to $P(\tau_k)\tau_k$. Observe that the first mode dominates $1/T_{1,2}$ at $\omega_0 = 0$ above $r_i/r_f \gtrsim 0.6$, and tends to 100\% at $r_i/r_f= 1$ (i.e., in the BPP limit).

We can also estimate the mode contributions to $1/T_1(\omega_0)$ dispersion (i.e., as a function of frequency) for like spins, such that \cite{valiyaparambathu:jpcl2023}:
\begin{align}
	\frac{1}{T_{1}(\omega_0)} = 2 \mathlarger{\mathlarger{\sum}}_k \,P(\tau_k) \left[\frac{\tau_k}{1+(\omega_0\tau_k)^2 } + \frac{4\tau_k}{1+(2\omega_0\tau_k)^2 }\right]. \label{eq:T1_dispersion}
\end{align}

\begin{figure}[!ht]
    \centering
        \includegraphics[width=0.60\textwidth]{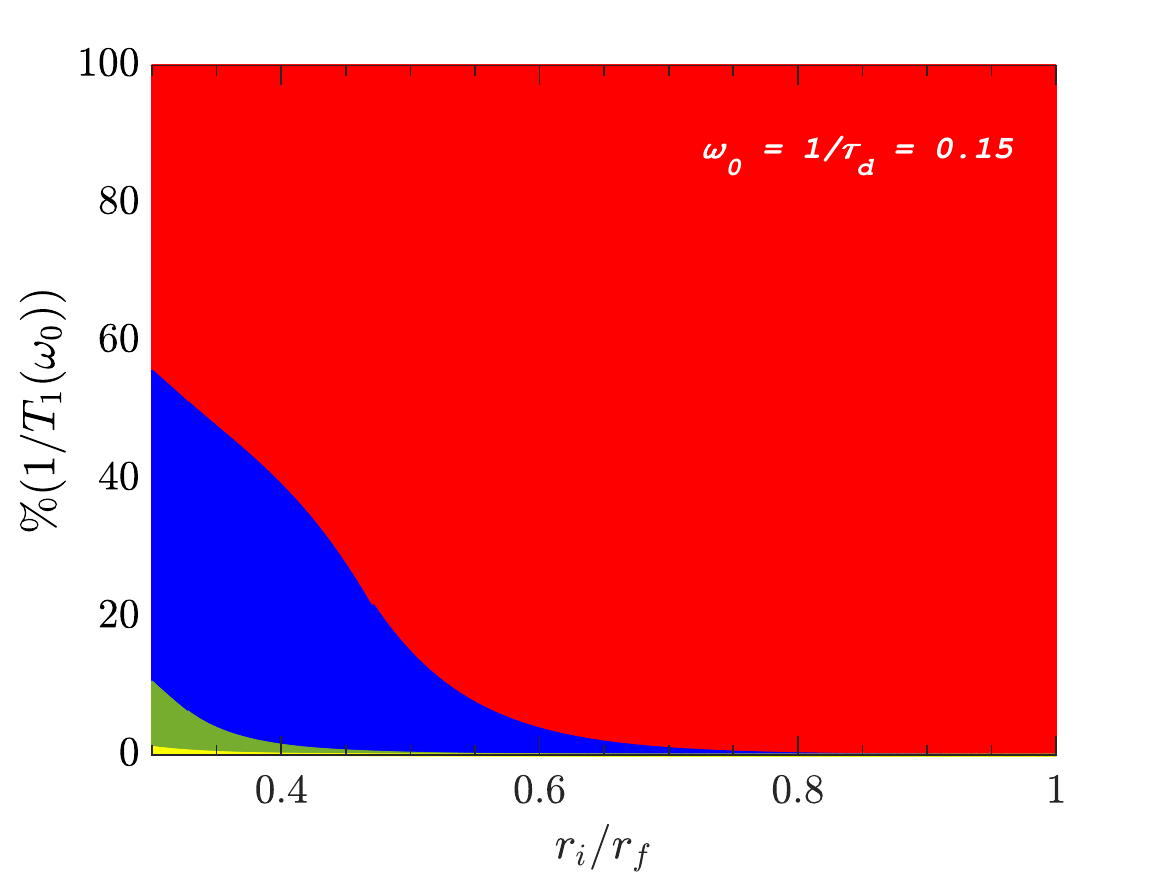}
        \caption{Percentage theoretical contribution to $1/T_{1}(\omega_0)$ at $\omega_0 = 1/\tau_d = 0.15$ that arise from the first molecular mode (red), second mode (blue), third mode (green), and fourth mode (yellow), with $r_f=2.00$. Within this range of $r_i/r_f$, the contribution of high order modes is negligible.}
    \label{fig:FigS7}
\end{figure}

\begin{figure}[!ht]
    \centering
        \includegraphics[width=0.60\textwidth]{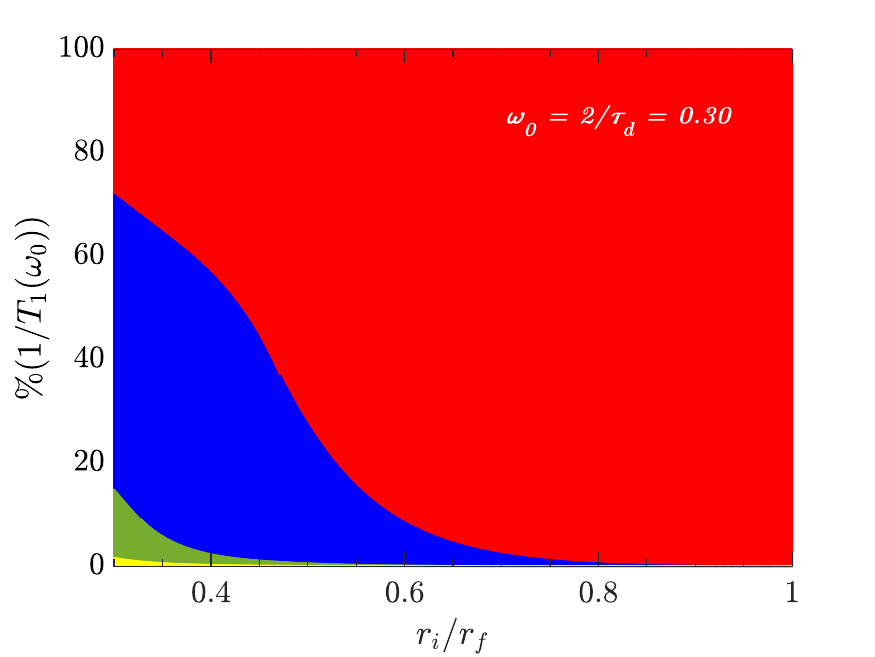}
        \caption{Percentage theoretical contribution to $1/T_{1}(\omega_0)$ at $\omega_0 = 2/\tau_d = 0.30$ that arise from the first molecular mode (red), second mode (blue), third mode (green), and fourth mode (yellow), with $r_f=2.00$. Within this range of $r_i/r_f$, the contribution of high order modes is negligible.}
    \label{fig:FigS8}
\end{figure}

Figures \ref{fig:FigS7} and \ref{fig:FigS8} show the theoretical percentage contribution from different molecular modes to $1/T_{1}(\omega_0)$ assuming two like-spin particles at $\omega_0=1/\tau_d=0.15$ and $\omega_0=2/\tau_d=0.30$, respectively, over a certain range of $r_i/r_f$ with $r_f=2.00$. Observe that as  $\omega_0$ increases, the relative importance of higher order modes ($k \geq 2$) to $1/T_{1}(\omega_0)$ also increases. 

Meanwhile $1/T_{2}(\omega_0)$ shows only minor dispersion. 

Note that for the case of unlike spins, the equivalent expression can be determined as such \cite{Pinheiro2022}
\begin{align}
	\frac{1}{T_{1}(\omega_0)} = 2 \mathlarger{\mathlarger{\sum}}_k \,P(\tau_k) \left[\frac{\tau_k}{1+(\omega_0\tau_k)^2 } + \frac{7/3 \,\tau_k}{1+(\omega_e \tau_k)^2 }\right], \label{eq:T1_dispersion_unlike}
\end{align}
where $\omega_e = 658 \omega_0$ is the resonance frequency of the electron spin.

\clearpage

%